\documentclass[apj,tighten,iop,twocolumn]{emulateapj}
\usepackage[colorlinks=true,citecolor=blue,linkcolor=blue]{hyperref}

\newcommand{\hunits}{$\rm km \ s^{-1} Mpc^{-1}$}
\newcommand{\kmps}{km s$^{-1}$}
\newcommand{\lsun}{$L_{\odot}$}
\newcommand{\lunits}{erg~s$^{-1}$}
\newcommand{\lnuunits}{erg~s$^{-1}$~Hz$^{-1}$}

\newcommand{\mstar}{$M_{\ast}$}
\newcommand{\msun}{$M_{\odot}$}
\newcommand{\msuny}{\msun yr$^{-1}$}

\newcommand{\runits}{cts s$^{-1}$}

\newcommand{\aeff}{$A_{\rm eff}$}
\newcommand{\aeffc}{$A_{\rm eff,r}$}
\newcommand{\airac}{$\alpha_{\rm IRAC}$}

\newcommand{\dflux}{$\Delta f_{\lambda}(uvm2 - NUV) / f_{\lambda} (NUV))$}

\newcommand{\iruv}{$\nu L_{\nu,\rm 24\mu{\rm m}} /\nu L_{\nu,\it uvw2}$}

\newcommand{\lamc}{$\lambda_{\rm c}$}
\newcommand{\lameff}{$\lambda_{\rm eff}$}

\newcommand{\lamr}{$\lambda_{\rm r}$}

\newcommand{\lnu}{$L_{\nu}$}
\newcommand{\lnuks}{$L_{\nu, K_s}$}

\newcommand{\lnuwtwo}{$L_{\nu,\it uvw2}$}
\newcommand{\lnutf}{$L_{\nu,\rm 24\mu{\rm m}}$}

\newcommand{\mhi}{$M_{\rm H I}$}

\newcommand{\sfrir}{SFR$_{\rm IR}$}
\newcommand{\sfrtot}{SFR$_{\rm TOTAL}$}
\newcommand{\sfrtf}{SFR$_{24\mu{\rm m}}$}
\newcommand{\sfruv}{SFR$_{\rm UV}$}
\newcommand{\sfrwtwo}{SFR$_{uvw2}$}

\newcommand{\x}{X-ray}

\newcommand{\zmed}{$z_{\rm med}$}

\newcommand{\ha}{H$\alpha$}

\newcommand{\hone}{H{\sc \,i}}
\newcommand{\htwo}{H{\sc \,ii}}

\newcommand{\chandra}{{\it Chandra}}
\newcommand{\fuv}{{\it FUV}}
\newcommand{\galex}{{\it GALEX}}

\newcommand{\iras}{{\it IRAS}}

\newcommand{\nuv}{{\it NUV}}

\newcommand{\spitzer}{{\it Spitzer}}
\newcommand{\swift}{{\it Swift}}

\newcommand{\wone}{{\it uvw1}}
\newcommand{\wtwo}{{\it uvw2}}
\newcommand{\wtwoc}{{\it uvw2$\,^{\prime}$}}
\newcommand{\mtwo}{{\it uvm2}}
\newcommand{\uvot}{{UVOT}}

\newcommand{\surphot}{{\sc surphot}}

\newcommand{\er}{Equation~\ref}
\newcommand{\fr}{Fig.~\ref}
\newcommand{\scr}{Sec.~\ref}
\newcommand{\tr}{Table~\ref}

\newcommand{\exi}{\begin{equation}}
\newcommand{\exo}{\end{equation}}

\newcommand{\ten}[2]{$#1\times 10^{#2}$} %{some number} x 10^{some power}

\begin{document} 

\slugcomment{Accepted by ApJ}

\shorttitle{UV+IR Star Formation in HCGs}
\shortauthors{Tzanavaris et al.}

\title{UV+IR Star Formation Rates: Hickson Compact Groups 
with \swift\ and \spitzer}

\author{
P.~Tzanavaris\altaffilmark{1,2},
A.~E.~Hornschemeier\altaffilmark{1}, 
S.~C.~Gallagher\altaffilmark{3}, 
K.~E.~Johnson\altaffilmark{4}, 
C.~Gronwall\altaffilmark{5},
S.~Immler\altaffilmark{1,6}, 
A.~E.~Reines\altaffilmark{4}, 
E.~Hoversten\altaffilmark{5},
J.~C.~Charlton\altaffilmark{5} 
}

\altaffiltext{1}{Laboratory for X-ray Astrophysics, 
NASA/Goddard Spaceflight Center,Mail Code 662, Greenbelt, Maryland, 20771}
\altaffiltext{2}{Department of Physics and Astronomy,
The Johns Hopkins University, Baltimore, MD 21218}
\altaffiltext{3}{Department of Physics and Astronomy,
The University of Western Ontario, London, ON N6A 3K7, Canada}
\altaffiltext{4}{Department of Astronomy, University of Virginia, P.O. Box 400325,
Charlottesville, VA 22904}
\altaffiltext{5}{Department of Astronomy and Astrophysics, 
The Pennsylvania State University, University Park, PA 16802}
\altaffiltext{6}{Department of Astronomy, University of Maryland,
College Park, MD 20742}

\begin{abstract}
We present \swift\ \uvot\ ultraviolet (UV, $1600 - 3000$\AA)
data with complete 3-band UV photometry
for a sample of 41 galaxies in 11 nearby ($<4500$~\kmps)
Hickson Compact Groups (HCGs) of galaxies.
We use UVOT \wtwo-band (2000\AA) photometry to estimate the dust-unobscured
component, \sfruv, of the total star-formation rate, \sfrtot.
We use \spitzer\ MIPS 24\micron\ photometry
to estimate \sfrir, the component of \sfrtot\ which suffers dust-extinction
in the UV and is re-emitted in the IR. 
By combining the two components, we obtain \sfrtot\ estimates for
all HCG galaxies. 
We obtain total stellar mass, \mstar, estimates by means of 2MASS $K_s$ band
luminosities, and use them to calculate specific star-formation rates,
SSFR~$\equiv$~\sfrtot/\mstar.
SSFR values show a clear and significant bimodality, 
with a gap between low
($\la$~\ten{3.2}{-11} yr$^{-1}$) and high SSFR ($\ga$~\ten{1.2}{-10}
yr$^{-1}$) systems.
We compare this bimodality to the
previously discovered bimodality in \airac, the MIR activity index from a power-law fit
to the \spitzer\ IRAC $4.5-8$\micron\ data for these galaxies.
We find that {\it all} galaxies with 
\airac~$\le 0$ ($>0$) are in the high- (low-) SSFR locus, as expected if
high levels of star-forming activity power MIR emission from
polycyclic aromatic hydrocarbon molecules
and a hot dust continuum.
Consistent with this finding, all elliptical/S0 galaxies are in
the low-SSFR locus, while 22 out of 24 spirals/irregulars are in the high-SSFR
locus, with two borderline cases. We further 
divide our sample into three subsamples (I, II and III) 
according to decreasing \hone-richness of the parent
galaxy group to which a galaxy belongs. Consistent with the SSFR and \airac\
bimodality, 12 out of 15 type-I (11 out of 12 type-III) 
galaxies are in the high- (low-) SSFR locus, while type II
galaxies span almost the full range of SSFR values. 
We use the Spitzer
Infrared Nearby Galaxy Survey (SINGS) to construct a 
comparison sub-sample of galaxies that
(1) match HCG galaxies in $J$-band total galaxy luminosity, and
(2) are not strongly interacting and largely isolated. This
selection eliminates mostly low-luminosity dwarfs and
galaxies with some degree of peculiarity, providing
a substantially improved, quiescent control sample.
Unlike HCG galaxies, galaxies in the comparison SINGS sub-sample 
are continuously distributed both in SSFR and \airac, although
they show ranges in \sfrtot\ values, morphologies and stellar masses similar to 
those for HCG
systems.
We test the SSFR bimodality against a number of uncertainties, and
find that these can only lead to its further enhancement.  
Excluding galaxies belonging to HCGs with three giant galaxies (triplets)
leaves both the SSFR and the \airac\ bimodality completely
unaffected.
We interpret these results as further evidence 
that an
environment characterized by high galaxy number-densities and
low galaxy velocity-dispersions, such as the one
found in compact groups,
plays a key role in accelerating galaxy evolution by enhancing
star-formation processes in galaxies and favoring a fast
transition to quiescence.
\end{abstract}


\section{Introduction}

One of the main successes of the Cold Dark Matter (CDM)
paradigm is its prediction of hierarchical
structure formation,
a direct consequence of which is that
galaxies are more likely to be
clustered than isolated \citep{press1974,geller1983}.
Galaxy clustering spans scales from small groups to clusters to
super-clusters. Galaxy groups, including
the sub-class of Compact Groups (CGs) \citep{rood1994,kelm2004},
make up an important part
of this hierarchy \citep{geller1983,nolthenius1987}.
Poor groups are of particular interest, as it has been established
that in the nearby Universe they host the majority of galaxies
\citep[see][and references therein]{mulchaey2000}.

CGs are concentrations of small numbers of galaxies, which appear to
occupy a compact angular area in the sky.  By imposing limiting
magnitude and density contrast requirements, \citet{rose1977}
constructed the first sample of objectively selected CGs.
\citet{hickson1982} used a different set of criteria, which included a
maximum magnitude difference between four or more galaxies, limiting
surface brightness, as well as an encircling ring devoid of
galaxies. His catalog of 100 HCGs has been the most widely studied
nearby CG sample. 
\citet{sulentic1997} re-analyzed this catalog,
obtaining a revised HCG sample. In particular, this author
excludes groups which contain only three spectroscopically
confirmed members (triplets), as it is unclear
whether such systems share the properties of
groups with larger numbers of members.
HCGs harbor diverse populations of galaxies,
characterized by extreme morphological variety
\citep{mendesdeoliveira1994}, unusual rotation curves
\citep{rubin1991}, and a high fraction of (mostly faint) AGN
\citep{coziol1998a,gallagher2008}.  CG studies have recently been
extended to higher redshift \citep[e.g.][]{deCarvalho2005}.

Thanks to the spectroscopic survey of \citet{hickson1992},
a number of key results have been established for HCGs.
These authors obtained galaxy radial velocities,
showing that the great majority of HCGs are 
not chance projections but real concentrations
of galaxies:
Out of the full sample, 92 groups have three or more 
accordant members
(median redshift \zmed~= 0.03). 
As expected for compact groups, HCG member galaxies are 
a few galaxy radii from each other, with
median projected separations of $\sim 40 h^{-1}$~kpc.
HCGs are also characterized by low velocity dispersions
(radial median $\sim 200$~\kmps), 
high number-densities (as much as $10^8 h^2 \rm{Mpc}^{-2}$)
and short crossing times (median $0.016 H_0^{-1}$).
These conditions favor galaxy interactions and
mergers, e.g. as observed in
HCG 16 \citep{mendes1998}. Further evidence
for such processes is provided
by an observed correlation between
crossing times and the fraction of gas-rich
galaxies \citep{hickson1992,darocha2008}, as well
as the anti-correlation between crossing times
and the fraction of intra-group light
\citep{darocha2008}.
HCG environments are thus
ideal laboratories for studying processes related
to galaxy evolution and morphological transformation.
In particular, HCGs, as well as CGs in general, are the only nearby environments
that are closely similar to interaction environments
in the earlier universe ($z\sim 4$) when galaxies were
assembling hierarchically \citep[e.g.][]{baron1987}.

The link between this highly interaction-prone environment and
individual member galaxy properties remains controversial.  In
particular, it is unclear whether, and to what extent, star formation
and/or AGN activity is either enhanced or impeded. \citet{coziol1998a}
find that AGN are mostly associated with the most luminous, early-type
galaxies, as is the case in the field. However, they find that in HCGs
these galaxies are preferentially located in the denser cores. They
suggest an evolutionary scenario starting with merger-induced
starbursts which in the case of the more massive systems evolve to
become central AGN.  \citet{shimada2000} correct for the higher
fraction of early-type galaxies in HCGs with respect to the field,
finding no significant differences in the number of emission-line
galaxies.  \citet{verdes1998} compare FIR and CO emission in 
spiral galaxies from their
\iras\ HCG sample 
to isolated, Virgo cluster and weakly interacting systems.
They find that most HCG spirals show no enhanced 
FIR and CO emission, with 20\%
showing {\it reduced} CO emission. On the other hand, some early-type
galaxies in HCGs are detected in CO and FIR.

\citet{verdes2001} calculate the deficiency in \hone\ in 72 HCGs.
They find that HCGs with higher numbers of early-type galaxies 
are more deficient
in \hone\ and have a higher detection rate in the \x\ band.
Using a 109-member group sample not restricted to 
compact groups, \citet{mulchaey2003} find
evidence for diffuse \x\ emission and an intra-group medium (IGM) in
half of their sample, in particular for groups containing at
least one elliptical. \citet{ponman1996} find evidence for diffuse \x\
emission in more than 75\%\ of a sample comprising 85 HCGs. 

\citet{verdes2001} interpret their findings as evidence for an
evolutionary sequence proceeding from \hone-rich groups, (mainly
containing spiral and irregular, S/I, galaxies) to \hone-poor groups (mainly
hosting ellipticals).  
An expanded, more speculative, 
version of this scenario, taking into account the \x\
evidence might be as follows:
Initially, loose
groups contract to a more compact configuration
\citep{barton1998}. 
At this stage, most of
the \hone\ is found in galaxy disks, which constitute the prevailing
morphological type. As the effects of tidal interactions gain in
importance with time, an increasing fraction of the group \hone\ mass
is stripped from the interstellar medium of member galaxies and forms
tidal tails, bridges and intergalactic structures.  The atomic gas is
heated and ionized at an increasing rate, filling the space between
member galaxies, eventually giving rise to a hot \x-bright IGM
that may characterize a group's final state.  At this stage groups
are made up mostly of gas-poor ellipticals.  
As this sequence is
characterized by removal of gas from individual galaxies, where it can
fuel star-formation, it is natural to expect a correlation between a
group's \hone-richness and star-forming activity. Note though that
this picture may still be too simplistic: In their investigation
of the \x\ properties in a sample of eight 
highly \hone-deficient HCGs \citet{rasmussen2008} find that only
four groups have an IGM detectable in the \x s.

\begin{deluxetable*}{ccccccccc}
\tablecaption{HCG SAMPLE \label{tab-sample}}
\tablewidth{500 pt} %new
\tablecolumns{9}

\tablehead{
\colhead{}  
& \colhead{$\bar{v}$\tablenotemark{a}} 
& \multicolumn{3}{c}{Morphology\tablenotemark{b}} 
& \colhead{log \mhi\tablenotemark{c}}
& \colhead{}
& \colhead{}
& \colhead{}
\\
\colhead{HCG ID}  
& \colhead{(\kmps)}
 & \colhead{E/S0}
 & \colhead{S}
 & \colhead{Other}
& \colhead{(\msun)}
& \colhead{Evolutionary Stage\tablenotemark{d}}
& \colhead{\hone\ type\tablenotemark{e}} 
& \colhead{Triplet?\tablenotemark{f}} 
\\
\colhead{(1)}
& \colhead{(2)}
& \colhead{(3)}
& \colhead{(4)}
& \colhead{(5)}
& \colhead{(6)}
& \colhead{(7)}
& \colhead{(8)}
& \colhead{(9)}
}
\startdata
 02 & 4309 & 0 & 2 & 1 & 10.53                    & early     & I   & Y \\ 
 07 & 4233 & 0 & 4 & 0 & 9.68                     & early     & II  & N \\ 
 16 & 3957 & 0 & 2 & 2 & 10.42                    & int       & I   & N \\ 
 19 & 4245 & 1 & 1 & 1 & 9.31                     & early/int & II  & Y \\ 
 22 & 2686 & 1 & 2 & 0 & 9.13                     & early     & II  & Y \\ 
 31 & 4094 & 0 & 2 & 5 & 10.35                    & int       & I   & N \\ 
 42 & 3976 & 4 & 0 & 0 & 9.40                     & late      & III & N \\ 
 48 & 3162 & 2 & 2(0)\tablenotemark{g} & 0 & 8.52 & late      & III & N \\ 
 59 & 4058 & 1 & 2 & 1 & 9.49                     & early/int & II  & N\\ 
 61 & 3907 & 2 & 1 & 0 & 9.96                     & early/int & I   & Y \\ 
 62 & 4122 & 4 & 0 & 0 & 9.06                     & late      & III & N \\
\enddata 
\tablecomments{}
\tablenotetext{a}{Mean recession velocity for all known group member galaxies calculated from
\citet{hickson1992}.}
\tablenotetext{b}{Taken from \cite{hickson1989}.}
\tablenotetext{c}{Mass of neutral hydrogen from 
\cite{verdes2001}.}
\tablenotetext{d}{Qualitative determination
of group evolutionary stage from G08 as early, early/intermediate, intermediate or late.
This is motivated by the evolutionary scenario proposed by \citet{verdes2001} and 
takes into account member galaxy morphology, \hone\ deficiency and the presence of
an X-ray intragroup medium.
}
\tablenotetext{e}{\hone\ type as measured and defined by J07. (I) \hone­-rich
 (log \mhi/log $M_{\rm dyn} > 0.9$); (II) intermediate (log \mhi/log $M_{\rm dyn}  = 0.8­-0.9$);
 (III) \hone­-poor (log \mhi/log $M_{\rm dyn} < 0.8$).}
\tablenotetext{f}{Y if group is a triplet, N otherwise.}
\tablenotetext{g}{Galaxies 48b and c have somewhat discordant velocities.}
\end{deluxetable*}

Recent results in the IR provide further support to this scenario.
Until recently results in this wavelength regime relied
on low-sensitivity and/or
angular resolution data
\citep[e.g.][]{allam1995,verdes1998}. However,
\citet[][hereafter J07]{johnson2007}
were the first to present results on a set of 45 galaxies belonging to
a sample of 12 nearby HCGs observed with \spitzer\ (IRAC and MIPS) and
2MASS. They establish trends connecting group
\hone\ gas deficiency and the level of
active star formation, implied by IR colors of individual member
galaxies. They detect a \lq\lq gap\rq\rq\ in IR color-color
space between gas-rich and gas-poor groups, suggestive of rapid
evolution in galaxy properties. Using the same
HCG sample (hereafter JG sample), \citet[][hereafter G08]{gallagher2008}
find further evidence for this gap in the distribution of \airac, the
mid-IR (MIR) activity index (see \scr{subsec_uvir}) for the nuclei of
galaxies in the same HCG sample. Their findings strongly suggest a
connection between \airac, 24\micron\ activity, and \hone\ content.

In this paper we further explore the connection between
group gas-content, galaxy morphology and star-formation by
obtaining star-formation rate (SFR) and specific SFR (SSFR) estimates for
41 galaxies in the JG sample.
Although it is common to use single-band data, such as 
UV, \ha\ or 24\micron\ to obtain such estimates \citep[but see][]{kennicutt2009}, 
these usually require corrections due to the effects of dust, which, in
general, are difficult to quantify. In this respect, the UV and IR
wavelength regions are complementary, and we combine information from both
to obtain total SFRs consisting of a UV and an IR component.

This paper makes part of a collaborative,
multi-wavelength campaign to observe and
characterize the JG sample in several
wavelength bands, from the X-rays to the far-IR and radio.
Our main goal is to investigate whether, and to
what extent, the UV regime provides support to the
evolutionary scenario suggested by the IR work.  

The structure of the paper is as follows.
In \scr{sec_obs} we give details on 
sample selection, UV and IR data analysis,
as well as issues related to flux calculations in the UV.
UV and IR results, including SFRs and SSFRs, are
presented in \scr{sec_results}.
\scr{sec_discussion} discusses results and relevant
uncertainties.
We summarize and conclude in \scr{sec_final}.
We use $\Omega_M=0.3$,
$\Omega_\Lambda = 0.7$, $H_0 = 70$~\hunits\ throughout.

\section{Observations and Data Analysis}\label{sec_obs}
\subsection{HCG sample selection}\label{subsec_sample}
The JG sample was chosen from the original Hickson Compact Group
catalog \citep{hickson1992} by application of criteria based
on membership (a minimum of 3 giant galaxies with accordant redshifts,
i.e. within 1000 \kmps\ of the group mean), distance ($\lesssim 4500$~\kmps)
and angular extent ($\lesssim 8$\arcmin\ in diameter).  
The analysis in the present paper does not include HCG 90, as explained
in sections \ref{subsec_uvdata} and \ref{subsec_irdata}.

Note that a sample based on these criteria 
includes triplets, and these are indicated in
\tr{tab-sample}.
There is some evidence that triplets may be different as a class
compared to groups with four or more members:
They do not show significant \hone\ deficiency
\citep{verdes2001} and their galaxies show larger velocity
dispersions \citep{sulentic2000}, making it more likely that they
are unbound systems \citep{sulentic2001}. 
Indeed the original HCG catalog criteria
did not include triplets, and
the revised catalog of \citet{sulentic1997} excludes triplets 
altogether. We chose to keep triplets in our samples for several
reasons. We are engaged on a long-term spectroscopic campaign to
identify new, fainter members in our HCG galaxies. 
New members will affect the overall dynamics, which remains an open question.
The dark matter mass also plays a role, but it is difficult to
measure without more members. Further, there is no consensus that
triplets should be excluded. Indeed, \citet{sulentic2000} argue for
HCG 92 that it is possible that three of its bright galaxies make up
a stable core, with the {\it fourth} causing it to be unstable. 
\citet{tovmassian1999} specifically address the issue of
reality of compact groups, including triplets which are not
considered to be a special case. 
\citet{barton1998} include several triplets in their redshift
survey of CGs, and \citet{darocha2005} include a triple, HCG 95,
in their investigation. We similarly prefer to
simply consider triplets as one extreme of groups of galaxies. Given
that CGs are relatively rare, this keeps restrictions at a minimum.
In any case, including triplets does not affect the
main results of this paper (see \scr{subsec_caveats}).

In
\tr{tab-sample} we present the group mean recession velocities, the
member galaxy morphology, \hone\ mass, qualitative evolutionary stage
based on the \citet{verdes2001} scenario, and \hone-richness type
from J07. \hone\ richness is defined as the ratio 
log \mhi/log $M_{\rm dyn}$, where
\mhi\ is the \hone\ mass 
and
$M_{\rm dyn}$ the dynamical mass,
as
described in J07.
\hone\ type I groups (gas-rich) have log \mhi/log $M_{\rm
dyn} > 0.9$; \hone\ type II (intermediate gas-rich) have log \mhi/log
$M_{\rm dyn} = 0.8­-0.9$; \hone\ type III (gas-poor) have log \mhi/log $M_{\rm
dyn} < 0.8$.

\subsection{UV data}\label{subsec_uvdata}
All galaxy groups in our HCG sample have UV data obtained with the
\swift\ UV/Optical telescope (\uvot).  \uvot\
\citep{2005SSRv..120...95R}, is one of
three telescopes on board NASA's international \swift\ mission
\citep{2004ApJ...611.1005G}. The mission's primary goal is detection
and characterization of gamma-ray bursts. 

\uvot\ has a $17\arcmin \times 17\arcmin$
field-of-view and six broadband filters covering the 1600 -- 8000\AA\
range with a spatial
resolution of $\sim 2.5$\arcsec\ (PSF FWHM
\footnote{\href{http://heasarc.gsfc.nasa.gov/docs/swift/analysis/uvot_digest.html}{http://heasarc.gsfc.nasa.gov/docs/swift/analysis/uvot\_digest.html}}).
Our data have been taken with the three UV filters and the
bluest optical filter, $u$. The characteristics of these filters 
are given in \tr{tab-uvfilt}. 

An observation log for our \swift\ \uvot\ data is given in \tr{tab-uvobs}.
Our targets were observed between August 2006 and November 2007 as
part of a Swift team fill-in program (P.I. C.~Gronwall).  The nominal
exposure times were 4000 sec in \wtwo, 3000 sec in \mtwo, 2000 sec in \wone,
and 1000 sec in $u$.  However because of the nature of the fill-in
program for Swift observations, sometimes the nominal exposure times
were not matched exactly.

\begin{deluxetable}{ccc} 
\tablecolumns{3} 
\tablewidth{200 pt} 
\tablecaption{UVOT AND {\it GALEX} BANDS\label{tab-uvfilt}}
\tablehead{
\colhead{Filter}
&\colhead{\lameff\tablenotemark{a}}  
&\colhead{Width\tablenotemark{b}}
\\
\colhead{}       
&\colhead{(\AA)}                
&\colhead{(\AA)}
}
\startdata
%\hline\\
&\uvot\tablenotemark{c}&\\
\hline
$u$ & 3501 & 785\\
\wone & 2634 & 693\\
\mtwo & 2231 & 498\\
\wtwo & 2030 & 657\\
\hline
&\galex\tablenotemark{d}&\\
\hline
\nuv  & 2271 & 1771--2831   \\
\fuv  & 1528 & 1344--1786   
\enddata
\tablecomments{}
\tablenotetext{a}{Filter effective wavelength.}
\tablenotetext{b}{Widths for \galex\ are bandpasses, and for UVOT full-width at
half-maximum values.}
\tablenotetext{c}{Data taken from \citet{poole2008}.}
\tablenotetext{d}{Data taken from \citet{morrissey2005}.}
\end{deluxetable} 
%\clearpage 

\begin{figure*}[h]
\epsscale{0.8}
\plottwo{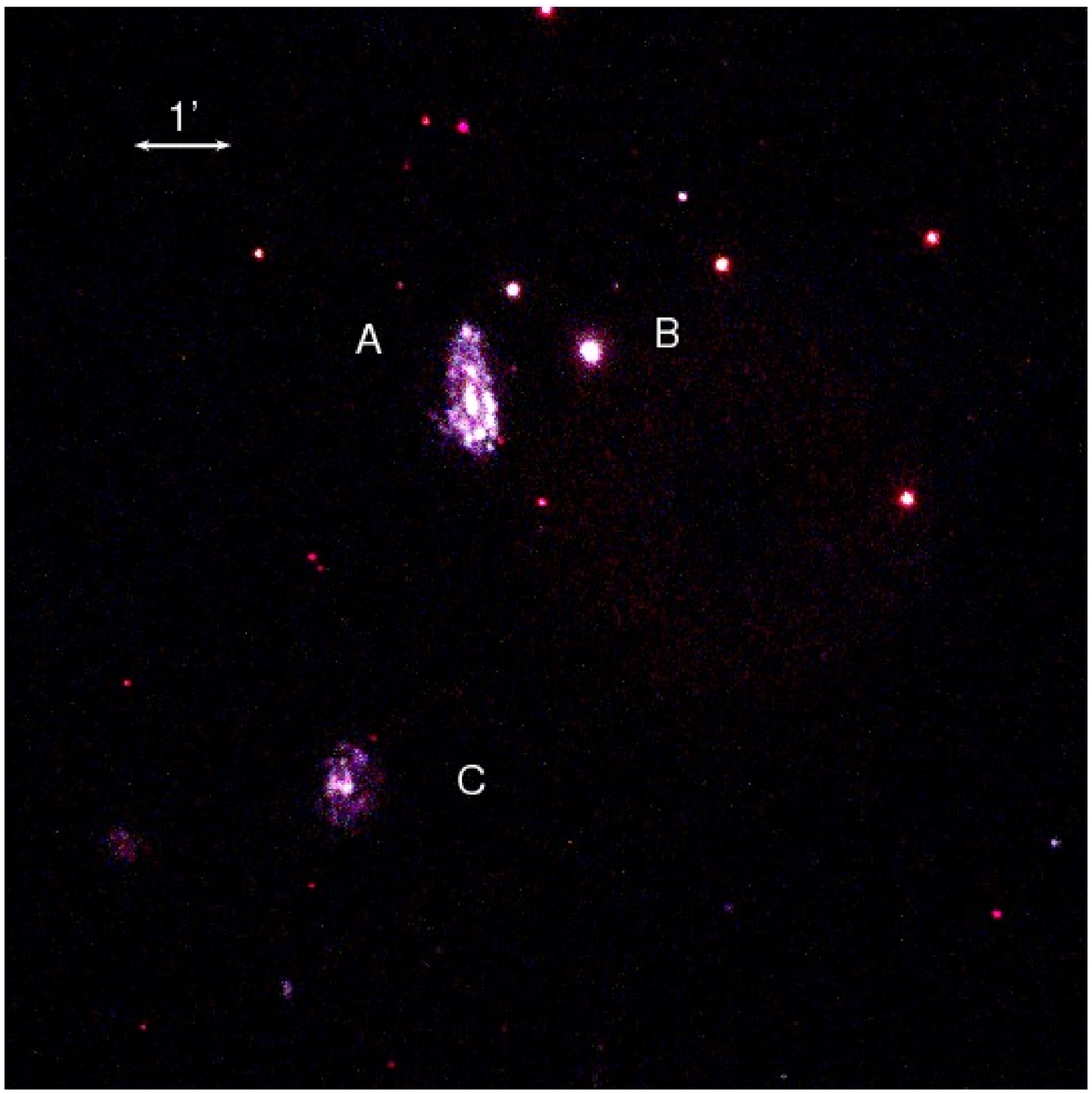}{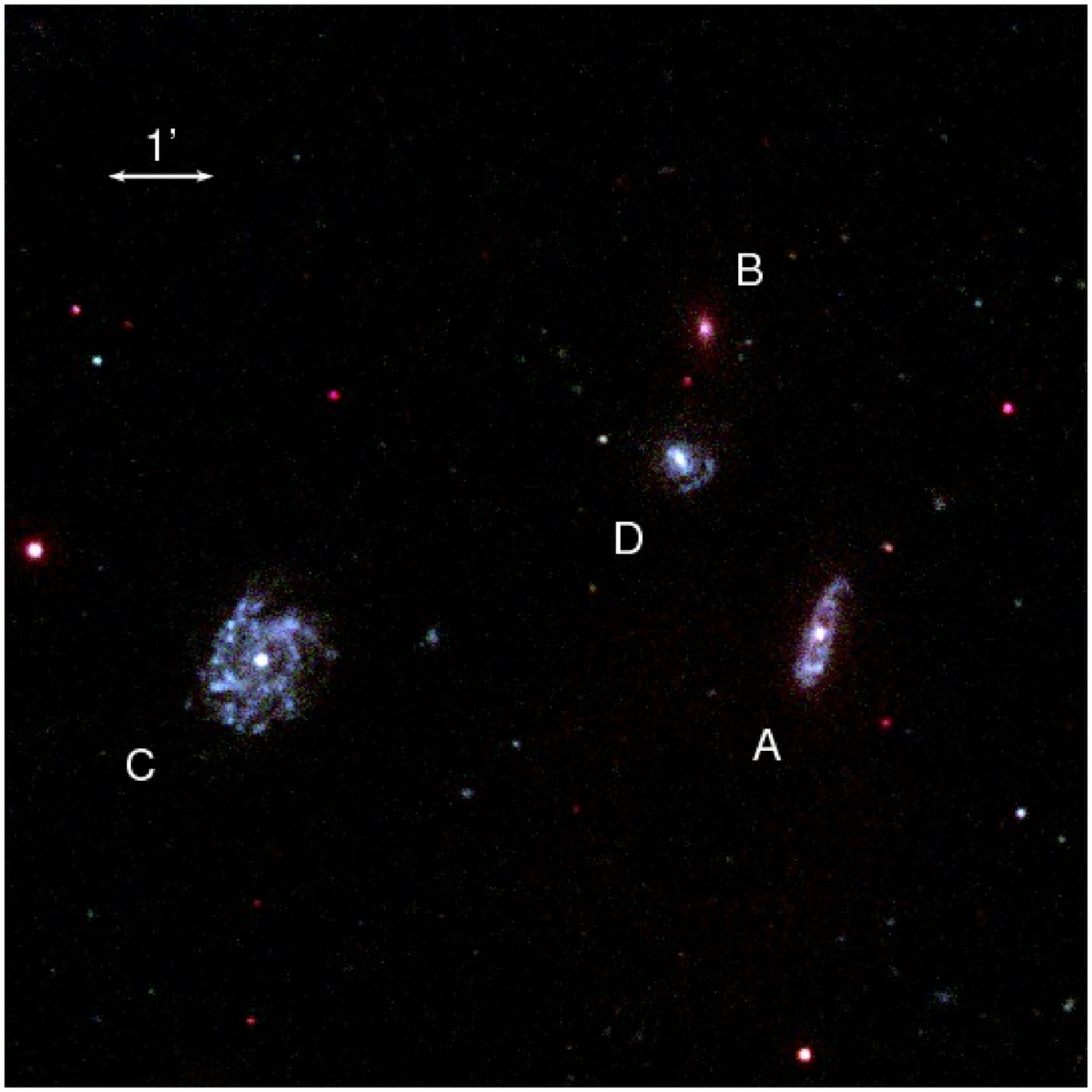}
\caption{\swift/UVOT 3-band images of Hickson Compact Groups in this
sample. Blue, green and red colors correspond to
the \wone, \mtwo, \wtwo\ filters, respectively. 
{\it Left:} HCG~2.{\it Right:} HCG~7.
\label{fig:uvot2-7}}
\end{figure*}

\begin{figure*}
\epsscale{0.8}
\plottwo{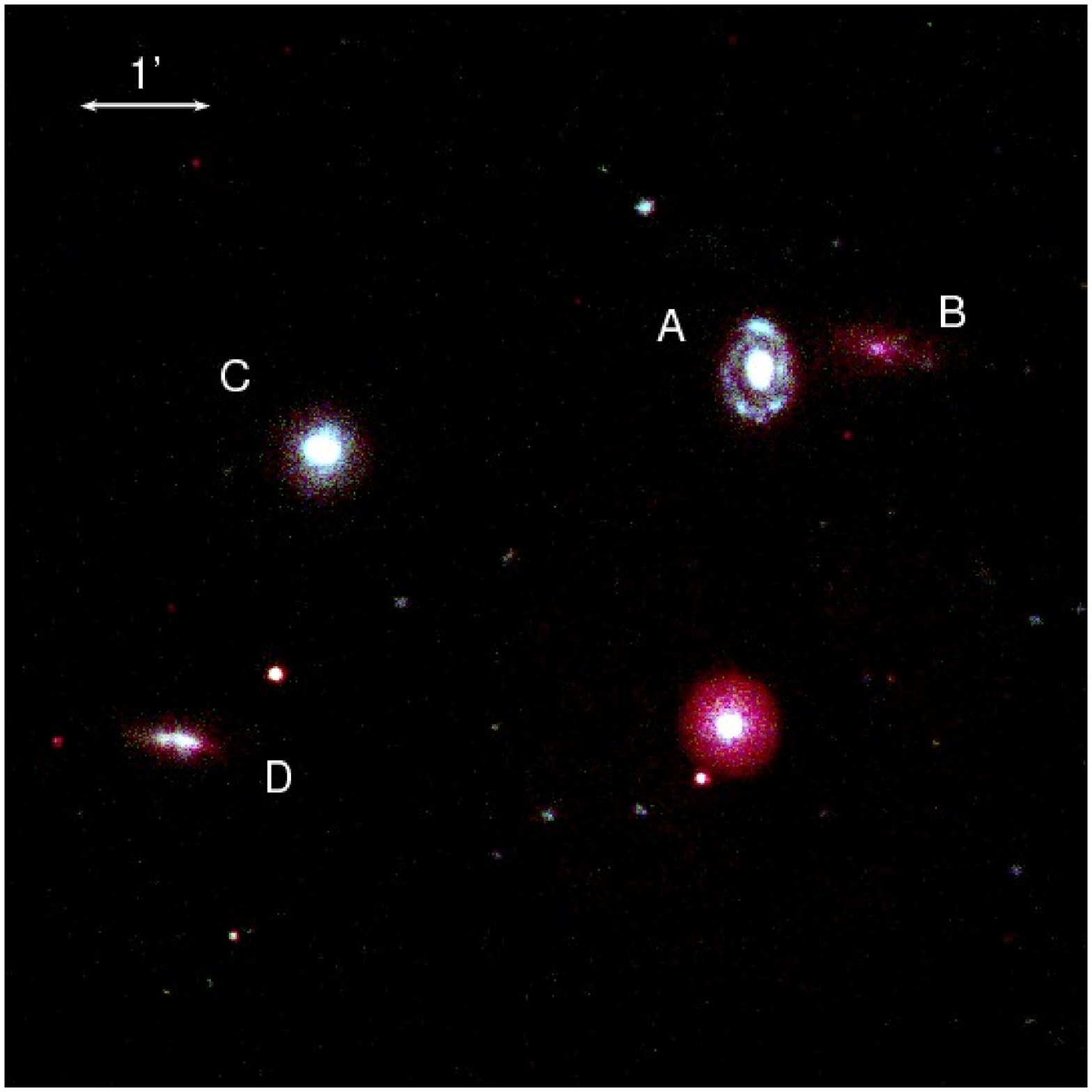}{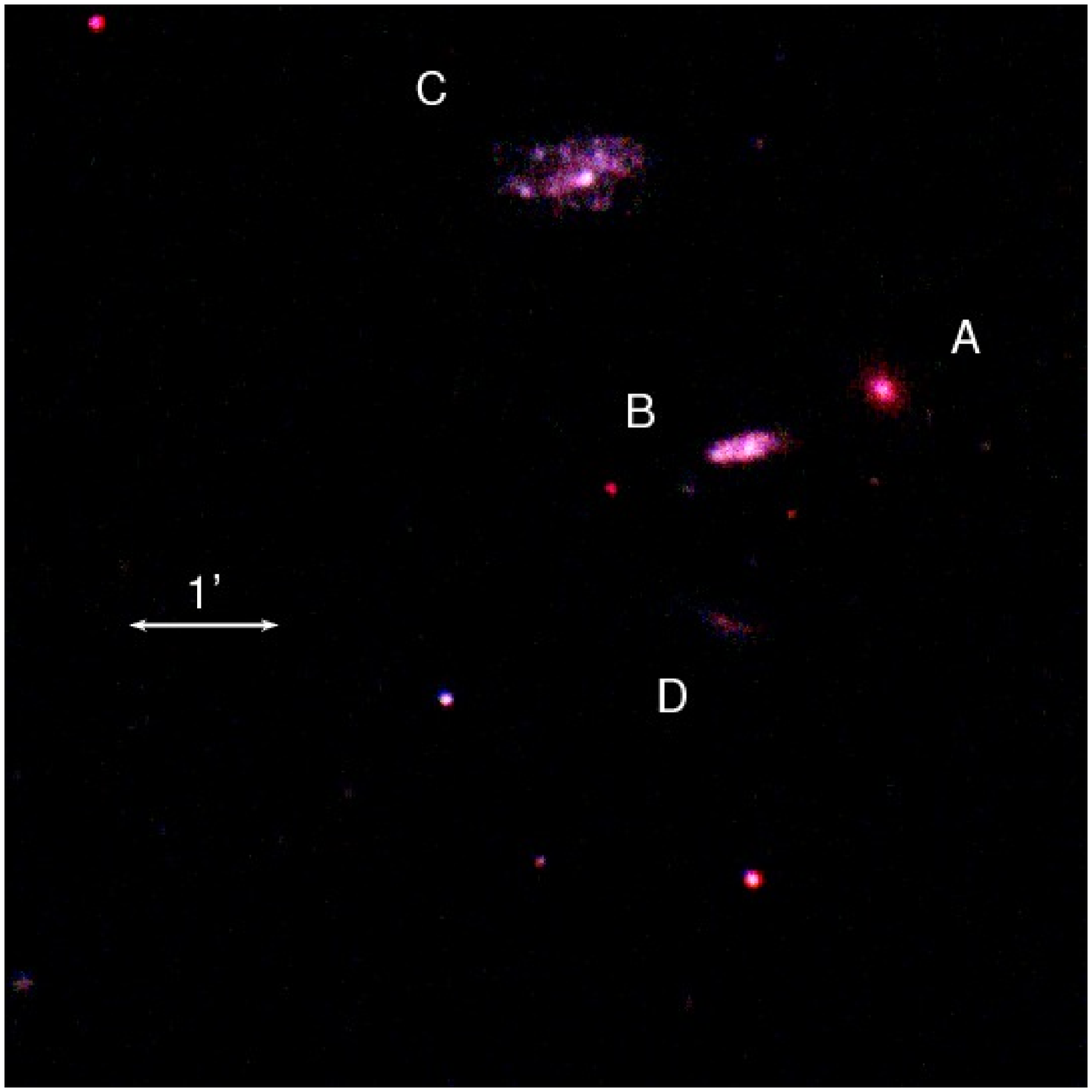}
\caption{ \swift/UVOT color images as in \fr{fig:uvot2-7}.
{\it Left:} HCG~16.{\it Right:} HCG~19. Galaxy D 
is a background object.
\label{fig:uvot16-19}}
\end{figure*}

\begin{figure*}
\epsscale{0.8}
\plottwo{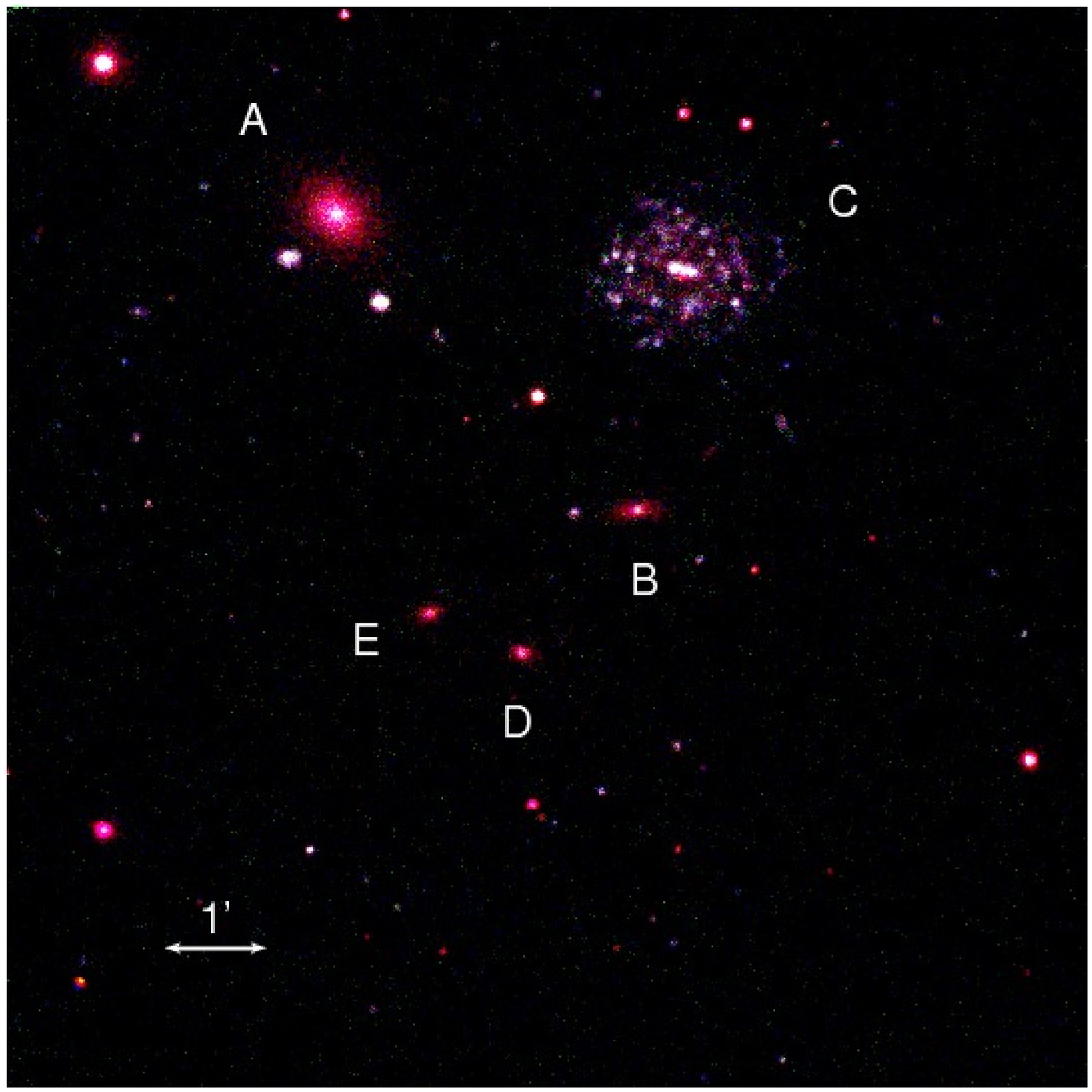}{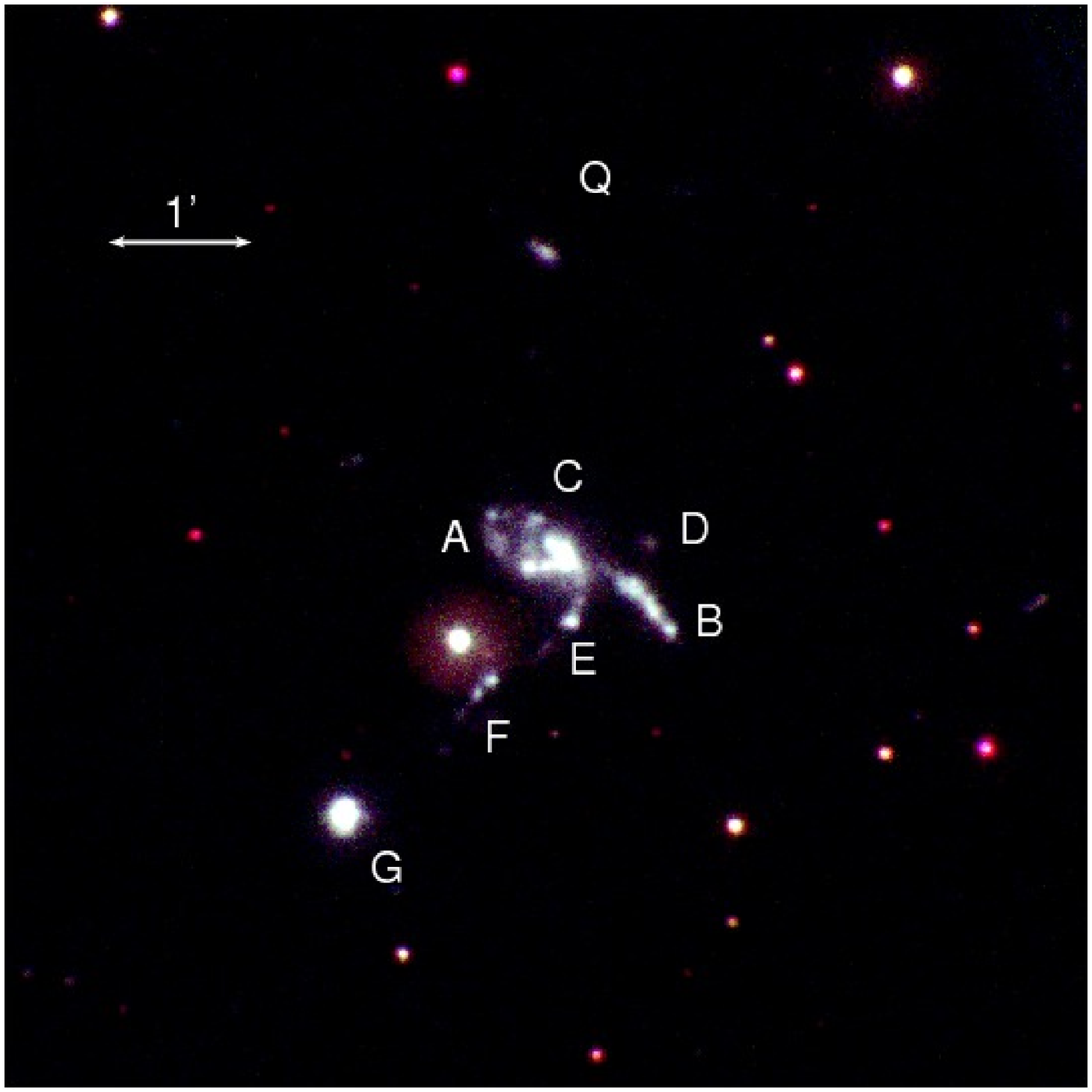}
\caption{ \swift/UVOT color images as in \fr{fig:uvot2-7}.
{\it Left:} HCG~22. D and E are background objects. {\it Right:} HCG~31.
D is a background object.
\label{fig:uvot22-31}}
\end{figure*}

\begin{figure*}
\epsscale{0.8}
\plottwo{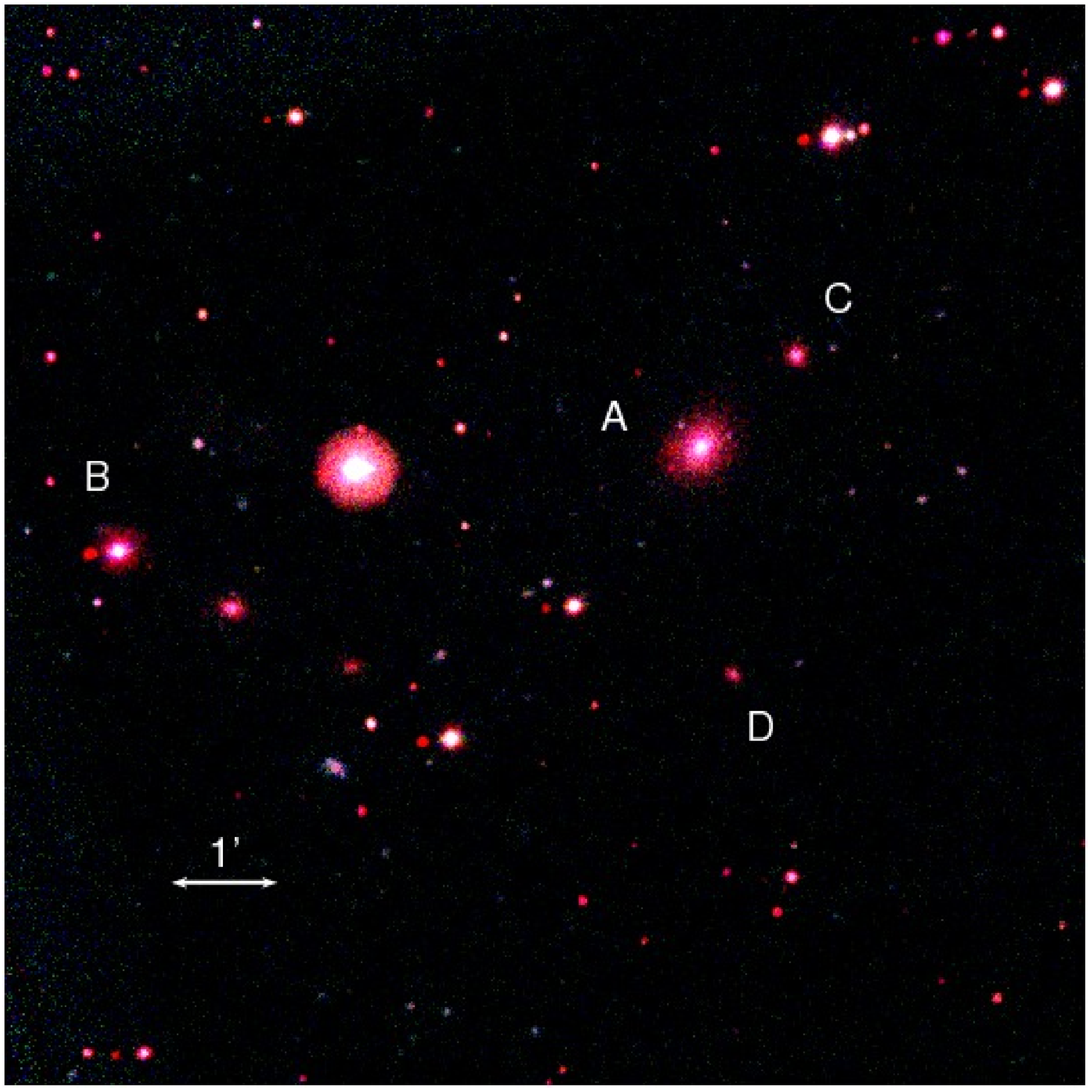}{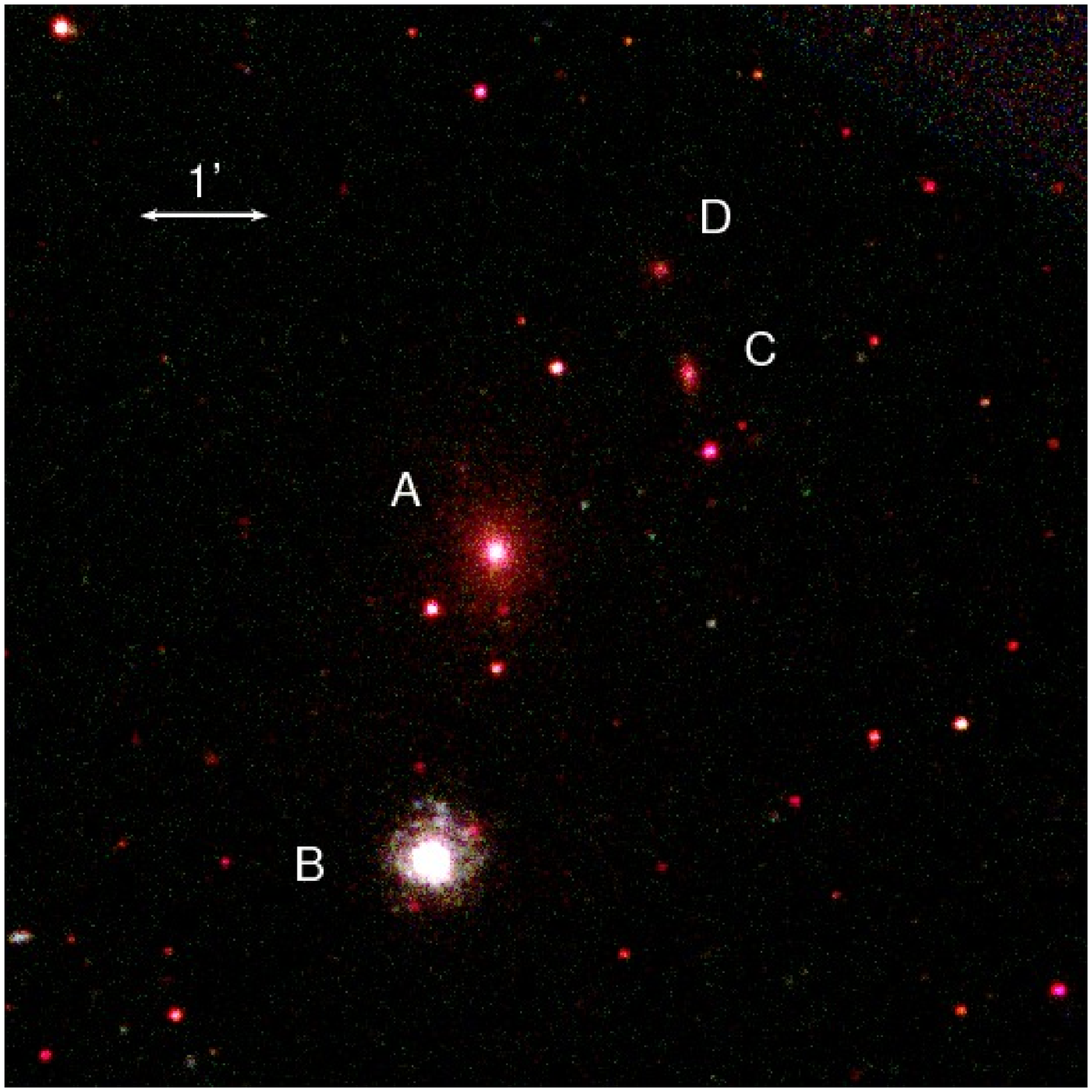}
\caption{ \swift/UVOT color images as in \fr{fig:uvot2-7}.
{\it Left:} HCG~42.{\it Right:} HCG~48.
\label{fig:uvot42-48}}
\end{figure*}

\begin{figure*}
\epsscale{0.8}
\plottwo{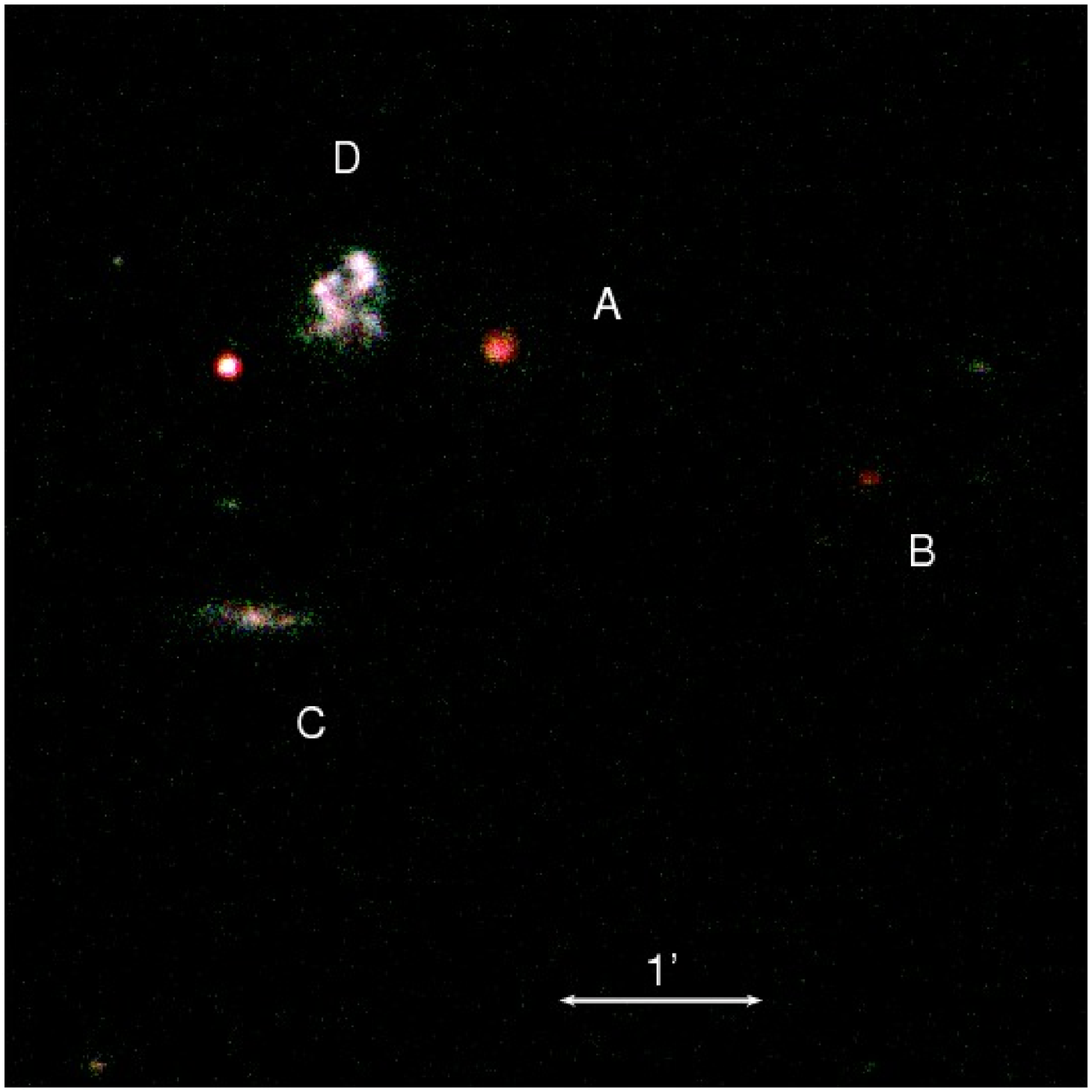}{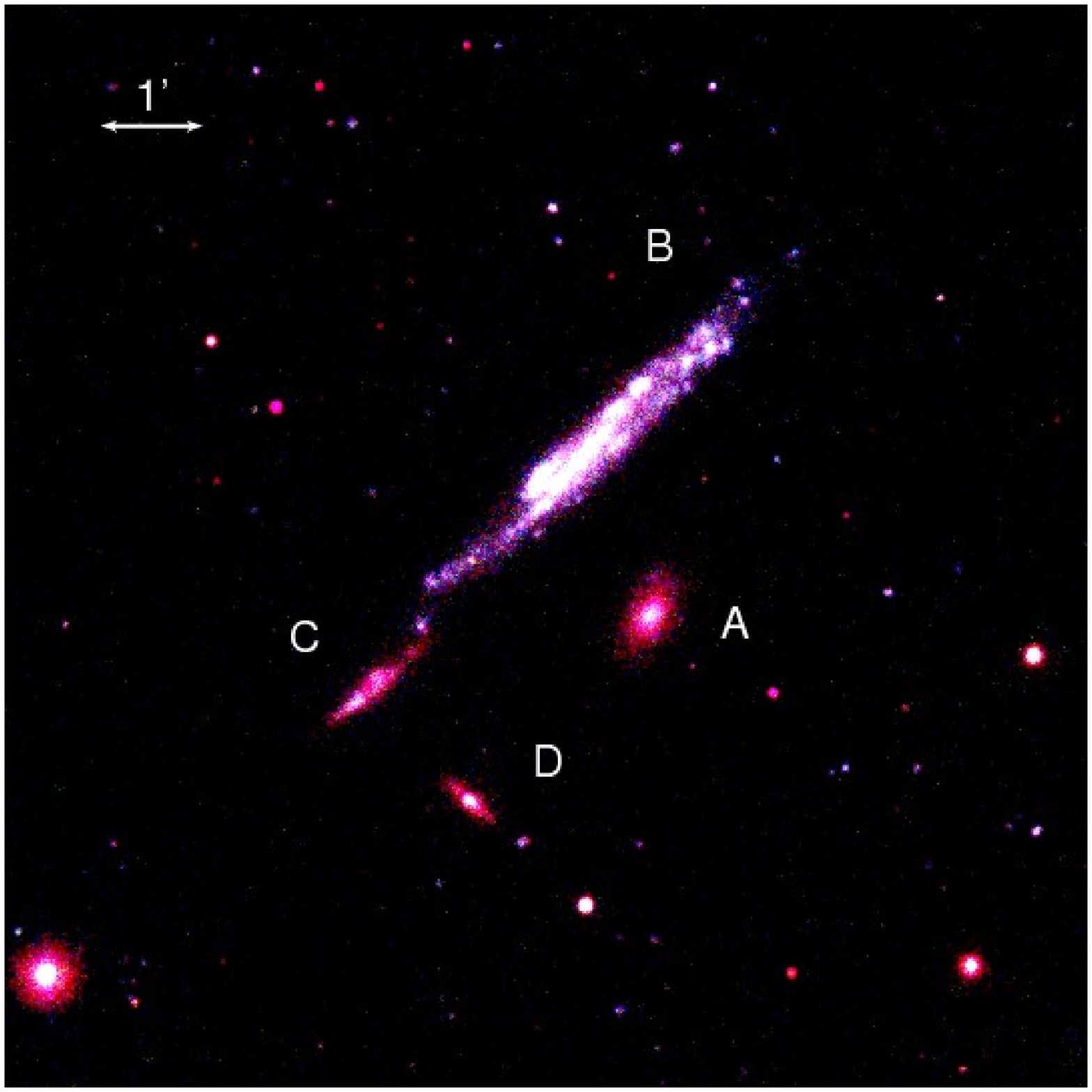}
\caption{ \swift/UVOT color images as in \fr{fig:uvot2-7}.
{\it Left:} HCG~59. {\it Right:} HCG~61. B is a foreground object.
\label{fig:uvot59-61}}
\end{figure*}

\begin{figure*}
\epsscale{0.8}
\plottwo{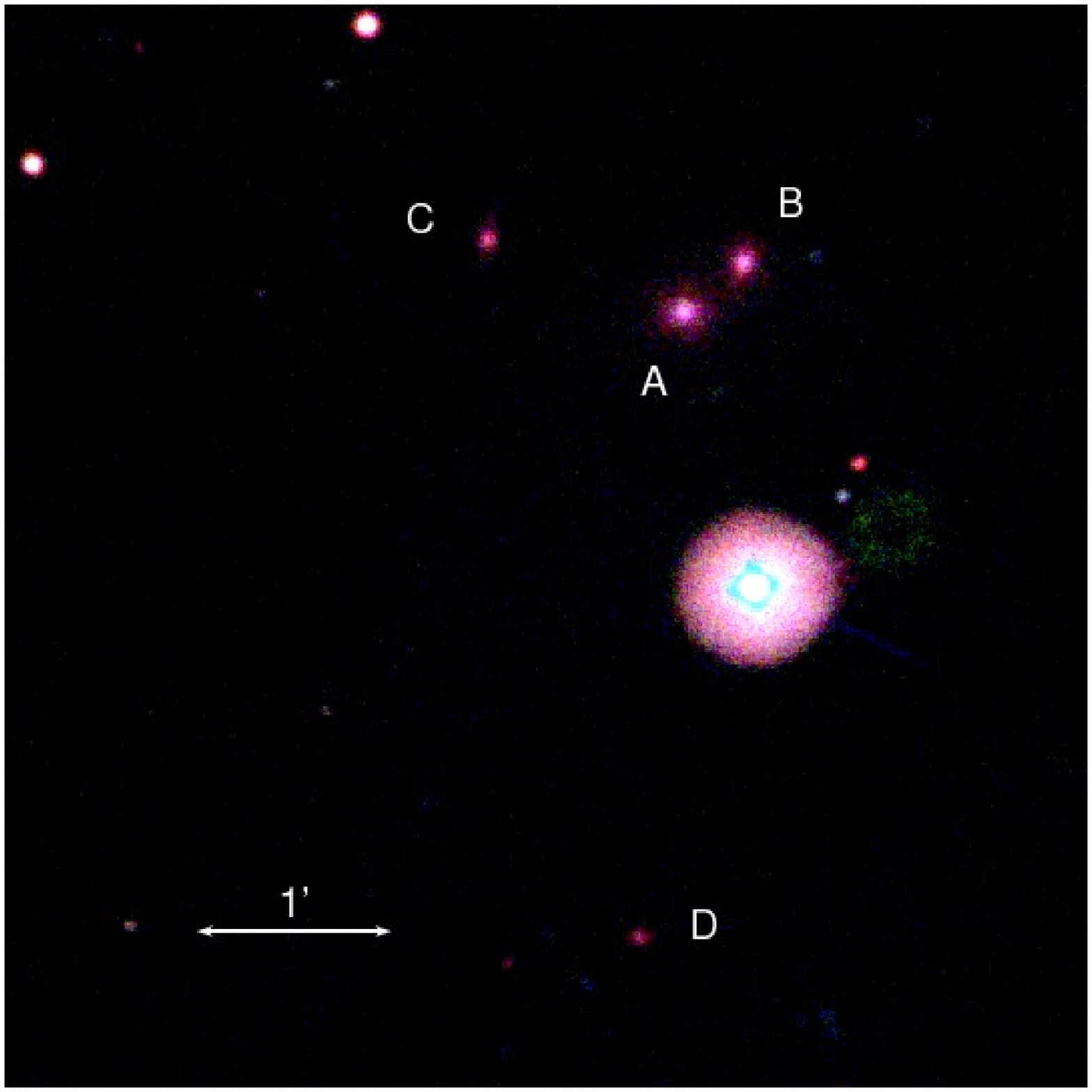}{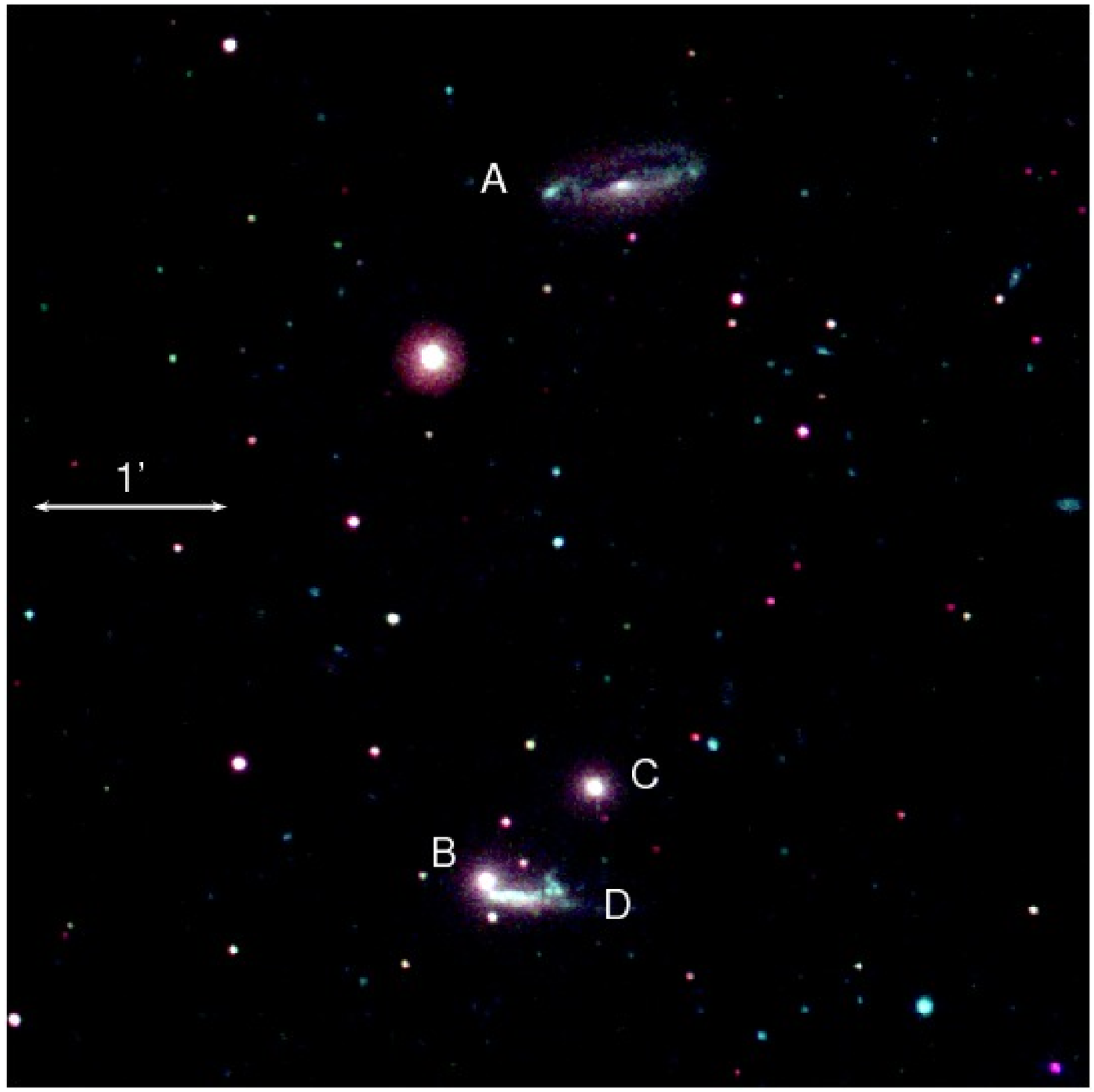}
\caption{ \swift/UVOT color images as in \fr{fig:uvot2-7}.
{\it Left:} HCG~62. {\it Right:} HCG~90.
\label{fig:uvot62-90}}
\end{figure*}

The data were reduced using dedicated \uvot\ pipeline tasks which
form part of the HEASOFT package.
\uvot\ sky images were prepared from raw images and event files, 
(task {\tt uvotimage}) and aspect corrected ({\tt uvotskycorr}).
Final images and exposure maps were produced by combination
of distinct exposures for the same observation ({\tt uvotimsum}).
Three-color composites of HCG \uvot\ images are shown in
Figures \ref{fig:uvot2-7} to \ref{fig:uvot62-90}.
Note that galaxies HCG 90 B, C and D
fall on the edges of the UVOT stacked image for this HCG.
This causes different parts of
the same galaxy to have different exposure times and total counts.
In this case calculating total count rates is not straightforward.
As HCG 90 is also not detected in the 24\micron\ band 
(see \scr{subsec_irdata}),
all of the HCG 90 galaxies are excluded from the analysis in this paper.

For five HCGs in our sample, GALEX guest investigator data 
(PI: J.~Paramo) are also available.
GALEX is a small-size NASA mission, with a UV telescope providing
images in both a far-UV (1528\AA) and near-UV (2271\AA) 
band. Filter characteristics for GALEX are listed in \tr{tab-uvfilt}. GALEX
has a circular field of view of 1.2\degr\ in diameter and a resolution
of $\sim 6$\arcsec\citep[PSF FWHM,][]{morrissey2005}. In comparison,
\uvot\ has better spatial and color resolution, but GALEX can probe
deeper in the FUV region both in terms of range and in terms of
effective wavelength, and has higher sensitivity. 
We compare the GALEX and UVOT data in order to perform
a \lq\lq sanity check\rq\rq\
for ultraviolet flux estimates (\scr{subsec_galuv}).

\subsection{IR data}\label{subsec_irdata}
We make use of the \spitzer\ [IRAC (3.6 -- 8.0\micron) and MIPS (24\micron)]
and 2MASS ($J, K_s$) data presented in J07.
For details on the IR observations and data reduction we refer the
reader to that paper and references therein.
The data are incomplete for HCG 31 F, which is
too faint to be detected in the 2MASS $K_s$ band,
and HCG 90 A, which is outside the MIPS field of view. We only perform
part of the analysis for 31 F. HCG 90 is excluded from our sample.

\subsection{Source detections}\label{subsec_sources}
The photometric properties of our HCG galaxies
in the near-to-mid IR 
were explored in detail
by
J07. 
For consistency with that work, we obtained count rates in
the same apertures used by these authors.
J07 defined apertures by determining
contour levels of 1.5-2$\sigma$ on wavelength weighted, combined IRAC
images. As these images were convolved to the
MIPS 24\micron\ PSF, we also convolved our
\uvot\
images to the 24\micron\ PSF, which is significantly broader
\citep[FWHM $\sim 6$\arcsec, e.g. ][vs. $\sim 2.5$\arcsec\
for \uvot]{dole2006}.    
Photometry on all UV images,
was carried out using \surphot,
a set of routines
written specifically to
allow simultaneous, net-counts-in-regions  calculations on
several images \citep{reines2008}. 
We also reproduced the 24\micron\ results of
J07 to ensure use of the same apertures for
the UV and IR datasets.
We used the same background
annuli with inner and outer sizes 2 and 2.5 times the size of the
apertures.  We note that differences in background values obtained
using these
annular regions and random, source-free,  regions in the same image
are insignificant ($\la 2$\%).

%
%Produced with input from Caryl but
%confirming this info online is messy (a couple of minor mistakes on consecutive dates):
%HEASARC ->
%MASTER SWIFT catalog -> 
%click on UVOT entries; each one gives date and obsid 
%NB: UVOT instrument catalog incomplete - MUST USE MASTER.
%
%in each UVOT directory  ( eg with alias command uvot2 ): 
%   guvex 
%command produces exposure times from fits files in tex format
%
%
\begin{deluxetable*}{c c c   c c c c} 
\tablecolumns{7} 
\setlength{\tabcolsep}{0.2cm}
\tablewidth{500 pt} 
\tablecaption{OBSERVATION LOG FOR SWIFT UVOT HCG DATA (P.I. C.~Gronwall)\label{tab-uvobs}}
\tablehead{
\colhead{}
&\colhead{}%\tablenotemark{a}  
&\colhead{}
&\multicolumn{4}{c}{Total Exposure Time}
\\
\colhead{}
&\colhead{}
&\colhead{}       
&\colhead{\wtwo}                
&\colhead{\mtwo}
&\colhead{\wone}
&\colhead{$u$}
\\
\colhead{HCG ID}
&\colhead{Observation IDs}
&\colhead{Dates}
&\colhead{(s)}                
&\colhead{(s)}
&\colhead{(s)}
&\colhead{(s)}
\\
\colhead{(1)}
&\colhead{(2)}
&\colhead{(3)}
&\colhead{(4)}                
&\colhead{(5)}
&\colhead{(6)}
&\colhead{(7)}
}
\startdata
2   &  00035906001  & 	2007 Feb 11, 2007 Feb 12	 &  2449  &  2431  &  1623  &  811  \\
    &  00035906002  & 	2007 Nov 01			 &   &  &  &  \\     
\hline 
7   &  00035907001  & 	2006 Oct 28, 2006 Oct 30	 &  4236  &  4633  &  2867  &  1406 \\
    &  00035907002  & 	2006 Nov 10			 &   &  &  &  \\        
    &  00035907003  & 	2007 Jan 29			 &   &  &  &  \\ 
    &  00035907004  & 	2007 Feb 10, 2007 Feb 11	 &   &  &  &  \\  
\hline
16  &  00035908001  & 	2006 Nov 03			 &  4652  &  3894  &  2596  &  1292 \\
    &  00035908002  & 	2007 Feb 24			 &   &  &  &  \\   
    &  00035908003  & 	2007 Dec 03			 &   &  &  &  \\  
\hline
19  &  00035909001  & 	2006 Aug 19, 2006 Aug 20	 &  2613  &  3044  &  2026  &  1012 \\
    &   00035909002 & 	2006 Oct 30			 &   &  &  &  \\
\hline
22  &  00035910001  & 	2006 Oct 30			 &  4304  &  3798  &  2524  &  1268 \\
    &  00035910002  & 	2007 Mar 04, 2007 Mar 05	 &   &  &  &  \\
    &  00035910003  & 	2007 Mar 17			 &   &  &  &  \\ 
\hline
31  &  00035911001  & 	2006 Aug 20, 2006 Aug 21	 &  3486  &  3222  &  2138  &  1066 \\
\hline
42  &  00035912001  & 	2007 Jan 30			 &  3326  &  3027  &  2017  &  1003 \\	
    &  00035912002  & 	2007 Feb 01			 &   &  &  &  \\ 
\hline
48  &  00035913001  & 	2006 Nov 12, 2006 Nov 13	 &  5478  &  4665  &  3145  &  1560 \\
    &  00035913002  & 	2006 Dec 29, 2006 Dec 30	 &   &  &  &  \\ 
\hline
59  &  00035914001  & 	2006 Nov 19			 &  2346  &  1755  &  1399  &  700  \\	
    &  00035914002  & 	2006 Nov 22			 &   &  &  &  \\ 
    &  00035914004  & 	2007 Jan 05			 &   &  &  &  \\   
    &  00035914005  & 	2007 Jun 24			 &   &  &  &  \\ 
\hline
61  &  00035915001  & 	2007 Jan 30			 &  3037  &  3387  &  2162  &  1078 \\
    &  00035915002  & 	2007 Feb 26			 &   &  &  &  \\ 
    &  00035915003  & 	2007 Mar 27			 &   &  &  &  \\
    &  00035915004  & 	2007 Mar 29			 &   &  &  &  \\
\hline
62  &  00035916001  & 	2006 Aug 18			 &  2656  &  4225  &  1727  &  807  \\
    &  00035916003  & 	2006 Dec 31			 &   &  &  &  \\   
    &  00035916004  & 	2007 Jan 07			 &   &  &  &  \\    
\hline
90  &  00053602001  & 	2007 Apr 24			 &  8601  &  7946  &  6663  &  963  \\
    &  00035917001  & 	2007 Apr 27			 &   &  &  &    
\enddata
\tablecomments{
Observational data are separated according to group HCG ID
(column 1). In all cases there are several observation IDs corresponding
to a given HCG ID.
Observation IDs (column 2) and corresponding observing dates
(column 3) appear on the same row. Exposure times 
in each filter (columns 4-7) are totals for each HCG.
} 
\end{deluxetable*}

% \onecolumn   %not compatible with emulateapj

%\clearpage

%\clearpage

%\clearpage

%\clearpage

%\clearpage

%\clearpage

\subsection{GALEX and \uvot\ fluxes}\label{subsec_galuv}
UVOT and GALEX flux densities for our HCG sample are given
in \tr{tab-flux}.
These were obtained by
multiplying measured count-rates 
by the instrument-specific 
flux conversion factors given
in \tr{tab-fcf}.
These were taken from
\citet{poole2008}
for \uvot\ and the GALEX website for \galex\
\footnote{\href{http://galexgi.gsfc.nasa.gov/docs/galex/FAQ/counts_background.html}{http://galexgi.gsfc.nasa.gov/docs/galex/FAQ/counts\_background.html}}.
The data were corrected
for Galactic extinction, using the maps of \citet{schlegel1998}
and the extinction curve of \citet{Cardelli1989}.
For those groups with both GALEX and \uvot\ data,
we 
compare the
\uvot\ \mtwo\ (\lameff~=~2231\AA) and GALEX \nuv\ (\lameff~=~2271\AA)
measured fluxes, as these two
filters are closest both in effective and central wavelength.

In \fr{fig:sep_GalUV}
we plot the fractional difference
in flux between the two filters, \dflux,
against \mtwo\ flux.
On average, the points appear to be
distributed around zero, with the exception of two
prominent outliers (shown as diamonds in \fr{fig:sep_GalUV}).
The topmost one (HCG 31 F) lies
close to a bright, saturated star. 
Because of GALEX's lower resolution,
part of this emission inevitably leads to an overestimate of 
the background for HCG 31 F in the GALEX image, and thus
an overestimate of the
\mtwo $-$ \nuv\ flux difference.
Otherwise, the differences are likely due to the
differences in shape between the two response functions.
In spite of the similarity both in 
effective and central wavelength, the \nuv\ filter
is more sensitive to radiation from longer wavelengths.
For instance, the outlier at the bottom left of the plot 
(HCG 42 D) is 
a faint, early-type system. The redder color of
this system likely leads to a higher flux estimate in the \nuv\ filter.
In fact, 4 out of 5 galaxies
for which \dflux~$<0$ are E/S0s, and
12 out of 14 galaxies for which \dflux~$>0$
are S/Is.

If we exclude the two most extreme outliers
we obtain a mean fractional difference $0.03\pm0.11$.  Although no
general statements can be made from such a small sample
(19 sources), this
comparison, taken at face value, supports the 
publicly available flux conversion factors
for \swift\ and
\galex. 
Further, this comparison provides some evidence that \lq\lq coincidence
loss\rq\rq\ (see \scr{subsec_coi}) is not a major concern for \mtwo,
even in cases of fairly bright sources.
As explained in \scr{subsec_coi},
three sources, indicated by stars in \fr{fig:sep_GalUV}, namely
HCG 2 B, 31 G and 31 ACE, may suffer from
this effect, with HCG 31 ACE being the brightest and most likely to be
affected. If there were significant coincidence losses for these
in the \mtwo\ filter,
\nuv\ flux densities should be systematically higher, which is
clearly not the case.

\subsection{Coincidence loss}\label{subsec_coi}
UVOT is a photon counting detector.
It thus suffers from coincidence loss 
at high photon rates, when two or more
photons arrive at a similar location 
on the detector within the same CCD read out interval
\footnote{In X-ray work the term {\it pile-up} is
commonly used for this effect.}.
Since we use \uvot\ to make quantitative estimates, we
investigate coincidence loss in the \uvot\ filters in greater
detail. \citet{poole2008} have calculated coincidence loss corrections
in 5\arcsec\ radius circular apertures for point sources (see their
Figure 6) by comparing theoretical and observed count rates. 
They find that
coincidence losses start to become important at $\sim$~10~\runits.
At this count-rate level the true flux is estimated to be 5\%\ higher.
Guided by this result, 
we identified UV surface brightness peaks for all galaxies
in our sample and
obtained total, non-background subtracted
count-rates in all filters within such circular regions centered at
the surface brightness peaks
\footnote{We stress that these regions are {\it different} from
the ones used to obtain total count rates for individual {\it galaxies}.
(\scr{subsec_sources}).}.

The results of this investigation for our sample are
shown in \tr{tab-uvcoi}. This Table
aims to provide an estimate of the
possible importance 
of coincidence loss in the
ultraviolet UVOT filters for
specific galaxies in our sample.
We only present count rates for
those circular regions that either exceed 10~\runits\ or
have between 5 and 10~\runits, for the three UV filters.
Among all 5\arcsec\ circular source regions, there are 21, 4, 1 and 4
sources with total count rates higher than 10~\runits\ in the $u$,
\wone, \mtwo\ and \wtwo\ filters, respectively.
The $u$ filter has the highest count-rate
values (up to $\sim 90$~\runits), and thus clearly suffers from
coincidence loss.  In contrast, the shortest UV filter, \wtwo, which
is most relevant to our science results, is only modestly affected, and
only for very few sources.
For these sources such a result is not surprising, as these
are found in some of the most UV-bright galaxies and
\hone-rich groups, with high levels of star-formation.
For instance, the highest count rate level ($\sim 23$~\runits) in this filter
is found at the center of
the A-C merging galaxy complex in the highly disturbed central region
of HCG 31. This complex, classified as an \htwo\ region, is the most
luminous in our sample. 
However, this is an exceptional case in this filter.
The other three high-count-rate sources 
in \wtwo\ are
all very close to $\sim 10$~\runits, and we consider these
borderline-importance cases.
Regarding the range
$5<$~\runits~$\le10$ there are 10, 2, 3 and 2 sources for the $u$,
\wone, \mtwo\ and \wtwo\ filters, respectively.  The two sources in
\wtwo\ are both at $\sim6$~\runits. 

To summarize, coincidence loss probably only moderately affects
a few bright sources in our sample. In the \wtwo\ filter
it most likely only affects one source. This may have a minor
effect on our quantitative results; our qualitative results
remain completely unaffected (see \scr{subsubsec_results}).

%Production:
%[dhcg]
%hcgfx.bash > all_flux.txt
%gawk '{print $1, $2"\\pm"$3, $4"\\pm"$5, $6"\\pm"$7, $8"\\pm"$9, $10"\\pm"$11}' all_flux.txt > prov
\tabletypesize{\scriptsize} %outside for emulateapj
\begin{deluxetable*}{cccc cccc}
\tablecolumns{8}
%\tablewidth{0pc} 
\setlength{\tabcolsep}{0.22cm}
%%\tablewidth{0pt}
\tablecaption{ULTRAVIOLET AND INFRARED FLUX DENSITIES FOR ACCORDANT HCG GALAXIES\label{tab-flux}}
\tablehead{
\colhead{} 
& \multicolumn{3}{c}{\uvot}
& \multicolumn{2}{c}{\galex}
& \colhead{2MASS}
& \colhead{MIPS}
\\ 
\colhead{} 
& \multicolumn{3}{c}{\rule{6cm}{.02cm}}
& \multicolumn{2}{c}{\rule{4cm}{.02cm}}
& \colhead{}
& \colhead{}
\\
\colhead{}
& \colhead{\wtwo\tablenotemark{a}}
& \colhead{\mtwo}
& \colhead{\wone}
& \colhead{\fuv}
& \colhead{\nuv}
& \colhead{$K_s$}
& \colhead{}
\\ 
\colhead{}
& \colhead{2030\AA}
& \colhead{2231\AA}
& \colhead{2634\AA}
& \colhead{1528\AA}
& \colhead{2271\AA}
& \colhead{2.17\micron\tablenotemark{b}}
& \colhead{24\micron\tablenotemark{b}}
\\ 
\colhead{HCG ID}
& \colhead{(mJy)}
& \colhead{(mJy)}
& \colhead{(mJy)}
& \colhead{(mJy)}
& \colhead{(mJy)}
& \colhead{(mJy)}
& \colhead{(mJy)}
}
\startdata
  02a & 3.894 $\pm$ 0.152$^{\rm S}$ & 4.185 $\pm$ 0.141 & 4.203 $\pm$ 0.170 & 2.839 $\pm$ 0.274 & 3.927 $\pm$ 0.371 &   24.7 $\pm$    2.5 &  115.0 $\pm$   11.5 \\ 
  02b & 1.613 $\pm$ 0.067$^{\rm S}$ & 1.889 $\pm$ 0.070 & 1.772 $\pm$ 0.077 & 1.112 $\pm$ 0.111 & 1.832 $\pm$ 0.176 &   20.9 $\pm$    2.1 &  351.0 $\pm$   35.1 \\ 
  02c & 0.984 $\pm$ 0.043$^{\rm S}$ & 1.036 $\pm$ 0.042 & 0.991 $\pm$ 0.045 & 0.740 $\pm$ 0.075 & 0.996 $\pm$ 0.097 &   12.7 $\pm$    1.3 &   21.5 $\pm$    2.1 \\ 
  07a & 0.885 $\pm$ 0.037$^{\rm S}$ & 0.935 $\pm$ 0.036 & 1.714 $\pm$ 0.071 & 0.506 $\pm$ 0.051 & 0.883 $\pm$ 0.087 &  130.0 $\pm$    13.0 &  303.0 $\pm$   30.3 \\ 
  07b & 0.258 $\pm$ 0.013$^{\rm E}$ & 0.250 $\pm$ 0.013 & 0.736 $\pm$ 0.033 & 0.135 $\pm$ 0.015 & 0.292 $\pm$ 0.031 &   68.8 $\pm$    6.9 &   12.7 $\pm$    1.3 \\ 
  07c & 2.113 $\pm$ 0.083$^{\rm S}$ & 2.248 $\pm$ 0.076 & 2.513 $\pm$ 0.101 & 1.588 $\pm$ 0.155 & 2.212 $\pm$ 0.210 &   54.9 $\pm$    5.5 &   76.0 $\pm$    7.6 \\ 
  07d & 0.632 $\pm$ 0.027$^{\rm S}$ & 0.682 $\pm$ 0.028 & 0.876 $\pm$ 0.039 & 0.480 $\pm$ 0.049 & 0.662 $\pm$ 0.066 &   29.7 $\pm$    3.0 &   12.0 $\pm$    1.2 \\ 
  16a & 1.769 $\pm$ 0.070$^{\rm S}$ & 1.993 $\pm$ 0.069 & 2.917 $\pm$ 0.118 &       \nodata     &       \nodata     &  159.0 $\pm$   15.9 &  409.0 $\pm$   40.9 \\ 
  16b & 0.238 $\pm$ 0.012$^{\rm S}$ & 0.237 $\pm$ 0.013 & 0.560 $\pm$ 0.027 &       \nodata     &       \nodata     &   91.8 $\pm$    9.2 &   22.5 $\pm$    2.2 \\ 
  16c & 2.358 $\pm$ 0.092$^{\rm S}$ & 2.621 $\pm$ 0.088 & 3.254 $\pm$ 0.130 &       \nodata     &       \nodata     &   82.9 $\pm$    8.3 & 1412.0 $\pm$  141.2 \\ 
  16d & 0.374 $\pm$ 0.017$^{\rm S}$ & 0.389 $\pm$ 0.019 & 0.830 $\pm$ 0.037 &       \nodata     &       \nodata     &   72.0 $\pm$    7.2 & 1785.0 $\pm$  178.5 \\ 
  19a & 0.156 $\pm$ 0.010$^{\rm E}$ & 0.143 $\pm$ 0.011 & 0.400 $\pm$ 0.021 &       \nodata     &       \nodata     &   40.6 $\pm$    4.1 &    3.3 $\pm$    1.6 \\ 
  19b & 0.338 $\pm$ 0.017$^{\rm S}$ & 0.349 $\pm$ 0.018 & 0.374 $\pm$ 0.020 &       \nodata     &       \nodata     &   11.8 $\pm$    1.2 &   24.1 $\pm$    2.4 \\ 
  19c & 0.550 $\pm$ 0.025$^{\rm S}$ & 0.570 $\pm$ 0.026 & 0.520 $\pm$ 0.025 &       \nodata     &       \nodata     &    5.1 $\pm$    1.0 &    6.4 $\pm$    1.3 \\ 
  22a & 1.386 $\pm$ 0.057$^{\rm E}$ & 1.128 $\pm$ 0.045 & 2.635 $\pm$ 0.108 & 0.814 $\pm$ 0.084 & 1.018 $\pm$ 0.102 &  297.0 $\pm$   29.7 &   13.9 $\pm$    1.4 \\ 
  22b & 0.125 $\pm$ 0.009$^{\rm S}$ & 0.120 $\pm$ 0.011 & 0.218 $\pm$ 0.013 & 0.092 $\pm$ 0.012 & 0.139 $\pm$ 0.016 &   20.5 $\pm$    2.1 &    1.1 $\pm$    0.6 \\ 
  22c & 1.532 $\pm$ 0.063$^{\rm S}$ & 1.603 $\pm$ 0.060 & 1.525 $\pm$ 0.065 & 1.203 $\pm$ 0.122 & 1.315 $\pm$ 0.128 &   20.5 $\pm$    2.1 &   29.4 $\pm$    2.9 \\ 
31ace & 7.212 $\pm$ 0.301$^{\rm S}$ & 8.000 $\pm$ 0.254 & 6.955 $\pm$ 0.274 & 6.164 $\pm$ 0.593 & 7.407 $\pm$ 0.690 &   21.3 $\pm$    2.1 &  463.0 $\pm$   46.3 \\ 
  31b & 1.219 $\pm$ 0.064$^{\rm S}$ & 1.342 $\pm$ 0.052 & 1.058 $\pm$ 0.048 & 1.018 $\pm$ 0.105 & 1.330 $\pm$ 0.127 &    3.5 $\pm$    1.8 &   16.6 $\pm$    1.7 \\ 
  31f & 0.373 $\pm$ 0.026$^{\rm S}$ & 0.514 $\pm$ 0.026 & 0.731 $\pm$ 0.036 & 0.226 $\pm$ 0.027 & 0.277 $\pm$ 0.031 &    \nodata &    2.8 $\pm$    1.4 \\ 
  31g & 2.388 $\pm$ 0.112$^{\rm S}$ & 2.618 $\pm$ 0.092 & 1.905 $\pm$ 0.081 & 2.370 $\pm$ 0.234 & 2.526 $\pm$ 0.237 &    7.9 $\pm$    1.6 &   38.9 $\pm$    3.9 \\ 
  31q & 0.121 $\pm$ 0.013$^{\rm S}$ & 0.131 $\pm$ 0.011 & 0.083 $\pm$ 0.007 & 0.136 $\pm$ 0.017 & 0.169 $\pm$ 0.017 &    0.9 $\pm$    0.9 &    0.7 $\pm$    0.7 \\ 
  42a & 0.999 $\pm$ 0.042$^{\rm E}$ & 0.848 $\pm$ 0.036 & 2.663 $\pm$ 0.110 & 0.335 $\pm$ 0.038 & 0.835 $\pm$ 0.085 &  372.0 $\pm$   37.2 &   25.5 $\pm$    2.5 \\ 
  42b & 0.167 $\pm$ 0.010$^{\rm E}$ & 0.137 $\pm$ 0.011 & 0.408 $\pm$ 0.022 & 0.040 $\pm$ 0.007 & 0.157 $\pm$ 0.018 &   46.6 $\pm$    4.7 &    4.5 $\pm$    2.2 \\ 
  42c & 0.204 $\pm$ 0.012$^{\rm E}$ & 0.205 $\pm$ 0.014 & 0.587 $\pm$ 0.029 & 0.087 $\pm$ 0.013 & 0.213 $\pm$ 0.025 &   55.4 $\pm$    5.5 &    2.8 $\pm$    1.4 \\ 
  42d & 0.023 $\pm$ 0.004$^{\rm E}$ & 0.023 $\pm$ 0.006 & 0.056 $\pm$ 0.006 & 0.005 $\pm$ 0.002 & 0.039 $\pm$ 0.005 &    8.1 $\pm$    1.6 &    0.4 $\pm$    0.4 \\ 
  48a & 0.868 $\pm$ 0.037$^{\rm E}$ & 0.681 $\pm$ 0.031 & 2.316 $\pm$ 0.096 &       \nodata     &       \nodata     &  263.7 $\pm$   26.4 &   12.8 $\pm$    1.3 \\ 
  48b & 2.054 $\pm$ 0.082$^{\rm S}$ & 2.297 $\pm$ 0.081 & 2.551 $\pm$ 0.105 &       \nodata     &       \nodata     &   37.7 $\pm$    3.8 &   81.9 $\pm$    8.2 \\ 
  48c & 0.054 $\pm$ 0.006$^{\rm S}$ & 0.040 $\pm$ 0.008 & 0.149 $\pm$ 0.011 &       \nodata     &       \nodata     &   18.6 $\pm$    1.9 &    0.8 $\pm$    0.8 \\ 
  48d & 0.033 $\pm$ 0.005$^{\rm E}$ & 0.026 $\pm$ 0.007 & 0.081 $\pm$ 0.007 &       \nodata     &       \nodata     &    8.1 $\pm$    1.6 &    0.4 $\pm$    0.4 \\ 
  59a & 0.177 $\pm$ 0.011$^{\rm S}$ & 0.176 $\pm$ 0.013 & 0.345 $\pm$ 0.019 &       \nodata     &       \nodata     &   19.4 $\pm$    1.9 &  453.0 $\pm$   45.3 \\ 
  59b & 0.037 $\pm$ 0.005$^{\rm E}$ & 0.044 $\pm$ 0.007 & 0.076 $\pm$ 0.007 &       \nodata     &       \nodata     &   10.2 $\pm$    1.0 &    0.6 $\pm$    0.6 \\ 
  59c & 0.212 $\pm$ 0.012$^{\rm S}$ & 0.230 $\pm$ 0.015 & 0.192 $\pm$ 0.013 &       \nodata     &       \nodata     &    3.0 $\pm$    1.5 &    3.6 $\pm$    1.8 \\ 
  59d & 0.870 $\pm$ 0.038$^{\rm S}$ & 0.942 $\pm$ 0.041 & 0.831 $\pm$ 0.039 &       \nodata     &       \nodata     &    3.4 $\pm$    1.7 &   12.7 $\pm$    1.3 \\ 
  61a & 0.980 $\pm$ 0.044$^{\rm S}$ & 0.654 $\pm$ 0.029 & 1.100 $\pm$ 0.048 &       \nodata     &       \nodata     &  145.0 $\pm$   14.5 &   20.5 $\pm$    2.0 \\ 
  61c & 0.944 $\pm$ 0.042$^{\rm S}$ & 0.689 $\pm$ 0.031 & 0.818 $\pm$ 0.037 &       \nodata     &       \nodata     &   90.4 $\pm$    9.0 &  357.0 $\pm$   35.7 \\ 
  61d & 0.312 $\pm$ 0.017$^{\rm E}$ & 0.247 $\pm$ 0.015 & 0.351 $\pm$ 0.018 &       \nodata     &       \nodata     &   33.3 $\pm$    3.3 &    2.5 $\pm$    1.2 \\ 
  62a & 0.508 $\pm$ 0.024$^{\rm E}$ & 0.410 $\pm$ 0.021 & 1.816 $\pm$ 0.079 &       \nodata     &       \nodata     &  155.0 $\pm$   15.5 &    9.4 $\pm$    1.9 \\ 
  62b & 0.137 $\pm$ 0.010$^{\rm E}$ & 0.176 $\pm$ 0.013 & 0.670 $\pm$ 0.038 &       \nodata     &       \nodata     &   63.5 $\pm$    6.4 &    3.6 $\pm$    1.8 \\ 
  62c & 0.128 $\pm$ 0.010$^{\rm E}$ & 0.196 $\pm$ 0.014 & 0.459 $\pm$ 0.035 &       \nodata     &       \nodata     &   27.9 $\pm$    2.8 &    2.1 $\pm$    1.1 \\ 
  62d & 0.018 $\pm$ 0.004$^{\rm E}$ & 0.022 $\pm$ 0.006 & 0.065 $\pm$ 0.007 &       \nodata     &       \nodata     &    7.9 $\pm$    1.6 &    1.9 $\pm$    0.9  
\enddata
\tablecomments{}
\tablenotetext{a}{
Flux densities in this column have been obtained after multiplying original
flux densities in the \wtwo\ filter by a factor $\alpha$
to compensate for the filter's red tail (see \scr{subsec_redleak}).
Accordingly, entries are flagged either \lq\lq S\rq\rq\
($\alpha = 0.8$) or \lq\lq E\rq\rq\ ($\alpha = 0.5$).
}
\tablenotetext{b}{
Data from J07.
}
\tablecomments{
Flux densities are calculated from measured count rates by application
of the flux conversion factors given in \tr{tab-fcf}.
}
\end{deluxetable*}
%\clearpage

\subsection{\wtwo\ red leak}\label{subsec_redleak}
Another concern with the \wtwo\ filter is the very shallow slope of
the filter response function towards longer wavelengths, which 
in the case of very red sources can
potentially lead to significant contamination of the UV flux with
emission from longer wavelengths \citep[\lq\lq
red leak\rq\rq,][]{roming2009}
\footnote{This is also an issue with the \wone\ filter.}.
In order to investigate the effect of
this red tail on different types of galaxies, we folded the full \wtwo\
effective area (response function, \aeff) curve, as well as modified
effective area curves, \aeffc, with galaxy templates for different morphological
types. We used the publicly available templates produced with the
chemical evolution code GRAZIL \citep{silva1998}.
These models (three spirals [Sa, Sb, Sc],
and one elliptical) include the effects of dust and
provide good fits to local galaxy spectral
energy distributions (SEDs).
The \wtwo\ central wavelength, \lamc, defined as the midpoint
between the FWHM wavelengths, is at 1928\AA\ \citep{poole2008}.
To modify the area curve redwards of \lamc, we need to impose
an artificial cut-off wavelength, \lamr,
to \aeff\ redwards of the central wavelength,
so that \aeffc\ represents a new effective area curve,
corresponding to a modified filter \wtwoc.
Since the \aeff\ starting wavelength is at about \lamc\ -- 1$\times$FWHM,
by symmetry, one obvious choice
is \lamr\ = \lamc\ + 1$\times$FWHM 
$\simeq 2260$\AA. 

\begin{deluxetable}{ccc} 
\tablecolumns{3} 
\tablewidth{250 pt} 
\tablecaption{ULTRAVIOLET COUNT-RATE-TO-FLUX CONVERSION FACTORS\label{tab-fcf}}
\tablehead{
\colhead{Filter} 
&\colhead{Conversion factor}%\tablenotemark{b}} 
&\colhead{RMS}
\\
\colhead{}
&\colhead{(erg~cm$^{-2}$~\AA$^{-1}$)}
&\colhead{(erg~cm$^{-2}$~\AA$^{-1}$)}
}
\startdata
\multicolumn{3}{c}{\uvot\tablenotemark{a}} \\
\hline
$u$    & \ten{1.63}{-16} & \ten{2.5}{-18} \\
\wone\ & \ten{4.00}{-16} & \ten{9.7}{-18} \\
\mtwo\ & \ten{8.50}{-16} & \ten{5.6}{-18} \\
\wtwo\ & \ten{6.2}{-16}  & \ten{1.4}{-17} \\
\hline
\multicolumn{3}{c}{\galex\tablenotemark{b}} \\
\hline
\nuv\ & \ten{2.06}{-16}  & \nodata\tablenotemark{c} \\
\fuv\ & \ten{1.40}{-15}  & \nodata\tablenotemark{c} 
\enddata
\tablecomments{}
\tablenotetext{a}
{Data taken from \citet{poole2008}, Table 10.}
\tablenotetext{b}
{\href{http://galexgi.gsfc.nasa.gov/docs/galex/FAQ/counts_background.html}{\tiny http://galexgi.gsfc.nasa.gov/docs/galex/FAQ/counts\_background.html}}
\tablenotetext{c}
{No data available.}
\end{deluxetable}

In \fr{fig:aeffc} we illustrate 
how the integrated flux in a truncated \wtwoc\ filter
changes for several \lamr\ values.
Specifically, we show the trend of the ratio of the integrated
fluxes in \wtwoc\ and \wtwo\ as 
a function of \lamr. With \lamr\ = 2260\AA, the artificially 
\lq\lq narrowed\rq\rq\ filter contains $\sim 0.5$ of the flux in the
full filter for the elliptical template and $\sim 0.8$ for the
spiral templates. 
Conversely, this implies that $\sim 0.5$ (0.2) of the flux in the
original, full filter is due to emission redwards of 2260\AA\ for the elliptical
template (spiral templates).
Even if the cutoff is moved 200\AA\ further towards
the red, the difference in flux between full and narrowed filter
is significant, with filter-flux ratios of
$\sim 0.6$ and $\sim 0.9$ for the two main morphological classifications.
These results will likely vary depending on the particular SEDs of
real galaxies, reflecting a variety of star-formation history
and dust content. However, similar trends can be observed with
different templates, e.g. \citet{bruzual2003}, and so in order to
minimize contamination of UV flux by non-UV emission
from the red tail, we impose a cut-off at 2260\AA.
This cut-off assumes that, in reality, only a fraction, $\alpha$,
of the calculated flux density in the \wtwo\ filter comes
from the UV wavelength region. For the assumed cut-off, $\alpha = 0.5$ 
for E/S0 galaxies and 0.8 otherwise. 
Although this choice directly affects SFR estimates, 
the quantitative effect is small; qualitatively our results remain
entirely unaffected (see \scr{subsubsec_results} for details).

\subsection{Comparison sample}\label{subsec_comp}
For the purposes of comparison, 
we use the
\spitzer\ infrared nearby galaxy survey
\citep[SINGS,][]{kennicutt2003} to construct
a comparison sub-sample. As in G08, we
select galaxies which have
log~$L_{\nu, J} = 27.70 - 30.17$ (\lnuunits), in
order to better match the HCG sample
luminosity range. The HCG sample comprises
mostly 3-5 bright galaxies per group, and this
selection filters mostly low-luminosity 
SINGS dwarfs. In addition, we use the 
non-interacting (\lq\lq normal\rq\rq) 
SINGS catalog of \citet[][their Table 2]{smithb2007} to
remove galaxies 
for which there is evidence that they
are strongly interacting. 
According to \citet{smithb2007},
target galaxies are deemed interacting
if \lq\lq they have 
companions whose velocities differ from that
of the target galaxy by $\le1000$ \kmps, that
have an optical luminosity brighter than
0.1 of the target galaxy, and are separated
from the target galaxy by less than 10 times
the optical diameter of the
target galaxy or the companion, whichever is larger.\rq\rq\
These criteria still allow for some of these
galaxies to have distant or low-mass companions but,
as \citet{smithb2007} note,
they are guaranteed to be
less perturbed than the Milky Way.
This additional selection is in fact 
warranted by a
clear effect on the final sub-sample: It
leads to the elimination of all
SINGS galaxies which have some level of
peculiarity (their morphological designation
includes \lq\lq p\rq\rq).
We thus obtain a SINGS control sub-sample of 33 galaxies
which in the broadest sense are similar to our
HCG galaxies (due to the $L_{\nu, J}$ matching) but
are otherwise in relatively quiescent environments.
Although a few of these galaxies are members of the 
Virgo galaxy cluster, they still fulfill the non-interaction and
isolation criteria, and we do not exclude them
from the comparison sample.
We use the photometry in Tables 2 and 3 of
\citet{dale2007} to calculate the same parameters
as for the HCG galaxies.

\begin{figure}[h]
\epsscale{1.2}
\plotone{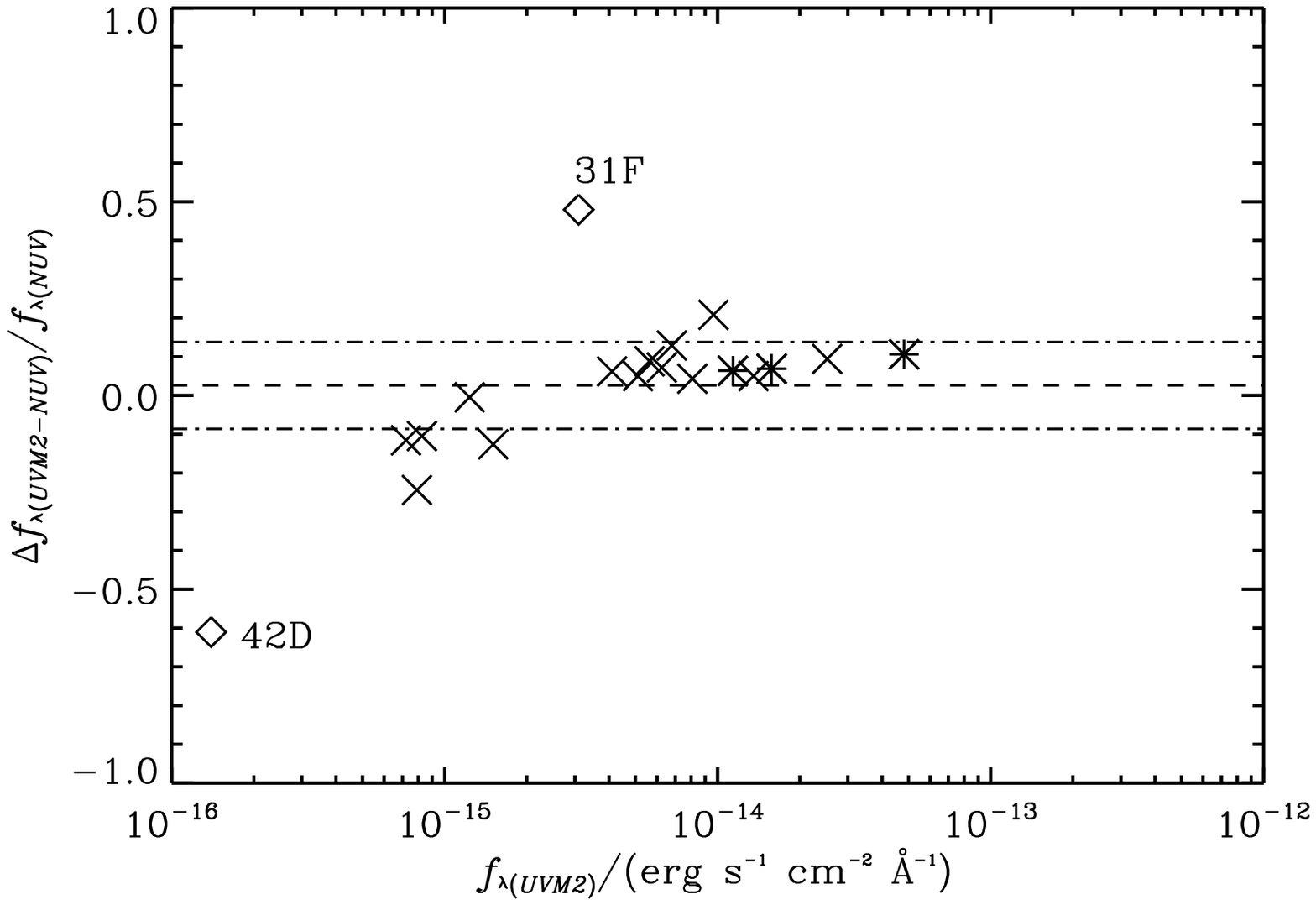}  
\caption{Fractional difference in flux between
the UVOT \mtwo\ and \galex\ \nuv\ filters
for HCG galaxies observed with both \galex\ and \swift.
The two extreme outliers, HCG 31F and HCG 42D, are indicated by diamonds.
The three sources which may be subject to some coincidence loss in the
\mtwo\ filter are shown by asterisks and are HCGs 2B, 31G and 31ACE in order
of increasing \mtwo\ flux.
The dashed line
indicates the mean fractional 
difference calculated from a subset of the points,
excluding the two outliers. The dash-dotted lines indicate
the standard deviation on the mean result.
\label{fig:sep_GalUV}}
\end{figure}

\begin{deluxetable}{cccc} 
\tablecolumns{4} 
\tablewidth{250 pt} 
\tablecaption{Importance of Coincidence Loss in UVOT HCG Data\label{tab-uvcoi}}
\tablehead{
\colhead{HCG\tablenotemark{a}} 
&\colhead{\wone}%\tablenotemark{b}} 
&\colhead{\mtwo}%\tablenotemark{b}} 
&\colhead{\wtwo}\\ %\tablenotemark{b}}\\
&\colhead{(\runits)}
&\colhead{(\runits)}
&\colhead{(\runits)}
}
\startdata
\multicolumn{4}{c}{$>10$~\runits} \\
\hline\hline
2 B      &12.33 &$\le10$  &10.43 \\
16 C     &16.46 &$\le10$  &12.85 \\
31 ACE   &22.6  &15.17    &23.40 \\
%%31 ACE 3 &57.4  &27.02 &31.20 \\ This is a star, partly in the background
31 G     &11.57 &$\le10$  &11.70 \\
\hline
\multicolumn{4}{c}{$5<$~\runits~$\le10$}\\
\hline\hline
2 A      &6.74              & $\le5$    &5.84 \\
2 B      &$\le5$            &6.70       &$\le5$     \\
16 A     &9.38              &$\le5$     &5.85 \\
16 C     &$\le5$            &8.71       &$\le5$    \\
31 G     &$\le5$            &7.47       &$\le5$     
\enddata
\tablenotetext{a}{Column gives the HCG galaxy
for which the total count rate in a 5\arcsec-radius circular aperture 
centered on 
an emission peak is
shown in the adjacent columns with no background subtracted.}
%\tablenotetext{b}{Column gives \runits\ in this filter.}
\tablecomments
{
Total count rates are grouped for sources with $>10$~\runits\ (top)
and those with $5<$~\runits~$\le10$ (bottom) in the three ultraviolet \swift\ 
UVOT filters. Upper limit entries imply a source has a count rate lower
than the lower limit of its tabulated group (less than either 10
or 5~\runits).
}
\end{deluxetable}

\begin{figure}
\epsscale{1.2}
\plotone{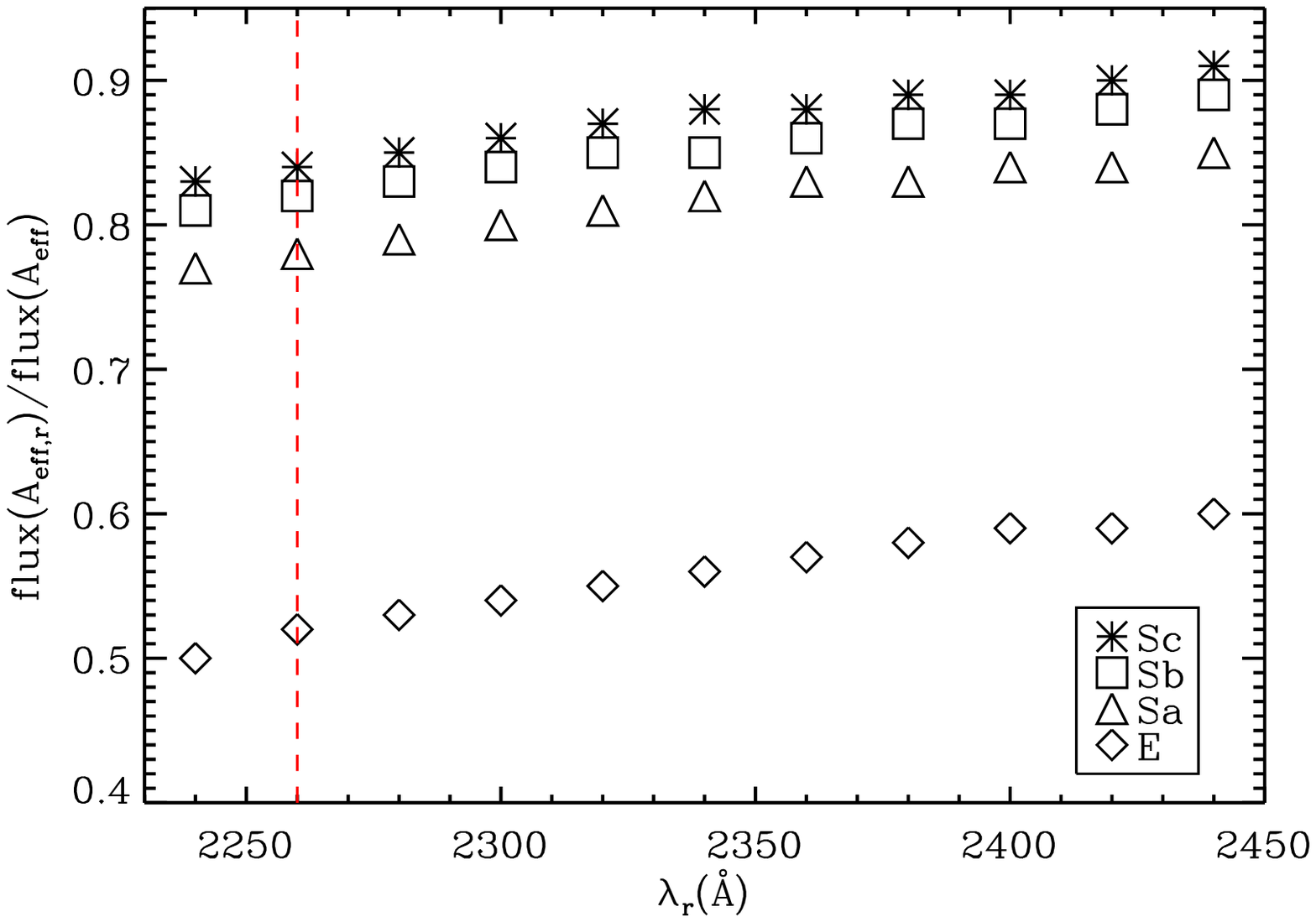}       
\caption{
Illustration of
effect of imposing an artificial cut-off to the 
\wtwo\ filter. 
The original full-filter effective
area, \aeff, is modified by a cut-off at \lamr,
so that the new effective area is \aeffc.
The original and modified filters
are folded with
model templates for different
galaxy types \citep{silva1998}, shown in the
legend, and the corresponding fluxes are
calculated.
The ratios of these fluxes are plotted
as a function of the imposed cut-off
wavelength. For all galaxy types,
it is clear that there is a
significant difference
between the flux level 
in the full filter and that
in the narrowed filter, i.e.
there is significant contribution
from emission captured by the filter's red tail
which does not originate in the UV wavelength region.
Thus, as explained in the text, 
for the purpose of calculations in 
this paper, a fraction
of the total filter flux corresponding
to the estimated red tail contribution
is excluded. The adopted value
for the cut-off wavelength is shown by the
dashed line.
}
\label{fig:aeffc}
\end{figure}

\section{Results}\label{sec_results}

\subsection{UV and IR comparisons}\label{subsec_uvir}
In \fr{fig:w2-24-lnu} we plot monochromatic luminosities, \lnu,
in the \wtwo\ and 24\micron\ bands for individual galaxies in 
the 11 HCGs of our sample.
In this and following plots, we use
different symbols to separate
galaxies according to morphological type
(either E/S0 or S/I) and
parent-group \hone-gas content (rich - type I,
intermediate - type II, poor - type III).
\wtwo\ luminosity correlates with
24\micron\ luminosity up to
log \lnu (24\micron) $\sim 30$ where
a turn-over and/or larger scatter dominate.
This is likely the effect of higher dust
attenuation at higher luminosities and
star-formation rates
\citep{hopkins2001,buat2007}.
E/S0 galaxies and gas-poor groups tend to 
be less luminous than S/I galaxies and
gas-rich groups both in the \wtwo\
and in the 24\micron\ band.  

\begin{figure*}     
\epsscale{1.1} 
\plottwo{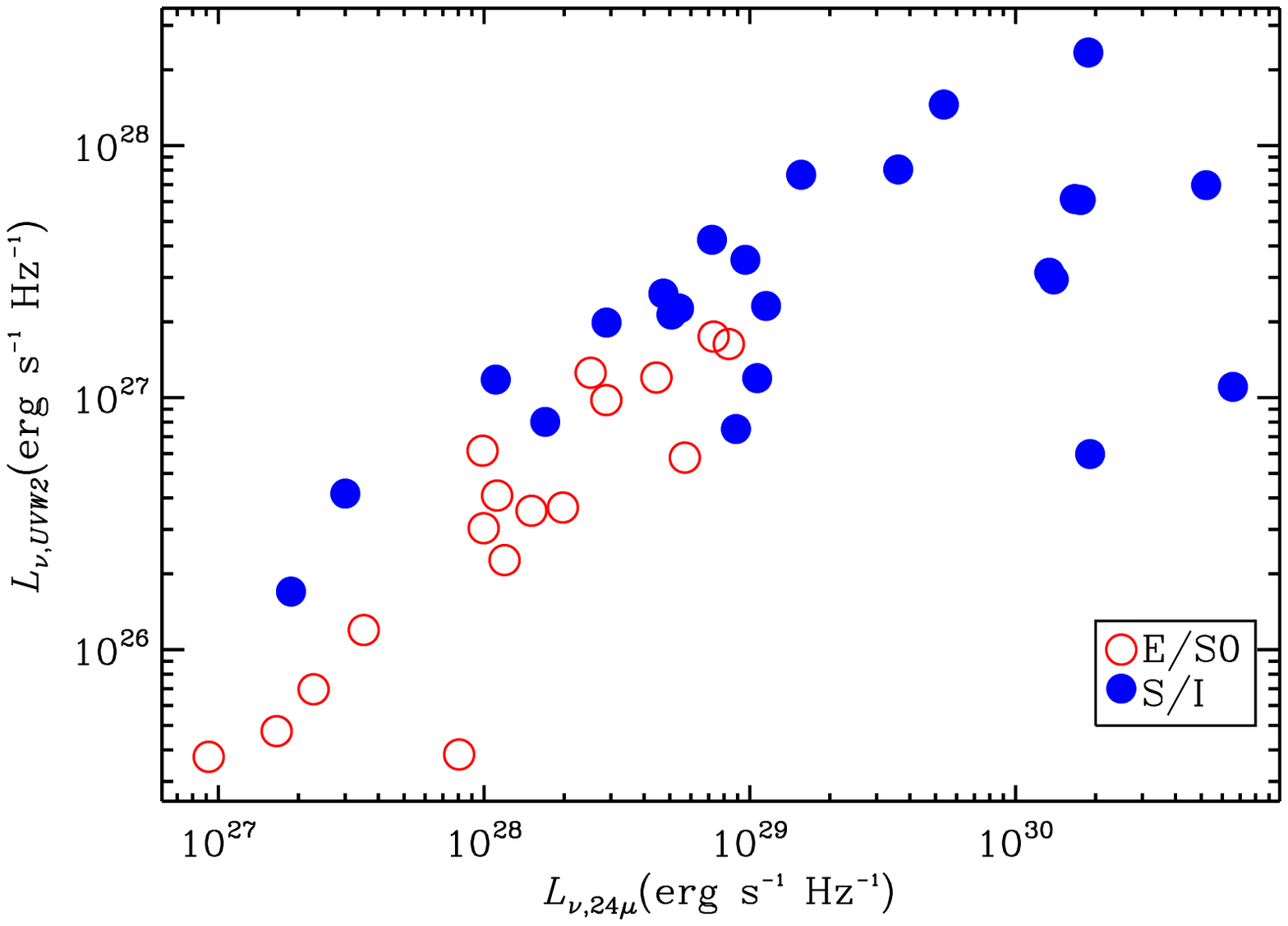}{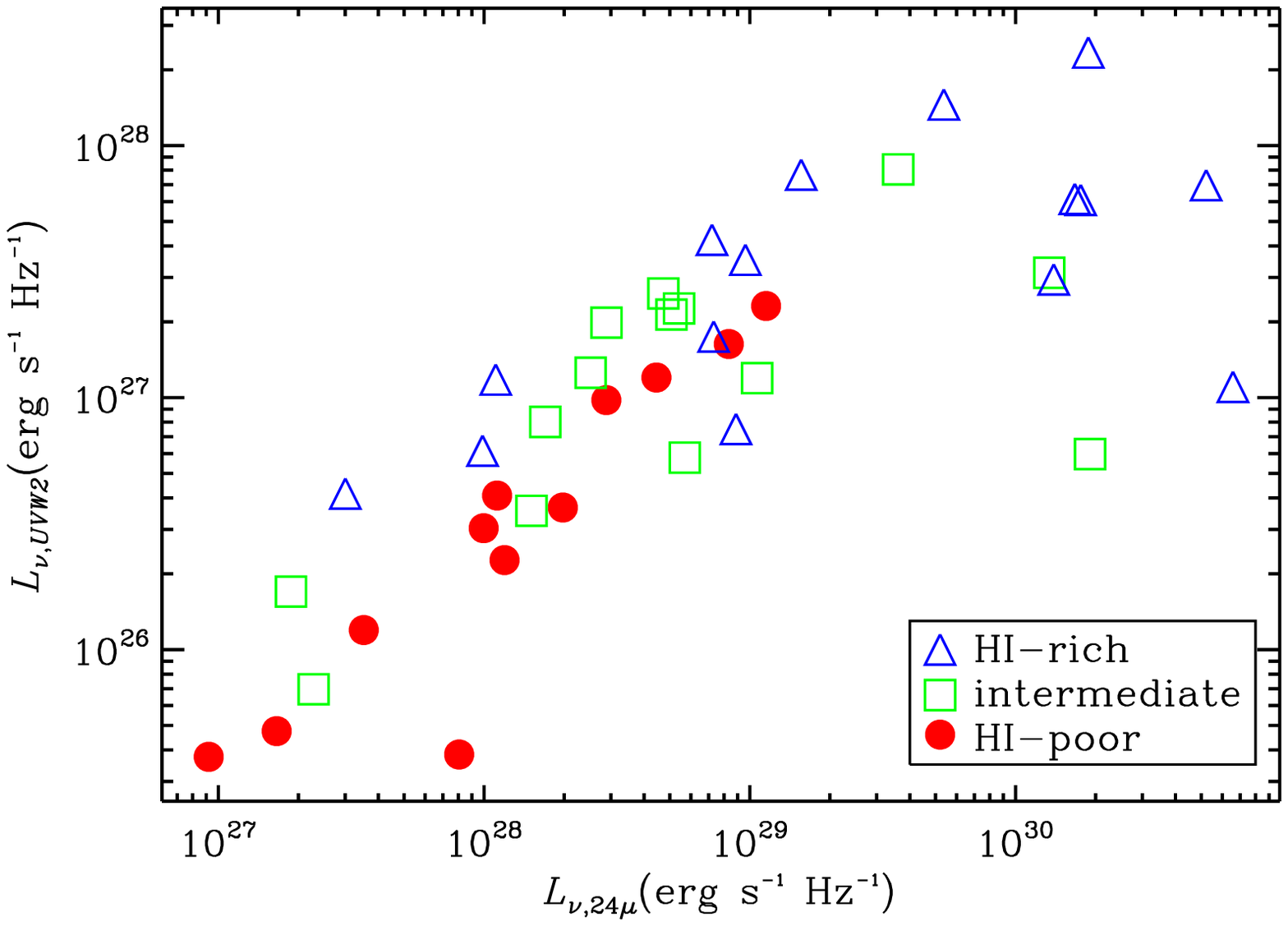}
\caption{\lnuwtwo\ versus \lnutf\ for HCG galaxies.
{\it Left:} E/S0s (S/Is) are shown with open (filled) circles.
{\it Right:} For each galaxy, parent-group gas-richness is indicated with
triangles (rich), squares (intermediate), filled circles (poor).
There is a good correlation of \wtwo\ and 24\micron\
luminosity up to log \lnu (24\micron) $\sim 30$.
}
\label{fig:w2-24-lnu}
\end{figure*}

This
luminosity segregation appears to be somewhat
more prononounced when galaxies are
classified by morphological type than it is
when galaxies are classified by
parent-group gas-richness.
We show later that this behavior characterizes other
star-formation related properties as well. 
The broad UV-IR correlation persists over 
$\sim 3$ dex in luminosity for both wavelength bands,
indicative, on average, of a consistent contribution from the UV.

\begin{figure}
\epsscale{1.3}
\hspace{-1.1cm}\plotone{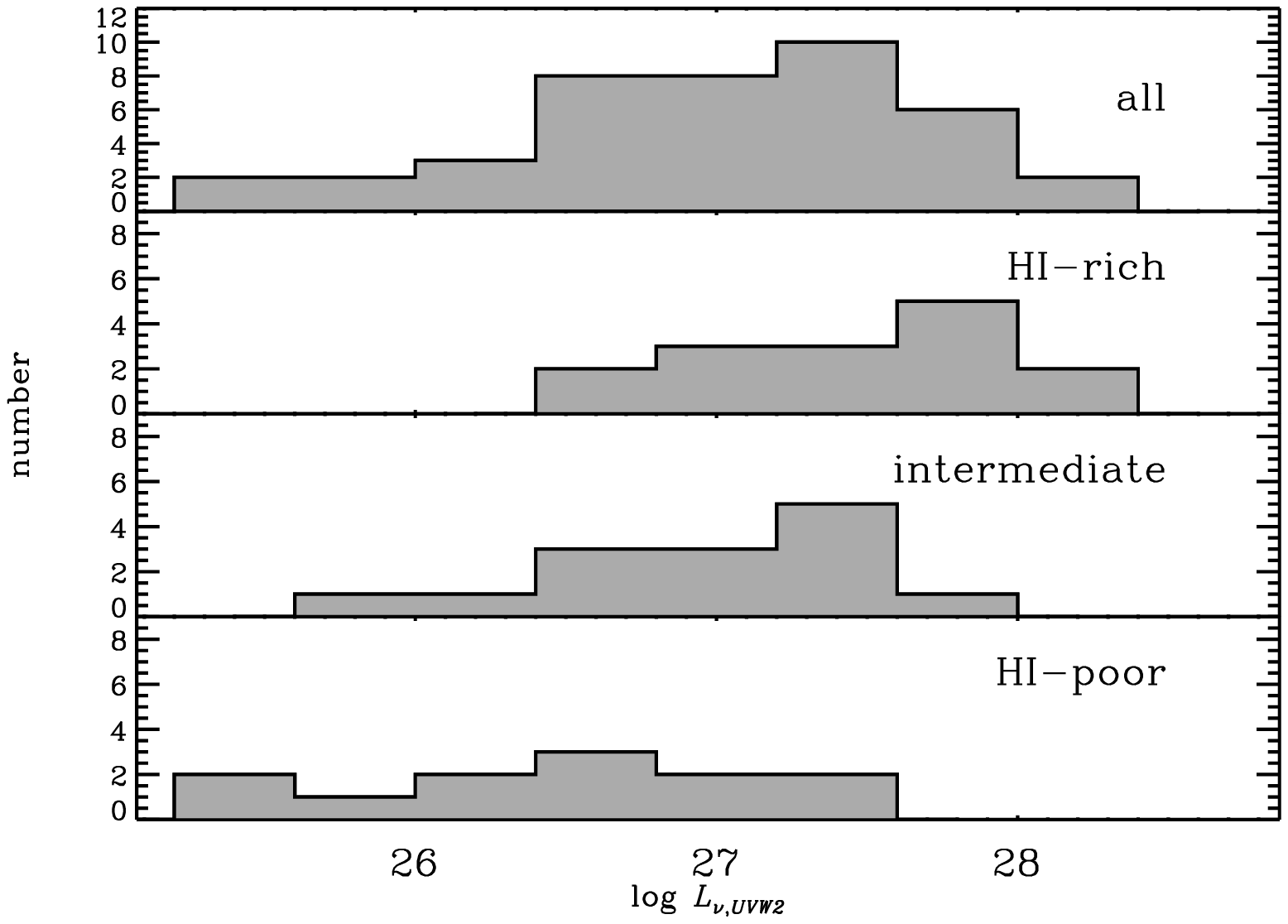}
%\plotone{w2-lnu_hist_I.eps}
%\plotone{w2-lnu_hist_II.eps}
%\plotone{w2-lnu_hist_III.eps}
\caption{\wtwo\ monochromatic luminosity distribution 
for HCG galaxies. Distributions are shown
for the full sample, as well as sub-samples according to group
\hone\ content. Group \hone-richness appears to correlate
with \lnuwtwo.}
\label{fig:w2-lnu_hist}
\end{figure}

\begin{figure}
\epsscale{1.3}
\hspace{-0.973cm}\plotone{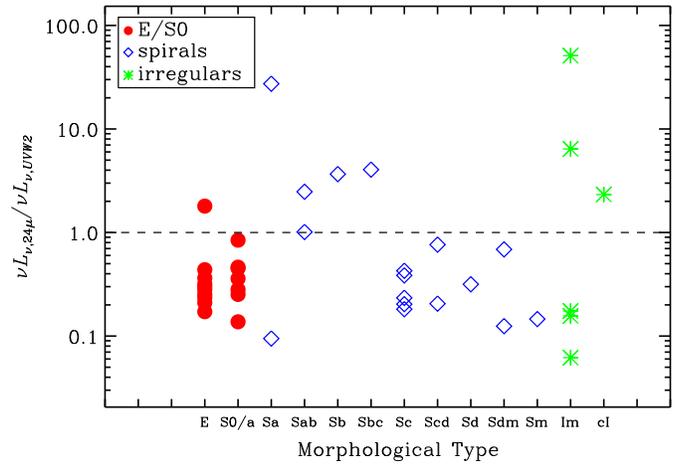} 
\caption{Ratio of 24\micron-to-\wtwo\ power, \iruv,
vs. galaxy optical morphology.
The horizontal dashed line marks the locus of equal contributions
from 24\micron\ and \wtwo\ power. There is a significant
contribution from the \wtwo\ band.}
\label{fig:24-w2-morph}
\end{figure}

%Made with:
%[uvir]
%This version made directly with SFR.pro
%Then add colhead commands and modify entries for HCG31F (nodata).
%\clearpage
%\begin{landscape}
\tabletypesize{\scriptsize} %outside for emulateapj
\begin{deluxetable*}{lccc cccc cccc c}
%\rotate
\tablecolumns{13}
\tablewidth{500 pt} 
\setlength{\tabcolsep}{.1 pt}
%%\tablewidth{0pt}
\tablecaption{UV And IR Properties Of HCG Galaxies\label{tab-sfr}}
\tablehead{
\colhead{}
& \colhead{}
& \colhead{}
& \colhead{}
& \multicolumn{4}{c}{\hspace{-1cm}log \lnu\ (\lnuunits)\tablenotemark{c}}
& \colhead{}
& \colhead{}
& \colhead{}
& \colhead{}
& \colhead{}
\\
\colhead{\hspace{-1.8cm}\sc HCG}  
& \colhead{\hspace{-2.8cm} Galaxy}
& \colhead{\hspace{-1cm}Nuclear}
 & \colhead{\hspace{-0.5cm}Group}
 & \multicolumn{4}{c}{\hspace{-1cm}\rule{4cm}{.02cm}} 
 & \colhead{\hspace{-1.4cm}\mstar\tablenotemark{d}}
 & \colhead{}
 & \colhead{\hspace{-0.7cm}\sc SFR\tablenotemark{f}}
 & \colhead{}
 & \colhead{\hspace{-0.5cm}\sc SSFR\tablenotemark{h}}
\\
\colhead{\hspace{-1.8cm}\sc ID}  
& \colhead{\hspace{-2.8cm} Morphology\tablenotemark{a}}
& \colhead{\hspace{-1cm}Type\tablenotemark{b}}
 & \colhead{\hspace{-0.5cm}\hone\ Type\tablenotemark{a}}
& \colhead{\hspace{-0.4cm}\wtwo}
& \colhead{\hspace{-0.4cm}\mtwo}
& \colhead{\hspace{-0.4cm}\wone}
& \colhead{\hspace{-1.0cm}24\micron}
& \colhead{\hspace{-1.4cm}($10^9$\msun)}
& \colhead{\hspace{-0.8cm}\airac\tablenotemark{e}}
& \colhead{\hspace{-0.7cm}(\msuny)}
& \colhead{\hspace{-0.7cm}$\frac{{\rm SFR}_{uvw2}}{{\rm SFR}_{\rm TOTAL}}$ \tablenotemark{g}}
& \colhead{\hspace{-0.5cm}($10^{-11}$ yr$^{-1}$)}
\\
\colhead{\hspace{-1.8cm}(1)}  
& \colhead{\hspace{-2.8cm}(2)}
& \colhead{\hspace{-0.9cm}(3)}
 & \colhead{\hspace{-0.5cm}(4)}
& \colhead{\hspace{-0.5cm}(5)}
& \colhead{\hspace{-0.5cm}(6)}
& \colhead{\hspace{-0.5cm}(7)}
& \colhead{\hspace{-0.9cm}(8)}
& \colhead{\hspace{-1.4cm}(9)}
& \colhead{\hspace{-0.8cm}(10)}
& \colhead{\hspace{-0.7cm}(11)}
& \colhead{\hspace{-0.7cm}(12)}
& \colhead{\hspace{-0.5cm}(13)}
}
\startdata
\multicolumn{13}{c}{\sc Actively Star-Forming}\\
\hline

   {2a}  &\hspace{-2.8cm} SBd  &\hspace{-1.1 cm}          R  &\hspace{-0.7cm}          I  &\hspace{-0.5cm}   28.16  &\hspace{-0.5cm}   28.29  &\hspace{-0.5cm}   28.29  &\hspace{-1cm}   29.73  &\hspace{-1.6cm}     3.32  &\hspace{-1.2cm} $ -2.89 \pm   0.03$  &\hspace{-0.7cm} $ 3.418 \pm  0.276$  &\hspace{-0.7cm}  0.60  &\hspace{-0.5cm} $ 102.89 \pm   16.31$ \\
   {2b}  &\hspace{-2.8cm}  cI  &\hspace{-1.1 cm}          R  &\hspace{-0.7cm}          I  &\hspace{-0.5cm}   27.79  &\hspace{-0.5cm}   27.95  &\hspace{-0.5cm}   27.93  &\hspace{-1cm}   30.22  &\hspace{-1.6cm}     2.86  &\hspace{-1.2cm} $ -3.76 \pm   0.18$  &\hspace{-0.7cm} $ 5.153 \pm  0.589$  &\hspace{-0.7cm}  0.17  &\hspace{-0.5cm} $ 179.95 \pm   31.97$ \\
   {2c}  &\hspace{-2.8cm} SBc  &\hspace{-1.1 cm}          ?  &\hspace{-0.7cm}          I  &\hspace{-0.5cm}   27.55  &\hspace{-0.5cm}   27.67  &\hspace{-0.5cm}   27.65  &\hspace{-1cm}   28.98  &\hspace{-1.6cm}     1.64  &\hspace{-1.2cm} $ -2.35 \pm   0.36$  &\hspace{-0.7cm} $ 0.741 \pm  0.060$  &\hspace{-0.7cm}  0.67  &\hspace{-0.5cm} $  45.27 \pm    7.23$ \\
   {7a}  &\hspace{-2.8cm}  Sb  &\hspace{-1.1 cm}        HII  &\hspace{-0.7cm}         II  &\hspace{-0.5cm}   27.50  &\hspace{-0.5cm}   27.62  &\hspace{-0.5cm}   27.88  &\hspace{-1cm}   30.13  &\hspace{-1.6cm}     16.6  &\hspace{-1.2cm} $ -2.23 \pm   0.09$  &\hspace{-0.7cm} $ 3.882 \pm  0.468$  &\hspace{-0.7cm}  0.11  &\hspace{-0.5cm} $ 23.46 \pm   4.25$ \\
   {7c}  &\hspace{-2.8cm} SBc  &\hspace{-1.1 cm}          ?  &\hspace{-0.7cm}         II  &\hspace{-0.5cm}   27.91  &\hspace{-0.5cm}   28.03  &\hspace{-0.5cm}   28.08  &\hspace{-1cm}   29.56  &\hspace{-1.6cm}     7.52  &\hspace{-1.2cm} $ -2.53 \pm   0.58$  &\hspace{-0.7cm} $ 2.056 \pm  0.169$  &\hspace{-0.7cm}  0.55  &\hspace{-0.5cm} $  27.33 \pm    4.34$ \\
   {7d}  &\hspace{-2.8cm} SBc  &\hspace{-1.1 cm}          ?  &\hspace{-0.7cm}         II  &\hspace{-0.5cm}   27.33  &\hspace{-0.5cm}   27.46  &\hspace{-0.5cm}   27.57  &\hspace{-1cm}   28.70  &\hspace{-1.6cm}     3.61  &\hspace{-1.2cm} $ -1.69 \pm   0.29$  &\hspace{-0.7cm} $ 0.429 \pm  0.035$  &\hspace{-0.7cm}  0.70  &\hspace{-0.5cm} $  11.89 \pm    1.90$ \\
  {16a}  &\hspace{-2.8cm}SBab  &\hspace{-1.1 cm}    LINER,X,R  &\hspace{-0.7cm}          I  &\hspace{-0.5cm}   27.78  &\hspace{-0.5cm}   27.93  &\hspace{-0.5cm}   28.10  &\hspace{-1cm}   30.25  &\hspace{-1.7cm}    19.68  &\hspace{-1.2cm} $ -2.85 \pm   0.26$  &\hspace{-0.7cm} $ 5.371 \pm  0.621$  &\hspace{-0.7cm}  0.16  &\hspace{-0.5cm} $  27.29 \pm    4.87$ \\
  {16c}  &\hspace{-2.8cm}  Im  &\hspace{-1.1 cm}       SBNG  &\hspace{-0.7cm}          I  &\hspace{-0.5cm}   27.84  &\hspace{-0.5cm}   27.99  &\hspace{-0.5cm}   28.08  &\hspace{-1cm}   30.72  &\hspace{-1.6cm}     8.81  &\hspace{-1.2cm} $ -4.39 \pm   0.46$  &\hspace{-0.7cm} $14.378 \pm  1.827$  &\hspace{-0.7cm}  0.07  &\hspace{-0.5cm} $ 163.12 \pm   30.39$ \\
  {16d}  &\hspace{-2.8cm}  Im  &\hspace{-1.1 cm}        LINER  &\hspace{-0.7cm}          I  &\hspace{-0.5cm}   27.04  &\hspace{-0.5cm}   27.16  &\hspace{-0.5cm}   27.49  &\hspace{-1cm}   30.82  &\hspace{-1.6cm}     7.64  &\hspace{-1.2cm} $ -3.79 \pm   0.07$  &\hspace{-0.7cm} $17.062 \pm  2.313$  &\hspace{-0.7cm}  0.01  &\hspace{-0.5cm} $ 223.34 \pm   43.01$ \\
  {19b}  &\hspace{-2.8cm} Scd  &\hspace{-1.1 cm}          ?  &\hspace{-0.7cm}         II  &\hspace{-0.5cm}   27.08  &\hspace{-0.5cm}   27.19  &\hspace{-0.5cm}   27.22  &\hspace{-1cm}   29.03  &\hspace{-1.6cm}     1.50  &\hspace{-1.2cm} $ -2.96 \pm   0.30$  &\hspace{-0.7cm} $ 0.441 \pm  0.041$  &\hspace{-0.7cm}  0.38  &\hspace{-0.5cm} $  29.37 \pm    4.85$ \\
  {19c}  &\hspace{-2.8cm} Sdm  &\hspace{-1.1 cm}        ELG  &\hspace{-0.7cm}         II  &\hspace{-0.5cm}   27.30  &\hspace{-0.5cm}   27.41  &\hspace{-0.5cm}   27.37  &\hspace{-1cm}   28.46  &\hspace{-1.6cm}     0.66  &\hspace{-1.2cm} $ -1.90 \pm   0.61$  &\hspace{-0.7cm} $ 0.352 \pm  0.033$  &\hspace{-0.7cm}  0.79  &\hspace{-0.5cm} $  53.13 \pm   12.51$ \\
  {22c}  &\hspace{-2.8cm}SBcd  &\hspace{-1.1 cm}        ELG  &\hspace{-0.7cm}         II  &\hspace{-0.5cm}   27.35  &\hspace{-0.5cm}   27.47  &\hspace{-0.5cm}   27.45  &\hspace{-1cm}   28.73  &\hspace{-1.6cm}     1.09  &\hspace{-1.2cm} $ -1.55 \pm   0.14$  &\hspace{-0.7cm} $ 0.455 \pm  0.040$  &\hspace{-0.7cm}  0.69  &\hspace{-0.5cm} $  41.88 \pm    7.04$ \\
{31ace}  &\hspace{-2.8cm} Sdm  &\hspace{-1.1 cm}        HII  &\hspace{-0.7cm}          I  &\hspace{-0.5cm}   28.37  &\hspace{-0.5cm}   28.51  &\hspace{-0.5cm}   28.45  &\hspace{-1cm}   30.27  &\hspace{-1.6cm}     2.49  &\hspace{-1.2cm} $ -3.51 \pm   0.05$  &\hspace{-0.7cm} $ 8.107 \pm  0.737$  &\hspace{-0.7cm}  0.40  &\hspace{-0.5cm} $ 325.78 \pm   53.05$ \\
  {31b}  &\hspace{-2.8cm}  Sm  &\hspace{-1.1 cm}        HII  &\hspace{-0.7cm}          I  &\hspace{-0.5cm}   27.63  &\hspace{-0.5cm}   27.77  &\hspace{-0.5cm}   27.66  &\hspace{-1cm}   28.86  &\hspace{-1.6cm}     0.44  &\hspace{-1.2cm} $ -2.07 \pm   0.86$  &\hspace{-0.7cm} $ 0.777 \pm  0.068$  &\hspace{-0.7cm}  0.76  &\hspace{-0.5cm} $ 177.78 \pm   94.20$ \\
  {31f}  &\hspace{-2.8cm}  Im  &\hspace{-1.1 cm}          ?  &\hspace{-0.7cm}          I  &\hspace{-0.5cm}   27.07  &\hspace{-0.5cm}   27.31  &\hspace{-0.5cm}   27.46  &\hspace{-1cm}   28.04  &\hspace{-1.6cm}  \nodata  &\hspace{-1.2cm} \nodata              &\hspace{-0.7cm} $ 0.194 \pm  0.024$  &\hspace{-0.7cm}  0.85  &\hspace{-0.5cm}   \nodata              \\
  {31g}  &\hspace{-2.8cm}  Im  &\hspace{-1.1 cm}          ?  &\hspace{-0.7cm}          I  &\hspace{-0.5cm}   27.88  &\hspace{-0.5cm}   28.02  &\hspace{-0.5cm}   27.88  &\hspace{-1cm}   29.19  &\hspace{-1.6cm}     0.91  &\hspace{-1.2cm} $ -2.15 \pm   0.30$  &\hspace{-0.7cm} $ 1.474 \pm  0.122$  &\hspace{-0.7cm}  0.73  &\hspace{-0.5cm} $ 161.56 \pm   38.28$ \\
  {31q}  &\hspace{-2.8cm}  Im  &\hspace{-1.1 cm}          ?  &\hspace{-0.7cm}          I  &\hspace{-0.5cm}   26.62  &\hspace{-0.5cm}   26.75  &\hspace{-0.5cm}   26.55  &\hspace{-1cm}   27.48  &\hspace{-1.6cm}     0.11  &\hspace{-1.2cm} $ -0.95 \pm   0.45$  &\hspace{-0.7cm} $ 0.066 \pm  0.011$  &\hspace{-0.7cm}  0.88  &\hspace{-0.5cm} $  59.34 \pm   60.44$ \\
  {48b}  &\hspace{-2.8cm}  Sc  &\hspace{-1.1 cm}          X  &\hspace{-0.7cm}        III  &\hspace{-0.5cm}   27.36  &\hspace{-0.5cm}   27.51  &\hspace{-0.5cm}   27.55  &\hspace{-1cm}   29.06  &\hspace{-1.6cm}     1.53  &\hspace{-1.2cm} $ -2.80 \pm   0.41$  &\hspace{-0.7cm} $ 0.619 \pm  0.055$  &\hspace{-0.7cm}  0.52  &\hspace{-0.5cm} $  40.59 \pm    6.79$ \\
  {59a}  &\hspace{-2.8cm}  Sa  &\hspace{-1.1 cm}          ?  &\hspace{-0.7cm}         II  &\hspace{-0.5cm}   26.78  &\hspace{-0.5cm}   26.87  &\hspace{-0.5cm}   27.16  &\hspace{-1cm}   30.28  &\hspace{-1.6cm}     2.35  &\hspace{-1.2cm} $ -2.42 \pm   0.46$  &\hspace{-0.7cm} $ 4.985 \pm  0.665$  &\hspace{-0.7cm}  0.02  &\hspace{-0.5cm} $ 212.01 \pm   40.11$ \\
  {59c}  &\hspace{-2.8cm}  Sc  &\hspace{-1.1 cm}          ?  &\hspace{-0.7cm}         II  &\hspace{-0.5cm}   26.90  &\hspace{-0.5cm}   27.03  &\hspace{-0.5cm}   26.96  &\hspace{-1cm}   28.23  &\hspace{-1.6cm}     0.41  &\hspace{-1.2cm} $ -2.00 \pm   0.48$  &\hspace{-0.7cm} $ 0.156 \pm  0.025$  &\hspace{-0.7cm}  0.72  &\hspace{-0.5cm} $  38.24 \pm   20.41$ \\
  {59d}  &\hspace{-2.8cm}  Im  &\hspace{-1.1 cm}          ?  &\hspace{-0.7cm}         II  &\hspace{-0.5cm}   27.41  &\hspace{-0.5cm}   27.54  &\hspace{-0.5cm}   27.49  &\hspace{-1cm}   28.67  &\hspace{-1.6cm}     0.36  &\hspace{-1.2cm} $ -2.34 \pm   0.32$  &\hspace{-0.7cm} $ 0.484 \pm  0.041$  &\hspace{-0.7cm}  0.75  &\hspace{-0.5cm} $ 132.84 \pm   68.46$ \\
  {61c}  &\hspace{-2.8cm} Sbc  &\hspace{-1.1 cm}      AGN,R  &\hspace{-0.7cm}          I  &\hspace{-0.5cm}   27.47  &\hspace{-0.5cm}   27.43  &\hspace{-0.5cm}   27.50  &\hspace{-1cm}   30.14  &\hspace{-1.7cm}    10.15  &\hspace{-1.2cm} $ -3.23 \pm   0.09$  &\hspace{-0.7cm} $ 3.990 \pm  0.486$  &\hspace{-0.7cm}  0.10  &\hspace{-0.5cm} $  39.32 \pm    7.14$ \\
 
\hline
\multicolumn{13}{c}{\sc Quiescent}\\
\hline

   {7b}  &\hspace{-2.8cm} SB0  &\hspace{-1.1 cm}          ?  &\hspace{-0.7cm}         II  &\hspace{-0.5cm}   26.76  &\hspace{-0.5cm}   27.05  &\hspace{-0.5cm}   27.52  &\hspace{-1cm}   28.76  &\hspace{-1.6cm}     8.88  &\hspace{-1.2cm} \phs$  1.14 \pm   0.18$  &\hspace{-0.7cm} $ 0.227 \pm  0.022$  &\hspace{-0.7cm}  0.36  &\hspace{-0.5cm} $   2.56 \pm    0.42$ \\
  {16b}  &\hspace{-2.8cm} Sab  &\hspace{-1.1 cm}        Sy2  &\hspace{-0.7cm}          I  &\hspace{-0.5cm}   26.88  &\hspace{-0.5cm}   26.97  &\hspace{-0.5cm}   27.34  &\hspace{-1cm}   28.95  &\hspace{-1.6cm}    10.42  &\hspace{-1.2cm} \phs$  0.12 \pm   0.30$  &\hspace{-0.7cm} $ 0.333 \pm  0.033$  &\hspace{-0.7cm}  0.32  &\hspace{-0.5cm} $   3.20 \pm    0.54$ \\
  {19a}  &\hspace{-2.8cm}  E2  &\hspace{-1.1 cm}        ABS  &\hspace{-0.7cm}         II  &\hspace{-0.5cm}   26.55  &\hspace{-0.5cm}   26.81  &\hspace{-0.5cm}   27.26  &\hspace{-1cm}   28.18  &\hspace{-1.6cm}     5.34  &\hspace{-1.2cm} \phs$  1.36 \pm   0.21$  &\hspace{-0.7cm} $ 0.089 \pm  0.020$  &\hspace{-0.7cm}  0.56  &\hspace{-0.5cm} $   1.66 \pm    0.44$ \\
  {22a}  &\hspace{-2.8cm}  E2  &\hspace{-1.1 cm}       dSy2  &\hspace{-0.7cm}         II  &\hspace{-0.5cm}   27.10  &\hspace{-0.5cm}   27.31  &\hspace{-0.5cm}   27.68  &\hspace{-1cm}   28.40  &\hspace{-1.6cm}    15.49  &\hspace{-1.2cm} \phs$  1.40 \pm   0.22$  &\hspace{-0.7cm} $ 0.240 \pm  0.020$  &\hspace{-0.7cm}  0.73  &\hspace{-0.5cm} $   1.55 \pm    0.25$ \\
  {22b}  &\hspace{-2.8cm}  Sa  &\hspace{-1.1 cm}        ABS  &\hspace{-0.7cm}         II  &\hspace{-0.5cm}   26.23  &\hspace{-0.5cm}   26.31  &\hspace{-0.5cm}   26.57  &\hspace{-1cm}   27.27  &\hspace{-1.6cm}     1.01  &\hspace{-1.2cm} \phs$  1.47 \pm   0.41$  &\hspace{-0.7cm} $ 0.029 \pm  0.004$  &\hspace{-0.7cm}  0.83  &\hspace{-0.5cm} $   2.84 \pm    0.54$ \\
  {42a}  &\hspace{-2.8cm}  E3  &\hspace{-1.1 cm}     dLINER,X  &\hspace{-0.7cm}        III  &\hspace{-0.5cm}   27.21  &\hspace{-0.5cm}   27.44  &\hspace{-0.5cm}   27.94  &\hspace{-1cm}   28.92  &\hspace{-1.6cm}    35.00  &\hspace{-1.2cm} \phs$  1.78 \pm   0.18$  &\hspace{-0.7cm} $ 0.443 \pm  0.037$  &\hspace{-0.7cm}  0.52  &\hspace{-0.5cm} $   1.26 \pm    0.20$ \\
  {42b}  &\hspace{-2.8cm} SB0  &\hspace{-1.1 cm}        ABS  &\hspace{-0.7cm}        III  &\hspace{-0.5cm}   26.56  &\hspace{-0.5cm}   26.78  &\hspace{-0.5cm}   27.25  &\hspace{-1cm}   28.30  &\hspace{-1.6cm}     5.90  &\hspace{-1.2cm} \phs$  1.42 \pm   0.50$  &\hspace{-0.7cm} $ 0.102 \pm  0.026$  &\hspace{-0.7cm}  0.50  &\hspace{-0.5cm} $   1.73 \pm    0.51$ \\
  {42c}  &\hspace{-2.8cm}  E2  &\hspace{-1.1 cm}        ABS  &\hspace{-0.7cm}        III  &\hspace{-0.5cm}   26.61  &\hspace{-0.5cm}   26.91  &\hspace{-0.5cm}   27.37  &\hspace{-1cm}   28.05  &\hspace{-1.6cm}     6.38  &\hspace{-1.2cm} \phs$  1.68 \pm   0.09$  &\hspace{-0.7cm} $ 0.086 \pm  0.016$  &\hspace{-0.7cm}  0.67  &\hspace{-0.5cm} $   1.35 \pm    0.31$ \\
  {42d}  &\hspace{-2.8cm}  E2  &\hspace{-1.1 cm}        ABS  &\hspace{-0.7cm}        III  &\hspace{-0.5cm}   25.68  &\hspace{-0.5cm}   25.98  &\hspace{-0.5cm}   26.36  &\hspace{-1cm}   27.22  &\hspace{-1.6cm}     0.97  &\hspace{-1.2cm} \phs$  1.69 \pm   0.47$  &\hspace{-0.7cm} $ 0.011 \pm  0.004$  &\hspace{-0.7cm}  0.61  &\hspace{-0.5cm} $   1.13 \pm    0.52$ \\
  {48a}  &\hspace{-2.8cm}  E2  &\hspace{-1.1 cm}        ABS  &\hspace{-0.7cm}        III  &\hspace{-0.5cm}   26.99  &\hspace{-0.5cm}   27.19  &\hspace{-0.5cm}   27.72  &\hspace{-1cm}   28.46  &\hspace{-1.6cm}    17.10  &\hspace{-1.2cm} \phs$  1.26 \pm   0.30$  &\hspace{-0.7cm} $ 0.211 \pm  0.018$  &\hspace{-0.7cm}  0.65  &\hspace{-0.5cm} $   1.23 \pm    0.20$ \\
  {48c}  &\hspace{-2.8cm} S0a  &\hspace{-1.1 cm}          ?  &\hspace{-0.7cm}        III  &\hspace{-0.5cm}   26.08  &\hspace{-0.5cm}   26.24  &\hspace{-0.5cm}   26.82  &\hspace{-1cm}   27.55  &\hspace{-1.6cm}     2.36  &\hspace{-1.2cm} \phs$  1.51 \pm   0.06$  &\hspace{-0.7cm} $ 0.026 \pm  0.009$  &\hspace{-0.7cm}  0.65  &\hspace{-0.5cm} $   1.09 \pm    0.43$ \\
  {48d}  &\hspace{-2.8cm}  E1  &\hspace{-1.1 cm}          ?  &\hspace{-0.7cm}        III  &\hspace{-0.5cm}   25.57  &\hspace{-0.5cm}   25.77  &\hspace{-0.5cm}   26.27  &\hspace{-1cm}   26.96  &\hspace{-1.6cm}     0.54  &\hspace{-1.2cm} \phs$  1.67 \pm   0.42$  &\hspace{-0.7cm} $ 0.008 \pm  0.003$  &\hspace{-0.7cm}  0.69  &\hspace{-0.5cm} $   1.42 \pm    0.59$ \\
  {59b}  &\hspace{-2.8cm}  E0  &\hspace{-1.1 cm}          ?  &\hspace{-0.7cm}         II  &\hspace{-0.5cm}   25.84  &\hspace{-0.5cm}   26.22  &\hspace{-0.5cm}   26.46  &\hspace{-1cm}   27.36  &\hspace{-1.6cm}     1.12  &\hspace{-1.2cm} \phs$  1.65 \pm   0.19$  &\hspace{-0.7cm} $ 0.016 \pm  0.006$  &\hspace{-0.7cm}  0.62  &\hspace{-0.5cm} $   1.40 \pm    0.58$ \\
  {61a}  &\hspace{-2.8cm} S0a  &\hspace{-1.1 cm}      Sy2,R  &\hspace{-0.7cm}          I  &\hspace{-0.5cm}   27.24  &\hspace{-0.5cm}   27.37  &\hspace{-0.5cm}   27.59  &\hspace{-1cm}   28.86  &\hspace{-1.6cm}    14.88  &\hspace{-1.2cm} \phs$  0.82 \pm   0.50$  &\hspace{-0.7cm} $ 0.432 \pm  0.036$  &\hspace{-0.7cm}  0.57  &\hspace{-0.5cm} $   2.91 \pm    0.46$ \\
  {61d}  &\hspace{-2.8cm}  S0  &\hspace{-1.1 cm}        ABS  &\hspace{-0.7cm}          I  &\hspace{-0.5cm}   26.79  &\hspace{-0.5cm}   26.99  &\hspace{-0.5cm}   27.14  &\hspace{-1cm}   27.99  &\hspace{-1.6cm}     3.78  &\hspace{-1.2cm} \phs$  1.24 \pm   0.48$  &\hspace{-0.7cm} $ 0.112 \pm  0.016$  &\hspace{-0.7cm}  0.77  &\hspace{-0.5cm} $   2.95 \pm    0.58$ \\
  {62a}  &\hspace{-2.8cm}  E3  &\hspace{-1.1 cm}     dLINER,X  &\hspace{-0.7cm}        III  &\hspace{-0.5cm}   27.08  &\hspace{-0.5cm}   27.29  &\hspace{-0.5cm}   27.93  &\hspace{-1cm}   28.65  &\hspace{-1.6cm}    21.13  &\hspace{-1.2cm} \phs$  1.29 \pm   0.11$  &\hspace{-0.7cm} $ 0.283 \pm  0.031$  &\hspace{-0.7cm}  0.60  &\hspace{-0.5cm} $   1.34 \pm    0.23$ \\
  {62b}  &\hspace{-2.8cm}  S0  &\hspace{-1.1 cm}      ABS,X  &\hspace{-0.7cm}        III  &\hspace{-0.5cm}   26.36  &\hspace{-0.5cm}   26.77  &\hspace{-0.5cm}   27.35  &\hspace{-1cm}   28.08  &\hspace{-1.6cm}     6.06  &\hspace{-1.2cm} \phs$  1.31 \pm   0.22$  &\hspace{-0.7cm} $ 0.062 \pm  0.016$  &\hspace{-0.7cm}  0.51  &\hspace{-0.5cm} $   1.03 \pm    0.30$ \\
  {62c}  &\hspace{-2.8cm}  S0  &\hspace{-1.1 cm}        ABS  &\hspace{-0.7cm}        III  &\hspace{-0.5cm}   26.48  &\hspace{-0.5cm}   26.97  &\hspace{-0.5cm}   27.34  &\hspace{-1cm}   28.00  &\hspace{-1.6cm}     3.81  &\hspace{-1.2cm} \phs$  1.51 \pm   0.54$  &\hspace{-0.7cm} $ 0.068 \pm  0.014$  &\hspace{-0.7cm}  0.62  &\hspace{-0.5cm} $   1.79 \pm    0.44$ \\
  {62d}  &\hspace{-2.8cm}  E2  &\hspace{-1.1 cm}        ABS  &\hspace{-0.7cm}        III  &\hspace{-0.5cm}   25.58  &\hspace{-0.5cm}   25.96  &\hspace{-0.5cm}   26.44  &\hspace{-1cm}   27.91  &\hspace{-1.6cm}     0.96  &\hspace{-1.2cm} \phs$  1.49 \pm   0.00$  &\hspace{-0.7cm} $ 0.026 \pm  0.011$  &\hspace{-0.7cm}  0.21  &\hspace{-0.5cm} $   2.70 \pm    1.26$

\enddata
\tablenotetext{a}{As in \tr{tab-sample}.} 
\tablenotetext{b}{Same as G08, Table 3. ABS: absorption-line galaxy; AGN: active galactic
nucleus; (d)Sy2: (dwarf) Seyfert 2; X: X-ray source; R: radio source; (d)LINER: (dwarf) 
low-ionization nuclear emission region; SBNG: starburst nucleated galaxy;
\htwo: strong \htwo\ emitter; ELG: emission line galaxy; ?: unknown.} 
\tablenotetext{c}{Logarithm of monochromatic luminosities for the three ultraviolet
\swift\ UVOT filters and the \spitzer\ MIPS 24\micron\ filter.}
\tablenotetext{d}{Total stellar mass estimated from 2MASS $K_s$-band luminosity
assuming \mstar/\msun~=~0.8~$\nu$~\lnuks/\lsun\ \citep{bell2003}.}
\tablenotetext{e}{MIR activity index calculated from power law fit, 
\lnu~$\propto \nu^{\alpha_{\rm IRAC}}$ to 4.5-8\micron\ \spitzer\ IRAC
luminosities.}
\tablenotetext{f}{Total SFR estimated using \er{equ_sfruvir}.}
\tablenotetext{g}{Fraction of \sfrtot\ due to \wtwo\ (2000\AA) emission.}
\tablenotetext{h}{SSFR calculated by normalizing \sfrtot\ by \mstar\ 
(column 11 divided by column 9).}
\tablecomments{
Galaxies with \airac~$\le0$ and high SSFR are presented 
separately (\lq\lq actively star-forming\rq\rq, upper table)
from those with \airac~$> 0$ and low SSFR
(\rq\rq quiescent\rq\rq, lower table).
These two groups of galaxies show a pronounced bimodality in
\airac\ and SSFR, as explained in the text.
}
\end{deluxetable*}
%\clearpage
%\end{landscape}  %emulateapj

In \fr{fig:w2-lnu_hist} we plot the distribution of
\wtwo\ luminosity for the full HCG sample, as well as sub-samples
based on parent-group \hone-richness. 
Overall, \lnuwtwo\ appears to correlate with \hone-richness, in the
sense that galaxies brightest in \lnuwtwo\ tend to belong to
more gas-rich groups.
This {\it trend} of the distributions by group \hone-type is broadly the same
as in the equivalent plot for the 24\micron\ data (J07, Fig.~14).
This is consistent with the general correlation between the \wtwo\ and
24\micron\ luminosities.

%Made with:
%[uvir]//home/pana/DATA/HCG/SFR-UVIR/SINGS_SMITH
%This version made directly with SFR.pro 
%Then add colhead commands and comments.
\tabletypesize{\scriptsize} %outside for emulateapj
\begin{deluxetable*}{cccc ccc}
\tablecolumns{7}
%\tablewidth{0pc} 
\setlength{\tabcolsep}{0.3cm}
%%\tablewidth{0pt}
\tablecaption{SINGS COMPARISON SUB-SAMPLE GALAXIES IN THE SSFR GAP\label{tab-sfr-sings}}
\tablehead{
\colhead{\sc SINGS}  
 & \colhead{Galaxy}
 & \colhead{\mstar}
 & \colhead{}
 & \colhead{SFR}
 & \colhead{SSFR}
 & \colhead{}
\\
\colhead{\sc ID}  
& \colhead{Morphology}
& \colhead{($10^9$\msun)}
& \colhead{\airac}
& \colhead{(\msuny)}
& \colhead{($10^{-11}$ yr$^{-1}$)}
& \colhead{Comments}
\\
\colhead{(1)}  
& \colhead{(2)}
& \colhead{(3)}
& \colhead{(4)}
& \colhead{(5)}
& \colhead{(6)}
& \colhead{(7)}
}
\startdata
\hline
   NGC1291 &      SB0/a &    12.74 &   1.22 &  0.450 &    3.53 & HIPASS J0317-41\tablenotemark{a}   \\ 
   NGC4450 &       SAab &    14.88 &   0.33 &  0.700 &    4.70 & \tablenotemark{b}    \\
   NGC2841 &        SAb &     8.83 &  -0.82 &  0.640 &    7.25 & \tablenotemark{c}   \\
   NGC4579 &       SABb &    25.08 &  -0.62 &  2.040 &    8.13 & \tablenotemark{c,b}   \\
   NGC4826 &       SAab &     5.70 &  -0.74 &  0.530 &    9.29 & \tablenotemark{c,d}   \\
   NGC4725 &      SABab &    24.27 &  -1.00 &  2.360 &    9.72 &    
\enddata
\tablecomments{
Galaxies are from the SINGS sub-sample obtained by selecting SINGS galaxies
that (1) have luminosities which fall within the luminosity range for
HCG galaxies, and (2) appear to be non-interacting and isolated (see \scr{subsec_comp} for details).
Morphology information (RC3) is taken from \citet{dale2007}. The rest of
the data are derived as in \tr{tab-sfr} based on \citet{dale2007}.
}
\tablenotetext{a}{\citet{koribalski2004}.} 
\tablenotetext{b}{Virgo cluster galaxies.}
\tablenotetext{c}{Also in \airac\ gap.}
\tablenotetext{d}{Evidence of counter-rotating gaseous disk \citep{braun1992}.} 
\end{deluxetable*}

\begin{figure}[h]
\epsscale{1.3}
\hspace{-1.1cm}\plotone{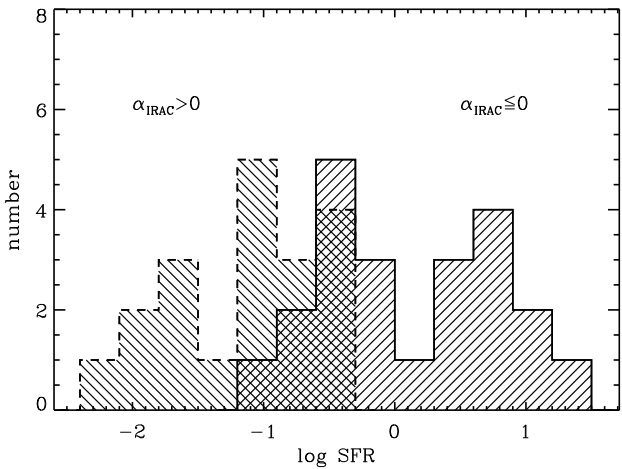}  
\caption{SFR distributions for the two HCG sub-samples of MIR active 
(\airac~$\le 0$) and \airac~$>0$ (MIR quiescent) galaxies. The total
distribution is continuous.}
\label{fig:sfr_hist_all}
\end{figure}

\begin{figure}
\epsscale{1.3}
\hspace{-1.1cm}\plotone{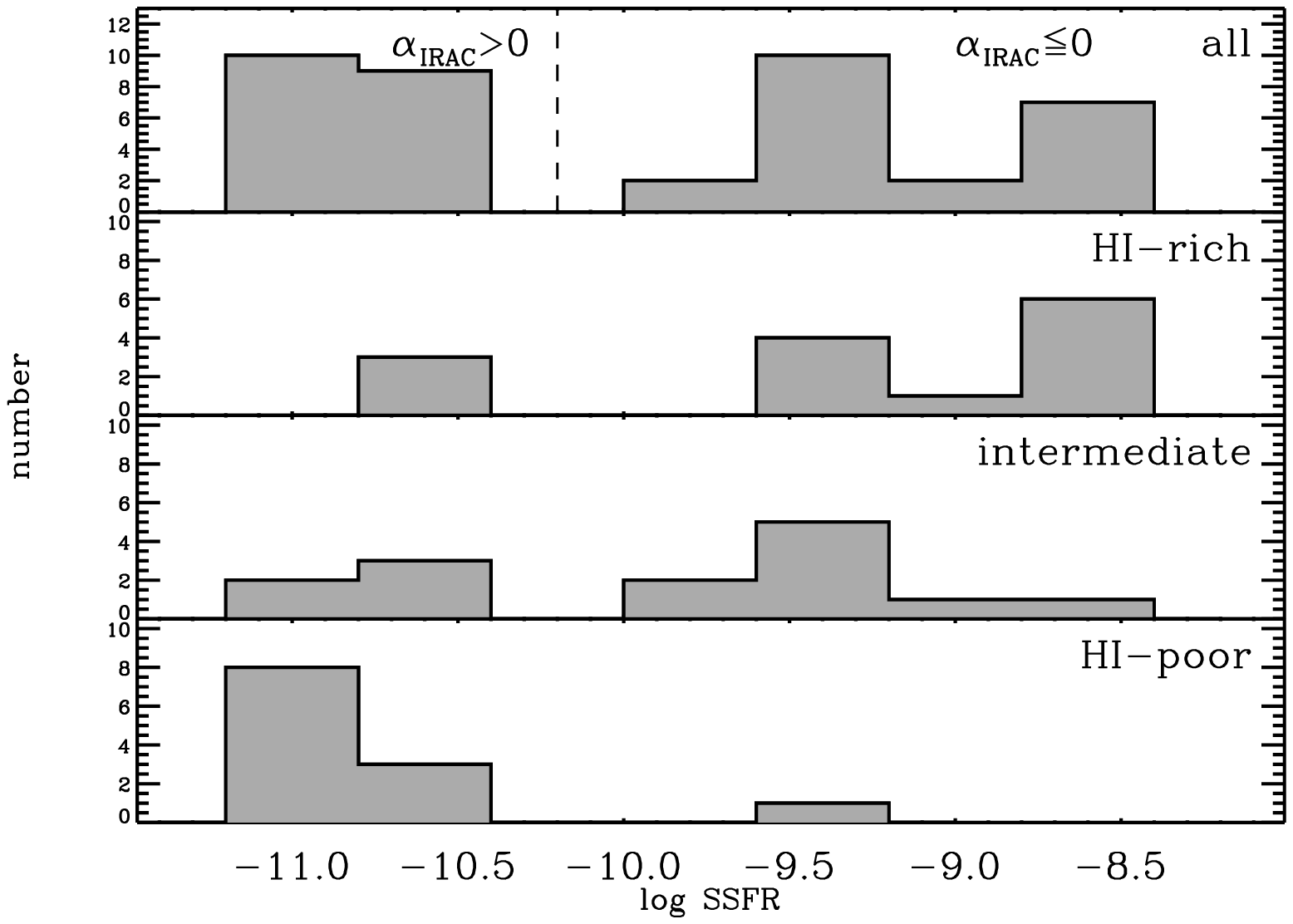}    
\caption{SSFR distributions for HCG galaxies.
Top: Full sample. 
The dashed vertical line marks the gap between
the high SSFR (right) and the low SSFR (left)
part of the distribution. 
All galaxies to the right of this gap have
\airac~$\le 0$ (MIR active).
All galaxies to the left of the gap have
\airac~$>0$ (MIR quiescent).
The bimodality is
clear (compare to \fr{fig:sfr_hist_all}).
Bottom three panels: 
Sub-samples of galaxies belonging to HCGs of
decreasing gas richness, as labeled.
SSFR values appear to correlate with \hone-gas
richness.
}
\label{fig:ssfr_hists}
\end{figure}

In \fr{fig:24-w2-morph}
we show the ratio of power in the
24\micron\ and \wtwo\ bands, 
\iruv,
as a function of Hubble type. 
This ratio only takes into account
one UV and one IR band, so it is
not an adequate quantitative estimate
of the overall relative contributions
from the UV and the IR.
Additionally, it suffers from small
number statistics.
However, at least qualitatively, there are some notable
patterns.
For 33 out of the full sample of 41 HCG galaxies ($\sim 80$\%)
\iruv~$\le 1$, i.e. 
the contribution to the total energy budget from
power emerging at 2000\AA\ is at least as important
as that at 24\micron. Only 8 galaxies emit more power
in the 24\micron\ band than they do at 2000\AA. Among these only one 
is an elliptical, with the rest being spirals or
irregulars. For ellipticals, the generally low values
of the \iruv\ ratios likely reflect the lower levels 
of star formation and associated dust production.
The trend for spirals seems to be different.
With the exception of HCG 22 B (a type Sa at \iruv~$\sim 0.1$),
the 24\micron\ contribution is highest for the earlier types (Sa's), getting
progressively lower for later types. As spirals are actively star-forming,
high dust levels and associated 24\micron\ emission are expected.
However, the UV contribution from the most actively star-forming
later types seems to be more than compensating for the presence of dust.
Finally, irregulars show a large range in values.
This is mainly due to
the presence of galaxies HCG 16 C, D, the two type Im
systems with the highest values of \iruv. 
Both galaxies have high SFRs ($\sim 14$ and 17\msuny, \tr{tab-sfr}),
and are likely recent merger remnants, 
with highly disturbed
velocity fields and double nuclei \citep{mendes1998}.

\subsection{Star formation rates}\label{subsec_sfr}
\subsubsection{Method}
One of the most important properties 
characterizing a galaxy is the rate at which 
it forms stars.
Active star formation is traced directly
or indirectly by  
young stellar populations ($10^6 - 10^8$ yr)
whose light is dominated by O 
and early B stars.
UV continuum emission (1250--2500\AA),
as well as \ha\ line emission, from massive stars provide a direct probe of these
populations. However, a significant part of this light
is often heavily absorbed by dust and re-emitted in the IR wavelength region (1--1000\micron).
UV-based SFR calibrations \citep[e.g.][]{kennicutt1998,salim2007} thus
require a correction for intrinsic extinction, to avoid
significant underestimates of the true SFR.
Conversely, IR-based calibrations \citep[e.g.][]{calzetti2007,rieke2009}
adopt the \lq\lq dusty starburst approximation\rq\rq,
by assuming that the bulk of the UV emission 
is re-emitted in the IR 
(also known as the \lq\lq calorimeter assumption\rq\rq).

Thus in general
the total SFR contains both a dust-obscured component, which can be measured
from the IR, and an unobscured component, measured directly from the UV
\citep{bell2003,hirashita2003,iglesias-paramo2006,dale2007}. 
In this paper we obtain the two components independently using
calibrations from the literature.
In particular, we use
\begin{equation} \label{equ_sfruvir}
{\rm SFR}_{\rm TOTAL} \equiv {\rm SFR}_{\rm UV+IR} \equiv {\rm SFR}_{UVW2}
+ {\rm SFR}_{24\mu{\rm m}} \, ,
\end{equation} 
where 
\begin{equation} \label{equ_kenni}
{\rm SFR}_{uvw2} (M_{\odot}\; {\rm yr}^{-1}) = 
9.5\times 10^{-44}\; \nu L_{\nu,uvw2} \; ({\rm erg \; s}^{-1})
\end{equation}
and
\begin{equation} \label{equ_rieke}
{\rm SFR}_{24\mu{\rm m}} (M_{\odot}\; {\rm yr}^{-1}) =
2.14\times 10^{-42} \; \nu L_{\nu, 24\mu{\rm m}} \; ({\rm erg \; s}^{-1}) \ \ .
\end{equation}

\er{equ_kenni} estimates \sfruv\ from the \wtwo\ luminosity using the
calibration by \citet{kennicutt1998}, valid for a Salpeter IMF and
continuous star formation over $\la 10^8$ years.  \wtwo\ luminosity
probes the spectral region around $\sim 2000$\AA, which is the
approximate center of the spectrally flat UV wavelength range,
1500-2800\AA. It thus provides a \lq\lq mean\rq\rq\ estimate of UV
emission properties \citep{hirashita2003,kennicutt1998}.

\er{equ_rieke} uses the 24\micron\ luminosity and the
\citet[][R09]{rieke2009} calibration (their
Equation 10) to estimate \sfrir.
This is based on the calorimeter assumption and 
is most appropriate for  
24\micron\ luminosities $>~10^{42.36}$~\lunits.
At lower luminosities,
significant numbers of UV photons escape and need
to be accounted for. Most of our galaxies fall in
this regime, and we account directly for 
non-negligible UV emission.
In essence, we use \sfrtf\ to obtain an effective extinction correction for \sfrwtwo.
Uncertainties related to the use of these calibrations are
addressed in 
\scr{subsubsec_sfr}.

Whereas SFRs provide an absolute measure of
current star-forming activity in a galaxy, 
specific SFRs (SSFRs) relate this
to the products of past star formation
\citep{guzman1997,brinchmann2000,feulner2005a}. 
SSFR is defined as
\begin{equation}\label{equ_ssfr}
{\rm SSFR} \equiv \frac{\rm SFR}{M_{\ast}} \, ,
\end{equation} 
where \mstar\ is the galaxy total stellar mass.  In general, 
star-forming galaxies are still actively increasing \mstar\
and have not yet reached their maximum values.
We expect the situation to be reversed for gas-poor
galaxies dominated by old stellar populations.
Thus this
normalization can help distinguish galaxies according to their star-formation
history.  
We use $K_s$ band luminosities as a proxy for 
\mstar, and normalize
SFRs calculated via \er{equ_sfruvir},
assuming \mstar/\msun~=~0.8~$\nu$~\lnuks/\lsun
\citep{bell2003}.

\subsubsection{Results}\label{subsubsec_results}
Total and specific SFR results for our
HCG galaxies are tabulated in \tr{tab-sfr}.  
For each galaxy we also give galaxy morphology,
nuclear classification, parent-group \hone-richness,
UVOT ultraviolet and MIPS 24\micron\ luminosities,
stellar masses, the ratio of \sfruv\ to \sfrtot, as
well as \airac\ values.

\airac\ is a MIR activity index, introduced by G08.
It is evaluated by means of a power-law fit to the
4.5, 5.7, and 8\micron\ data (\lnu~$\propto \nu^{\alpha_{\rm IRAC}}$).
The 3.6\micron\ band has been excluded as it is
dominated by the Rayleigh-Jeans tail of
stellar photospheres.
The MIR regime hosts several emission features
from polycyclic aromatic hydrocarbon (PAH) 
molecules, which are vibrationally excited upon absorption
of single UV/optical photons 
\citep{leger1984,allamandola1985}.
Among the three bands used for the
fit, the 4.5\micron\ is the one
least affected by PAH emission
\citep{draine2007}, while the
8\micron\ is the one most affected. Thus, if
there is significant PAH emission,
\airac\ will be negative (MIR SED decreasing with $\nu$).
Additionally, the MIR includes a thermal 
continuum component (\lq\lq hot dust\rq\rq) attributed
to very small grains \citep{li2001,draine2007}.
\airac\ will also be negative if this component is strong.

G08 detected a significant gap,
with no \airac\ values between $-0.95$ and 0.12.
As the origin of negative \airac\ values 
is ultimately star-forming and/or AGN activity,
they label galaxies with \airac~$\le 0$ as MIR-active,
and those with \airac~$> 0$ as MIR-quiescent.

\begin{figure*}            
\epsscale{1.1}
\plottwo{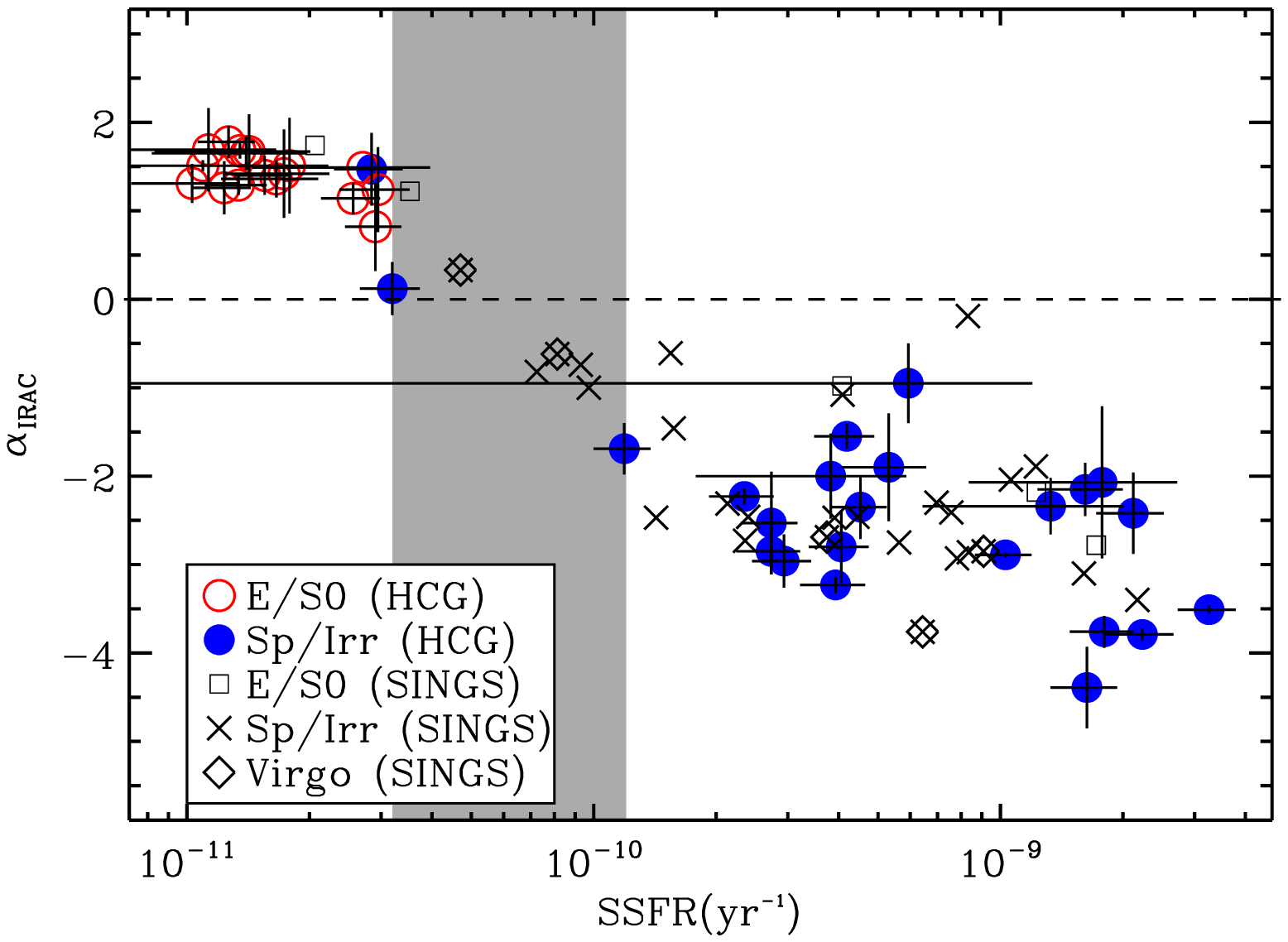}{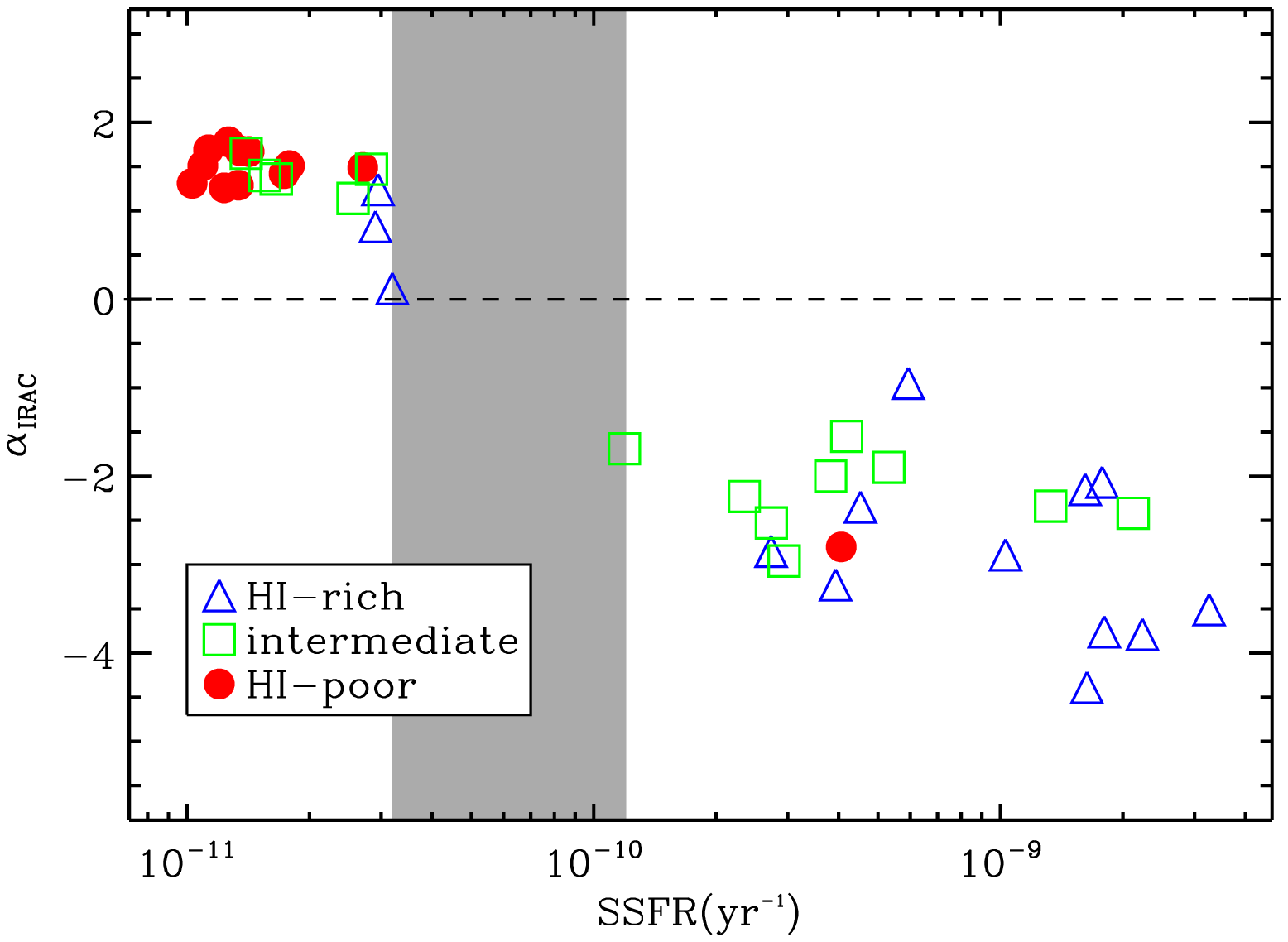}
\caption{\airac\ MIR activity index versus SSFR.
The horizontal dashed line indicates the \airac\ gap.
The SSFR gap is shown as a shaded region.
Error bars are only shown in the left-hand plot.
{\it Left:} 
Plotting symbols indicate morphological types, as
shown in the legend, for both HCG and SINGS subsample galaxies.
In addition, diamonds indicate SINGS
galaxies which are members of the Virgo cluster.
All SINGS galaxies have been selected to match the luminosity
range for HCG galaxies, as well as to be isolated and non-interacting.
The large error bar for the point at SSFR~$\sim$~\ten{6}{-10}~yr$^{-1}$
is due to the fact that this (HCG 31 Q) is the faintest galaxy
in the $K_s$ band with a 100\%\ fractional error in flux density.
The SSFR gap for HCG galaxies almost completely separates
quiescent E/S0 galaxies from actively star-forming S/I types, and is
populated by some SINGS subsample galaxies.
{\it Right:} 
Plotting symbols indicate parent-group \hone-gas richness
for HCG galaxies as shown in the legend. The SSFR gap for HCG
galaxies in general appears to separate \hone-rich from \hone-poor groups.
}
\label{fig:airac_ssfr}
\end{figure*}

In \fr{fig:sfr_hist_all} we show the SFR distributions
for MIR active and MIR quiescent galaxies in the same panel.
On average, MIR active galaxies have higher SFRs than
MIR quiescent ones, but, taken as a whole, the distribution for
the full galaxy sample presents a continuous aspect.
However, this picture changes when SSFRs, the distributions
for which are shown in \fr{fig:ssfr_hists}, are considered.
Here 
SSFR distributions for the total sample are shown in the top
panel. Distributions for
sub-samples, defined according to group gas-richness, are shown in
the three lower panels. The SSFR distribution for the full sample shows
a clear 
bimodality. High SSFR galaxies (\ten{1.2}{-10} $\la$ SSFR/yr$^{-1} \la$
\ten{5}{-9}) are separated from low SSFR ones (\ten{1}{-11} $\la$
SSFR/yr$^{-1} \la$ \ten{3.2}{-11}) by a gap of magnitude $\sim$
\ten{9}{-11} yr$^{-1}$. A two-sided Kolmogorov-Smirnov test gives a
probability $\sim$~\ten{4}{-9} for the low-SSFR and high-SSFR
distributions to come from the same parent population.  
Further, this bimodality coincides completely with
the bimodality in the \airac\ index:
All low SSFR galaxies have \airac~$> 0$ (and vice versa), 
and similarly for high SSFR galaxies and  \airac~$\le 0$.
 
On the other hand, the subsample distributions indicate that
gas-rich groups preferentially contain high SSFR
galaxies, groups of intermediate gas-richness have a broad
distribution of SSFRs, covering almost the full range
of SSFR values, and gas-poor groups show a pronounced peak in
low SSFRs.  
%Note though that in general all distributions according to
%group type are not particularly peaked; in a sense, 
%they all contain some \lq\lq outliers\rq\rq.

We plot \airac\ against SSFR in \fr{fig:airac_ssfr}.  In the left-hand
panel we use different symbols to denote E/S0 and S/I morphological
types.  For comparison, galaxies from the SINGS control subsample
are also shown.
Several patterns are apparent here. First, the bimodality
in SSFR is not shared by the SINGS subsample galaxies, similarly to the case of
the bimodality in \airac, also exclusive to HCG galaxies (G08).  A two-sided
Kolmogorov-Smirnov test gives a probability 
$\sim$~\ten{1.3}{-3} 
for the
SINGS-subsample and 
HCG-full-sample SSFR distributions to come from the same parent
population.  
The six SINGS galaxies that populate
the SSFR gap are shown in
\tr{tab-sfr-sings}. One is a lenticular and the rest
are spirals of relatively early type. The mean $K$-band derived
stellar mass for these galaxies is \ten{1.5}{10} \msun. For
higher SSFR SINGS galaxies outside the gap this value is
\ten{5.7}{9}. All galaxies are classified as LINER or LLAGN from the
MIR analysis of \citet{smithj2007}. Given the small numbers and the fact that
10/25 (40\%) of the SINGS subsample galaxies outside the gap, as well
as many HCG galaxies, are
also LINER/LLAGN, this is not significant. It is likely that these are
largely isolated galaxies, which have accumulated significant
stellar mass and are on their way to becoming quiescent ellipticals.

Second, in the HCG sample, with the exception of two
galaxies, the \airac~$>0$/low SSFR and \airac~$\le0$/high SSFR regions
are populated exclusively by E/S0s and S/Is, respectively. 
In this respect SINGS galaxies do not differ:
The lower SSFR systems are E/S0 and the great majority of
high SSFR systems are S/I. The three E/S0 systems with high SSFR are
NGC 0855, which is possibly an edge-on spiral mis-classified as E/S0
\citep{phillips1996}, NGC 3773, a high SFR dwarf irregular in the process of
becoming a dwarf elliptical
\citep{dellenbusch2008}, and NGC 1377, which, in spite of its
optical morphology, is a \lq\lq nascent starburst\rq\rq\
\citep{roussel2006}.

The right-hand panel in \fr{fig:airac_ssfr} shows HCG galaxies
with symbols indicating parent-group \hone-richness. 
Comparing with the left-hand panel, we note that the general
correspondence between, on the one hand, S/I morphology and high
gas-richness, and, on the other hand, E/S0 morphology and low
gas-richness is evident, as noted already by J07 and G08.  However,
the morphological segregation is somewhat more pronounced than the
segregation according to parent-group \hone-gas
content, as was the case when comparing \lnuwtwo\ to
\lnutf. Morphologically, there are only two S/I galaxies 
--an Sa and an Sab-- at the
high-SSFR edge of the low-SSFR group (\fr{fig:airac_ssfr},
left). In
contrast, in \fr{fig:airac_ssfr} (right), there is a gas-poor group galaxy
(HCG 48 B) in the high-SSFR locus, while intermediate group-richness
galaxies cover a broad range of both the low and high-SSFR regimes.

It is important to stress that the SSFR bimodality and gap are only
detected when scaling by total stellar mass. It is also only present
when a {\it total} (UV+IR) SFR is calculated.  The distributions of
\lnuks, \sfruv, and \sfrtf, are not bimodal and none of them alone
can give rise to the SSFR bimodality.  In contrast, the bimodality
can be traced to genuine {\it absolute} SFR differences, evident in
\fr{fig:sfr_hist_all} which are enhanced by
the normalization by total stellar mass.  

Two final points are in order regarding these results.
First, the comparison with the SINGS subsample shows that there is no
pronounced difference in the {\it level} of activity between
HCG and (luminosity-matched, non-interacting) SINGS galaxies. 
On the whole, HCGs show neither
significantly higher nor significantly lower SFRs.
Second, we note that the effect of the \wtwo\ filter cut-off imposed
(or, equivalently, the exclusion of a fraction of the measured
\wtwo\ flux)
with the aim of correcting for the effects of the filter's red tail
(\scr{subsec_redleak}) is minimal. If we impose no cut-off whatsoever,
and so calculate all results using the full flux in the \wtwo\ filter,
almost all SFR values increase by less than 0.3~\msuny. The only two
exceptions are HCGs 31 ACE and 2 A, which increase by 0.8 and
0.5~\msuny, respectively.  The SSFR gap remains prominent and,
although slightly shifted to higher values, its width remains
essentially the same at \ten{8.7}{-11}~yr$^{-1}$
(vs. \ten{8.8}{-11}~yr$^{-1}$). All qualitative results are completely
unaffected. Similar remarks apply to the effects of coincidence losses.
The effect is small, and, in any case, 
all galaxies that may be affected (\tr{tab-uvcoi}) belong to the high-SFR group.
Hence,
if coincidence losses were to be taken into account, we would expect these
galaxies to show even higher SFRs and SSFRs.

\section{Discussion}\label{sec_discussion}

\subsection{The evolution of Hickson Compact Groups}\label{subsec_evolution}
J07 were the first to detect a gap in MIR color space, separating
galaxies in gas-rich groups from those in gas-poor groups.
G08 found further evidence for a separation between these two
classes of HCG galaxies in the form
of a gap in \airac\ values.
The SSFR estimates in the present paper support and
extend these results. 
Our results show that S/I galaxies in our sample
populate preferentially groups that are gas-rich (Type I) 
or of intermediate gas-richness (Type II), have
high SSFRs and are MIR-active. E/S0 galaxies are seen to populate mostly 
groups that are gas-poor
(Type III) or Type II, have low SSFRs and are MIR-inactive. 
Thus, S/I and E/S0 HCG galaxies may constitute two distinct subclasses, 
consistent with being
the two extremes of a possible evolutionary sequence progressing
from the S/I -- high-SSFR subclass to the E/S0 -- low-SSFR one.
The overall amount of \hone\ gas in the group appears to correlate with this
bimodality. 
Intermediate \hone-richness groups play an important role within the context of
such an evolutionary scheme.
They contain all morphological types and cover the full
range of observed SSFRs (\fr{fig:ssfr_hists}). This
might suggest that they are representative 
of the environment where the morphological transformation
is most actively under way.
Evidence for a progression from
\hone-rich groups to \hone-poor groups is also
seen at the {\it group} scale. In \fr{fig:ssfr_mass}
we plot the {\it total} group SSFR against the
corresponding total group log \mhi/log $M_{\rm dyn}$ ratio.
The broad correlation observed (Spearman $\rho = 0.75$ for
a probability \ten{8}{-3} that the variables are
uncorrelated) suggests that SSFR broadly
tracks the depletion of the gas supply.
\fr{fig:ssfr_morpho_hi} shows that for a given
morphology, galaxies in \hone-rich groups have
higher SSFR than galaxies in \hone-intermediate and
poor groups. Note though that it is possible
for much of the \hone\ gas to be outside
the galaxies where most of the star-formation
may be taking place. As we lack high-resolution
\hone\ data for all of
our sample galaxies at the present time, we are unable to 
investigate this point any further.

This simple picture is useful only as a first approximation, in which
\hone-richness is a proxy for evolutionary stage. As can be seen in
\fr{fig:airac_ssfr} (right panel) and \tr{tab-sfr}, there are 3
galaxies belonging to \hone-rich groups at the {\it low} end of the
SSFR gap. Two of these (61 A and B) are type S0 and S0a, and the third
(16b) Sab.  For such systems, \hone-richness does not tell the whole
story.  In reality, an evolutionary scheme should include several
criteria.  Column 7 in \tr{tab-sample} is a qualitative estimate of
evolutionary stage, using galaxy morphologies and the presence or
absence of an \x\ bright IGM. According to this, the parent groups of
these galaxies are in late and intermediate stages, consistent with
their SSFR.

Thus HCG evolution needs to be examined on a case-by-case basis.
In some cases, a group may be \hone-poor perhaps mainly because gas has
been used up by member galaxies to fuel rapid star-formation.
In other cases, gas-stripping under the effect of violent interactions, will end
star-formation in the member galaxies, but the group may still be relatively
gas-rich. A complete picture requires studying
both individual galaxies and
all phases of the group IGM.
Although in this paper we focus exclusively on the bright HCG galaxies, it has been
shown that in a number of cases, star formation takes place outside of galaxies, 
as a result of tidal stripping of \hone\ gas from individual galaxies
\citep{demello2008,torres2009arxiv}. Such evidence provides a 
complementary viewpoint for HCG evolution. 
Indeed, the evolutionary
scheme proposed by \citet{verdes2001} takes into account \hone\
{\it distribution} as well as \hone\ deficiency. Future
work (Walker et al., in preparation) 
will be using high-resolution \hone\ data to explore
the links between \hone-content, \hone-distribution, SSFR
and \airac.

\begin{figure}           
\epsscale{1.3}
\hspace{-1.1cm}\plotone{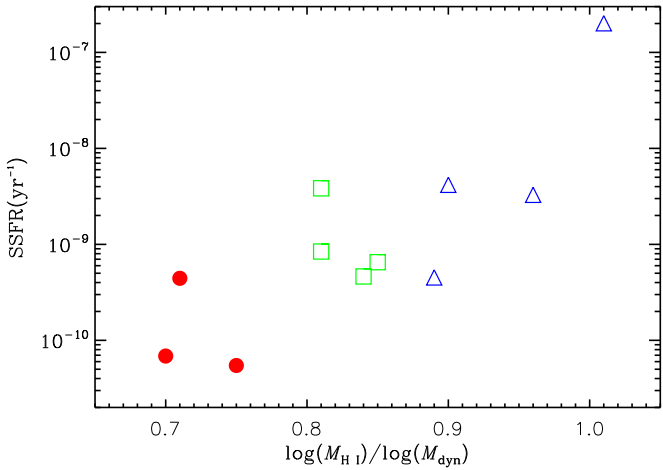}  
\caption{Total group SSFR vs. log \mhi/log $M_{\rm dyn}$ for each
galaxy {\it group} in our sample. 
Plotting symbols indicate parent-group \hone-gas richness as in
\fr{fig:airac_ssfr}. There is a broad trend for 
\hone-richer groups to have higher SSFRs.}
\label{fig:ssfr_mass}
\end{figure}

\begin{figure}           
\epsscale{1.3}
\hspace{-1.1cm}\plotone{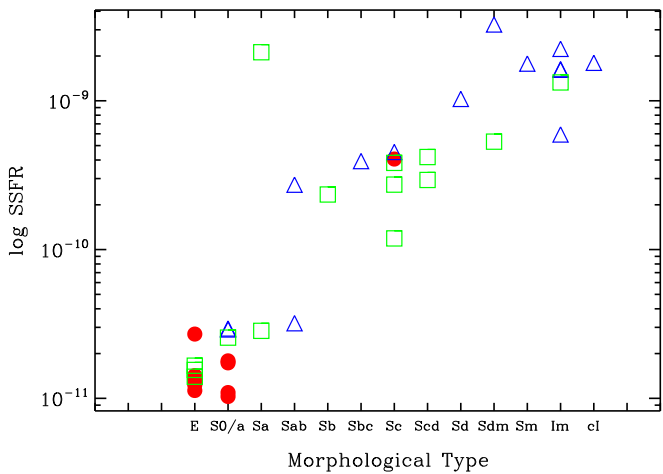}
\caption{SSFR vs. morphological type for {\it individual galaxies} in our
HCG sample. Plotting symbols indicate parent-group \hone-gas richness as in
\fr{fig:airac_ssfr}. For a given galaxy type, galaxies in groups with
higher \hone-gas richness tend to have higher SSFRs.}
\label{fig:ssfr_morpho_hi}
\end{figure}

In any case, the existence of significant gaps in both \airac\ and
SSFR suggests that this evolution is rapid.  This must be linked to
the special environment of HCGs.  No such gaps are found in the SINGS
comparison subsample.  
It must be noted that the original, full SINGS sample
was designed to include a range in
galaxy properties, and so does not favor the
detection of a SSFR gap. However,
from the full SINGS sample we have selected those
galaxies that match the HCG galaxy-luminosity range and appear
to be isolated and non-interacting.  Otherwise,
this SINGS sub-sample shows the same range in properties
such as morphologies, stellar masses, and absolute range of SFRs and
SSFRs as the HCG sample.
What appears to be different is the distribution of SFRs and
SSFRs, which, unlike in HCGs, appears to be continuous.  The general
level of star-formation also appears to be consistent with that in
other environments. In the \citet{verdes1998} HCG sample, FIR and
H$_2$ emission in HCG galaxies is comparable to that in isolated,
Virgo cluster as well as weakly interacting
systems. 

\citet{walker2009arxiv} compare MIR color distributions for
the JG sample to a set of diverse environments, including core
and infall regions of the Coma
cluster, interacting galaxies, as well
the combined Local Volume Legacy \citep{dale2009} and SINGS
samples.  They find evidence for another gap
in MIR color space in the Coma cluster \lq\lq infall\rq\rq\ region.
No gaps are found in any of the other environments.
This is further evidence that a
high-density, low velocity-dispersion
environment clearly plays a key role in accelerating
star-formation processes.

Finally, a note regarding the evolutionary hypothesis.
In this paper we present several lines of evidence for
an evolutionary sequence from gas-rich groups, mostly containing
S/I galaxies, to gas-poor groups, mostly containing E/S0 systems.
However, there is no independent evidence 
excluding the possibility that
observed gas-rich groups with S/I galaxies are {\it not} the
direct precursors of observed gas-poor groups with E/S0 systems.
In other words, it is still possible that gas-poor and
gas-rich groups are two distinct classes of compact groups.
The characteristics of intermediate gas-richness groups mitigate this possibility
but it cannot be excluded altogether.

\subsection{Caveats}\label{subsec_caveats}
\subsubsection{The origin of UV emission}\label{subsubsec_uv}
In interpreting our SFR results, we are
making two simplifying assumptions. The first is that
UV emission traces exclusively active star formation,
i.e. emission from young, massive stars.
In reality, low-mass He-burning stars which
undergo a post-asymptotic giant branch (AGB) phase
also emit in the UV. In star-forming galaxies, this
emission will make up for a small fraction of the
total UV emission, which is dominated by
massive OB stars. However, in quiescent galaxies
(E/S0) this is not necessarily the case.
Such stars are known to be responsible for a
\lq\lq UV excess\rq\rq\ (UVX) or \lq\lq UV-upturn\rq\rq,
sometimes observed in such systems
\citep{dorman1995,oconnell1999}.
However, it is only shortward of $\la 2000$\AA\
that UV elliptical-galaxy SEDs are starting to become
affected by the UV upturn.
As the \wtwo\ effective wavelength is at 2030\AA,
it is possible that the integrated flux
from this filter will at least not be dominated by any
UV upturn in ellipticals. Further,
UV emission from ellipticals can indeed
be due to
residual star formation. 
This is well-known for at least 
a number of nearby systems
\citep{oconnell1999}.
Usually these are galaxies that have undergone
recent interactions or mergers.
The HCG
environment is similarly highly interaction-prone. 
\citet{verdes1998} detect
CO emission in a number of HCG ellipticals.
This signifies the existence of significant 
reservoirs of cold gas in these
systems which they attribute
to the recent merger of a gas-rich companion.
In the HCG environment, gas transfer, either
via mergers between gas-rich and gas-poor
galaxies or via continuous infall from the intragroup medium,
can well be an on-going process that fuels star-formation.
Given that the evidence suggests a rapid morphological
evolution for HCG galaxies, it would not be
surprising that ellipticals show signs of some on-going star-formation.
In any case, 
correcting for the UV emission from
old stars will reduce the true SFR and SSFR values
at the low end of the observed SSFR gap, so that
this will be even larger.

Our second assumption is that 
we may neglect any AGN contribution
to the \wtwo\ and 24\micron\ emission used
to obtain SFR estimates. This is
reasonable for our sample which contains
no known Seyfert-1
types. \citet{coziol1998a}
also find that as much as 50\% of AGN detected
in their HCG sample are faint low-luminosity AGN
(LLAGN), either Seyfert-2 or low-ionization nuclear emission
regions (LINERs), hidden by a strong stellar continuum.
In our sample the majority of known active galactic nuclei are
classified as Seyfert-2 or LINER (\tr{tab-sfr}, column 3).
There are a few such cases among both actively star-forming
and quiescent HCG galaxies (top and bottom part of
\tr{tab-sfr}, respectively). 
In the case of star-forming galaxies these are not likely
to dominate emission at UV or IR wavelengths. In the case
of quiescent galaxies, an AGN contribution would 
lead to an overestimate of emission due to
low-level star-formation.
All of our quiescent galaxies
have \airac~$> 0$, suggesting reduced contributions
from PAH emission and hot dust in the MIR band.
MIR quiescence has been associated with the presence of LLAGN
\citep{roche1991,smithj2007}.
However, as in the case of contribution to the
UV emission from old stars, correcting for
this effect would lead
to reduced SFR and SSFR values, making the
observed SSFR gap larger. 

\subsubsection{SFR estimates}\label{subsubsec_sfr}
The errors for SFR and SSFR values
presented in \tr{tab-sfr} reflect
uncertainties in the \wtwo\ and
24\micron\ photometry and flux conversion factors. In addition,
the \wtwo\ to \sfruv\ calibration
(\er{equ_kenni}) includes an uncertainty of at least
0.3 dex in the calculated \sfruv, depending
on which stellar population synthesis models
are used for deriving it
\citep{kennicutt1998}.

The 24\micron\ to \sfrir\ calibration is most
appropriate for 24\micron\ bright galaxies.
According to the estimates of R09,
for galaxies with 
$\nu$ \lnutf\ $> 10^{42.36}$ \lunits, this
calibration includes a small correction (2.5\%) for
UV photons not re-emitted in the IR, which are
thus detectable in the UV. Some of
our brightest galaxies fall in this
luminosity regime, so there is a possibility
that \sfrtot\ may be overestimated for these systems. 
However, the actual level of UV leakage for
specific galaxies is difficult to assess, as it depends as much
on luminosity as it does on geometry.
This correction is also within the 10\%\ uncertainty margin
introduced by the conversion of 24\micron\ luminosity
to total IR luminosity  
(Equations 4 and 5, R09).
We thus attempt no further corrections for this
effect.

Further, the R09 calibration
uses an IMF which leads to \sfrtf\ values $\sim 0.66$ times
those obtained with a standard Salpeter IMF, assumed by
Kennicutt's \sfruv\ formula. The choice of the most
appropriate IMF is not
a settled issue, and, given the
wide use of the Salpeter IMF, we choose to use the R09 
24\micron\ calibration together with the \citet{kennicutt1998} 2000\AA\ calibration,
with no further modifications, in spite of the IMF discrepancy.
We have tested that, if the R09 calibration is adjusted
so that this disagreement disappears,
on average \sfrtot\ values increase by $\sim 0.7$ \msuny, and
SSFR values by \ten{17}{-11} yr$^{-1}$. The SSFR gap undergoes
a slight increase in width to \ten{9.4}{-11} yr$^{-1}$.

\subsubsection{HCG triplets}\label{subsubsec_triplets}
As already mentioned, there are claims that HCGs with only three
accordant members may need to be treated separately. 
We have checked that this issue is irrelevant for the present 
paper. There are 4
triplets in our sample, containing 12 out of the total 41 galaxies.
However, these appear to be randomly distributed in \airac, SFR and
SSFR space, so that excluding them leaves the main results of this paper
completely unaffected. In particular, there is no change in the
magnitude both of the \airac\ and of the SSFR gap.

\subsubsection{\hone\ richness}\label{subsubsec_richness}
There is no particular physical reason for the 
log \mhi/log $M_{\rm dyn}$ boundaries selected to classify
HCGs into \hone-richness types (\scr{subsec_sample}). 
These boundaries were chosen arbitrarily by J07,
to obtain roughly equal numbers of galaxies in each
richness category. A different choice would lead to
somewhat different versions for all of our figures which
show galaxies according to parent-group \hone-richness,
but the overall picture does not change.
Additionally, our definition of group \hone-richness does
not explicitly take into account member galaxy morphology.
This is particularly problematic for groups which have
no S/I galaxies, but there are only 2 such groups in our
sample.
However, in spite of this arbitrariness,
it is clear from Figures~\ref{fig:ssfr_mass} and \ref{fig:ssfr_morpho_hi}
that \hone-richness, as defined here, does track star-formation activity
and gas depletion.

\section{Summary and Future Prospects}\label{sec_final}
We have presented the UV data for 41 galaxies from the JG HCG
bright-galaxy sample. 
We combined \swift\ UVOT \wtwo\ (2000\AA) photometry and
\spitzer\ MIPS 24\micron\ photometry to obtain
SFR and SSFR estimates.

These are the main results of this paper:
\begin{enumerate}
\item \lnuwtwo\ and \lnutf\ are significantly correlated
up to \lnutf~$\sim 10^{30}$\lnuunits, where dust extinction
of UV emission becomes important. 

\item When SFRs are normalized by stellar mass to calculate SSFRs,
the HCG galaxies have a clear bimodal distribution that indicates
galaxies are either actively star-forming 
(SSFR $\ga$~\ten{1.2}{-10} yr$^{-1}$) or almost entirely
quiescent ($\la$~\ten{3.2}{-11} yr$^{-1}$). 
From previous work, the index \airac, a measure
of the strength of dust emission from 4.5--8\micron,
is also known to be
strongly bimodal. SSFR correlates significantly with \airac\ 
for the HCG galaxies, i.e., the galaxies with the reddest
MIR SEDs have the highest SSFR.

\item The bimodality in SSFR is mirrored closely by galaxy morphology.
All elliptical/S0 galaxies have low SSFR values, and
22 out of 24 spiral/irregular galaxies have high SSFR values.

\item The bimodality in SSFR is also mirrored by the \hone-gas 
richness of a galaxy's parent-group. 12 out of 15 galaxies belonging to groups
with high levels of \hone-gas have high SSFR values, and
11 out of 12 galaxies belonging to groups with
little \hone-gas have low SSFR values.
Galaxies in HCGs with intermediate amounts of \hone-gas span
almost the entire range of calculated SSFR values. 

\item When compared to a sub-sample of SINGS galaxies selected to
be non-interacting, isolated and to match HCG galaxies in near-IR luminosity,
HCG galaxies have the same range of SFRs and SSFRs. However,
the SINGS sub-sample galaxies do not demonstrate the same bimodality
in either \airac\ or SSFR. This difference
is interpreted as a consequence of the compact group
environment accelerating galaxy evolution by
enhancing star-formation and leading to rapid gas
depletion, followed by a quick transition to
quiescence.
\end{enumerate}

In the near future, 
better understanding of the HCG environment will come by 
combining information from several wavelength regions.

A major step 
is to obtain broad-band galaxy SEDs ranging from the UV to the IR.
We have completed a ground-based imaging campaign 
with the aim of carrying out multi-filter ($B$, $V$, $R$, $I$) 
photometry for the JG sample.
We have also complete NIR wide-field imaging with WIRCam
at the Canada-France-Hawaii telescope.

The \swift, \spitzer, WIRCam and
ground-based photometry provides
coverage from $\sim 2000$\AA\ to 24\micron.
SED fitting will provide independent and more complete
estimates of SFRs, SSFRs and dust attenuation.
UV-optical photometry can also better probe
the UV upturn in early-type galaxies
\citep[e.g.][]{schawinski2007}.

From what we know at present,
AGNs are not prevalent in this sample.
Seven of the HCGs in our sample have been
observed with \chandra\ (P.I. Gallagher as well as
archival observations), and
these data will allow us to better explore this issue.
At the same time we will investigate
the \hone--\x\ anticorrelation 
via the detection (or otherwise) of
diffuse \x\ emission. 
Further, SFRs and \x\ luminosities
will also be used to investigate the
\x-SFR correlation, which has been established in the
field, but has not yet been investigated in the HCG
environment.

An increase in galaxy sample size is desirable for 
improving statistics. We are engaged in a comprehensive
optical spectroscopic campaign with the aim of detecting new,
lower luminosity members of HCGs. 
It will be instructive
to re-examine the SFR and SSFR results when
more HCG member galaxies can be included.

\acknowledgments

We thank the anonymous referee for useful comments that helped to
improve the paper.  We would also like to thank Harry Ferguson, Derek
Hammer, Stephen Holland, Leigh Jenkins, Wayne Landsman and Duilia de
Mello for useful discussions and suggestions.  S.~C.~G. acknowledges
support from the National Science and Engineering Research Council of
Canada. K.E.J. gratefully acknowledges support for this paper provided
by NSF through CAREER award 0548103 and the David and Lucile Packard
Foundation through a Packard Fellowship.

{\it Facilities:} \facility{GALEX}, \facility{Swift},
\facility{2MASS}, \facility{Spitzer}

\vspace{3cm}

\bibliographystyle{apj} 
\bibliography{hsfr}

\end{document}